\documentclass[twocolumn, 10pt]{aastex63}
\usepackage{mathtools}
\usepackage{amsmath}
\usepackage{amsfonts}
\usepackage{amsthm}
\usepackage{CJK}
\usepackage[figuresright]{rotating}
\usepackage{color}
\usepackage{lineno}

\newcommand{\uat}[2]{\href{http://astrothesaurus.org/uat/#2}{#1 (#2)}}
\received{XXXX}
\revised{XXXX}
\accepted{XXXX}
\submitjournal{ApJ}

\shorttitle{UDG Detection through GCs with Log-Gaussian Cox Process}
\shortauthors{Li et al.}
\graphicspath{{./}{figures/}}

\begin{document}
\begin{CJK*}{UTF8}{gbsn}

\title{Light from the Darkness: Detecting Ultra-Diffuse Galaxies in the Perseus Cluster through Over-densities of Globular Clusters with a Log-Gaussian Cox Process}

\correspondingauthor{Dayi Li}
\email{dayi.li@mail.utoronto.ca}

\author[0000-0002-5478-3966]{Dayi (David) Li (李大一)}
\affiliation{Department of Statistical Sciences, University of Toronto, Ontario Power Building,
700 University Avenue, 9th Floor,
Toronto, ON M5G 1Z5,
Canada}

\author[0000-0003-3734-8177]{Gwendolyn M. Eadie}
\affiliation{Department of Statistical Sciences, University of Toronto, Ontario Power Building,
700 University Avenue, 9th Floor,
Toronto, ON M5G 1Z5,
Canada}
\affiliation{David A. Dunlap Department of Astronomy and Astrophysics, University of Toronto, 50 St George St,
Toronto, ON M5S 3H4,
Canada}

\author[0000-0002-4542-921X]{Roberto Abraham}
\affiliation{David A. Dunlap Department of Astronomy and Astrophysics, University of Toronto, 50 St George St,
Toronto, ON M5S 3H4,
Canada}
\affiliation{Dunlap Institute for Astronomy and Astrophysics, University of Toronto, 50 St George St,
Toronto, ON M5S 3H4,
Canada}

\author[0000-0003-2541-3744]{Patrick E. Brown}
\affiliation{Department of Statistical Sciences, University of Toronto, Ontario Power Building,
700 University Avenue, 9th Floor,
Toronto, ON M5G 1Z5,
Canada}
\affiliation{Center for Global Health Research, St. Michael's Hospital, 30 Bond St, Toronto, ON M5B 1W8
Canada}

\author[0000-0001-8762-5772]{William E. Harris}
\affiliation{Department of Physics and Astronomy, McMaster University, Hamilton, ON L8S 4M1,
Canada}

\author[0000-0003-0327-3322]{Steven R. Janssens}
\affiliation{David A. Dunlap Department of Astronomy and Astrophysics, University of Toronto, 50 St George St,
Toronto, ON M5S 3H4,
Canada}
\affiliation{Centre for Astrophysics and Supercomputing, Swinburne University, Hawthorn VIC 3122, Australia}

\author[0000-0003-2473-0369]{Aaron J. Romanowsky}
\affiliation{Department of Physics \& Astronomy, San Jos\'{e} State University, One Washington Square,
San Jos´e CA 95192,
USA}
\affiliation{University of California Observatories,
1156 High Street,
Santa Cruz CA 95064,
USA}
\affiliation{Department of Astronomy and Astrophysics, University of California Santa Cruz, 1156 High Street, Santa Cruz, CA 95064, USA}

\author[0000-0002-8282-9888]{Pieter van Dokkum}
\affiliation{Department of Astronomy, Yale University, New Haven, CT 06511, USA}

\author[0000-0002-1841-2252]{Shany Danieli}
\altaffiliation{NASA Hubble Fellow}
\affiliation{Department of Astrophysical Sciences, 4 Ivy Lane, Princeton University, Princeton, NJ 08544, USA}



\begin{abstract}

We introduce a new method for detecting ultra-diffuse galaxies by searching for over-densities in intergalactic globular cluster populations. Our approach is based on an application of the log-Gaussian Cox process, which is a commonly used model in the spatial statistics literature but rarely used in astronomy. This method is applied to the globular cluster data obtained from the PIPER survey, a \textit{Hubble Space Telescope} imaging program targeting the Perseus cluster. We successfully detect all confirmed ultra-diffuse galaxies with known globular cluster populations in the survey. We also identify a potential galaxy that has no detected diffuse stellar content. Preliminary analysis shows that it is unlikely to be merely an accidental clump of globular clusters or other objects. If confirmed, this system would be the first of its kind. Simulations are used to assess how the physical parameters of the globular cluster systems within ultra-diffuse galaxies affect their detectability using our method. We quantify the correlation of the detection probability with the total number of globular
clusters in the galaxy and the anti-correlation with increasing half-number radius of the globular cluster system. The S\'{e}rsic index of the globular cluster distribution has little impact on detectability.
\end{abstract}



\section{Introduction} \label{sec:intro}

Ultra-diffuse galaxies (UDGs) are a class of extended low-surface brightness systems first defined in \cite{VanDokkum2015} using data from the Dragonfly Telephoto Array \citep{Abraham2014}. These authors noted the existence of a large number of extended low-surface brightness galaxies in the Coma cluster, and they proposed a definition based on half-light radius ($R_e > 1.5$~kpc), and $g$-band central surface brightness ($\mu_{0,g} > 24.5$ mag arcsec$^{-2}$).  Subsequent papers have identified thousands of UDGs in rich clusters  \citep[e.g.][]{Yagi2016,Wittmann2017,Janssens2019, Lim2020} but also in smaller groups and sparser environments \citep[e.g.,][]{Martinez-Delgado2016,Roman+2019,Forbes+2019,Forbes2020,Danieli+2020}. This progress has led to a growing appreciation of the high abundance of UDGs, and also to the intriguing discovery that these objects exhibit a rich variety in their dark matter properties. For example, the iconic UDG Dragonfly~44 was found to be dominated by dark matter \citep{VanDokkum2015, VanDokkum2019}, while the qualitatively similar NGC 1052-DF2 shows little evidence for any dark matter \citep{VanDokkum2018, Danieli2019, Shen2021}.

Investigations of their globular clusters (GCs) have added to the mystery.
While many UDGs have normal GC populations consistent with dwarfs \citep{Amorisco2018}, several UDGs in Coma exhibit an excess in their GC populations \citep{vanDokkum2016,vanDokkum2017,Lim2018,Forbes+2020} --- though this is under debate \citep[e.g.][]{Amorisco2018,Saifollahi2022}.
Several other UDGs host extremely peculiar GC systems.
The dark matter free UDGs NGC~1052-DF2 and DF4 have a large population of overluminous GCs \citep{VanDokkum2018a, Shen2021}, which may be the result of these galaxies forming from a high-speed galactic collision \citep{vanDokkum2022}.
Meanwhile, NGC~5846-UDG1 is the only known galaxy to date with a stellar mass fraction in GCs exceeding 10\%, and this fraction could plausibly have been nearly 100\% at formation---a pure GC galaxy---where the present day stellar body is composed of disrupted GCs \citep{Danieli2022}.

Some UDGs have been known for decades \citep[e.g.,][]{Binggeli1984}, but substantial recent progress has largely been driven by new approaches to detecting low surface brightness galaxies, which have revealed the existence of huge numbers of these objects. It is by no means clear that the current detection process has hit its limit. In fact, there are strong suggestions that selection biases continue to play an important role in defining samples of UDGs. For example, the majority of UDGs currently identified are in distant galaxy clusters, with the nearest examples being at distances around that of the Virgo cluster ($d = 16.5$~Mpc, \citealt{Mei_2007}). Finding a very nearby UDG would be highly significant, since a dark matter-dominated UDG at $\lesssim 5$ Mpc would be an ideal testbed for probing dark matter models, most notably ultra-light axionic (`fuzzy') dark matter models \citep{Hu2000, Walker2011, Hui2017} which predict the existence of very compact soliton cores in massive ultra-diffuse galaxies \citep{Wasserman2019, Burkert2020}.
    
UDGs have an extended, faint, and diffuse appearance. For this reason, a challenge to finding UDGs in the nearby Universe is distinguishing them from imaging systematics and artefacts such as internal reflections and flat-fielding imperfections. Another issue is distinguishing them from small patches of Galactic cirrus, which are quite common in wide-field surveys \citep{Wittmann2017, Greco2018}. This challenge is most significant at distances from a few Mpc out to 30 Mpc or so. At larger distances, individual UDGs are small enough that they no longer resemble most types of cirrus, while at distances less than a few Mpc (just beyond the Local Group), the stellar content of UDGs becomes resolved in ground-based data. At such distances, resolved star-counts are probably the most effective tool for finding UDGs. 

There is also no physical theory requiring a UDG --- or galaxy in general --- to have a faint, diffuse appearance. While UDGs could represent the low surface brightness end of the normal galaxy population, they might instead represent the {\em high} surface-brightness end of an even more diffuse population --- the most diffuse of which are composed almost entirely of dark matter except perhaps for a population of a few old stars including globular clusters. The total number or mass of the GCs appears to be correlated almost linearly with the dark matter content of galactic halos \citep[e.g.][]{Blakeslee1997,Spitler_Forbes2009,Harris+2017,Forbes2020, Gannon2021}. Although the linear relation is not well-suited for low-mass galaxies, it does not exclude the existence of such an ultra-low-luminosity system. Such systems have not yet been discovered, though \cite{Peng2016}  explored the idea in the context of the Coma cluster UDG Dragonfly 17. Clearly, such `pure' dark matter halos would be nearly impossible to find using conventional approaches.
    
In this paper, we introduce a new method for finding GC-rich UDGs.
The idea is simple --- if a UDG has an associated population of GCs, these GCs should appear clustered compared to a random GC population from its surrounding intergalactic region. Therefore, if we detect a clustering signal of GCs, we can search for the presence of diffuse stellar light at the same location.  A `dark' UDG or one with extremely low stellar light might be found in this way, assuming of course that such objects would have GCs present. Such a discovery would have important implications for the empirical relation between GC system mass and galaxy stellar mass \citep{eadie+2021}. If a UDG has no diffuse component and no associated population of globular clusters, it is undetectable via starlight (though other approaches, such as using gravitational lensing, or searching for diffuse line emission from cooling gas, may succeed) \footnote{Although one could argue if such an object can even be classified as a UDG since it is a pure dark matter halo with some gas content.}.

We describe our approach using concepts and methods from spatial statistics, using the formalism of \emph{spatial point process theory}. Several of our key ideas are widely used in other subject areas (most notably ecology, geology, and epidemiology). We will adopt the common formalism used in these subject areas. While point processes are ubiquitous in astronomy, the terminology used is specific to astronomy and unfamiliar to most working statisticians, which limits cross-talk between disciplines. In this paper we strive to map several of the key ideas of spatial statistics into an astronomical context. At the same time, we aim to remind statisticians that some key ideas in spatial statistics originated in astronomy, before evolving to become widely used in other disciplines. A good example of this is the Neyman--Scott process \citep{Neyman1958} originally developed to better understand the clustering of galaxies. 

Our approach follows in the footsteps of other papers which apply spatial point process models to astronomy. For example, \cite{Tempel2016} developed the Bisous model for detecting galaxy filaments, and \cite{Li2021} employed a Gibbs point process model to study the formation of young star clusters in M33. 

In the present paper, we focus on a class of point process models called the log-Gaussian Cox process (LGCP) and apply it to UDG detection within the Perseus cluster of galaxies. Interestingly, the LGCP also has its roots in astronomy, where the closely related log-Gaussian random field is used to model the distribution of matter in the Universe by \cite{Coles1991}, who notably provided some examples on simulating galaxy distribution arising from a LGCP. The theoretical properties and applications of LGCP were developed and refined later in the lens of point process theory by \cite{Moller1998}, and the method has since garnered a wide range of applications in ecology \citep{Raynaud2014}, epidemiology \citep{Li2012}, neuroscience \citep{Samartsidis2019}, environmental science \citep{Serra2013}, etc. But the use of the LGCP in astronomy seems to have remained restricted to investigations of the large-scale structure of the Universe. Here, we attempt to broaden the application of the LGCP in astronomy, and describe developments in the LGCP made by statisticians working in other areas over the past two decades.

The outline of this paper is as follows. In Section \ref{sec:data} we describe the data used in our study. Section \ref{sec:method} contains a general introduction to the LGCP methodology and the associated techniques for cluster detection. Details on model construction for our specific problem are also included. Section \ref{sec:result} presents our results on UDG detection using the LGCP and Section \ref{sec:CDG1} contains a preliminary analysis on a potential pure GC galaxy detected using our method. Section \ref{sec:simulation} describes a set of simulations used to test the performance and reliability of our method. Section \ref{sec:conclusion} offers some concluding remarks and describes our plan for future research.

\section{Data}\label{sec:data}

The data used in our study come from the Program for Imaging of the PERseus cluster (PIPER; \citealt{Harris2020}). The images for this survey were obtained with the \textit{Hubble Space Telescope} (\textit{HST}) Advanced Camera for Surveys ACS/WFC and WFC3/UVIS cameras,
under program GO-15235. The target region is the rich Perseus cluster (Abell 426) at $d = 75$ Mpc \citep{Gudehus1995, Hudson1997}. A series of fields, including many positioned through the Intracluster Medium (ICM) away from major galaxies, was imaged in the F475W and F814W filters for ACS, and F475X and F814W for WFC3. The data used in our study consist of 10 pairs of pointings (with the WFC3 camera as the Primary and the ACS as the Parallel camera) of these ICM regions; they avoid any of the most giant Perseus members but include many less massive giant early-type galaxies (ETGs), numerous dwarfs, and previously known UDGs \citep[see Figure 1 in][]{Harris2020}.

GCs in these pointings were extracted for analysis. As described in \cite{Harris2020}, Perseus is distant enough that its GCs are virtually all unresolved (star-like) and so a relatively clean list of candidates can be constructed by rigorous elimination of non-stellar objects.  For analysis in this paper, we select the GC candidates from this catalog of star-like objects based on the following additional criteria for magnitude and color range, which include the GC population:
\[
22.0 < \text{F814W}_0 < 25.5 \ \text{and}
\]
\[
1.0  < (\text{F475W} - \text{F814W})_0 < 2.4,
\]
where the subscript $0$ denotes the de-reddened values of the magnitudes with foreground Galactic extinction removed \citep[see][]{Harris2020}. For comparison, the turnover or peak point of the normal globular cluster luminosity function (GCLF) at the distance modulus of Perseus ($(m-M)_I = 34.62$) would be at an apparent magnitude F814W$_{to} \simeq 26.2$, about 0.7 mag fainter than the imposed selection limit above \citep{Harris2020}. Our search for GC populations around UDGs is therefore biased towards using fairly luminous GCs.

Note also that the GC candidates selected will contain some fraction of contaminating objects that are either foreground stars or faint, compact background galaxies. We do not expect these to strongly influence our results as the number of contaminants should be fairly low.  In the selected color and magnitude ranges used here, \citet{Harris2020} show that the contaminants will make up no more than about 20\% of the total (see their Figures 7, 10, and 14 particularly and the accompanying discussion).

As a final note on the photometry, the WFC3 F475X band is slightly different from ACS F475W, so a transformation is applied to the WFC3 magnitudes to obtain the equivalent magnitude in ACS F475W \citep[see][]{Harris2020}.  All our final data are then based on the ACS (F475W, F814W) Vegamag system.

Aside from the PIPER Survey, for cross-checking our detected UDGs we also utilize the low-surface brightness galaxy catalog from \cite{Wittmann2017} obtained from the Prime Focus Imaging Platform at the William Herschel Telescope through the Opticon programme, as well as a more recent unpublished UDG catalog obtained by A.~Romanowsky.  The latter was based on archival Canada--France--Hawaii Telescope (CFHT)/MegaCam imaging of the Perseus cluster, and was constructed for the PIPER survey \textit{HST} proposal.  This has been superseded by much better data from Subaru Hyper Suprime-Cam (see initial results in \citealt{Gannon2021}). From here on, we refer to the UDGs from \cite{Wittmann2017} as WUDGs and UDGs from Romanowsky as RUDGs. 
The WUDG catalog contains a total of 89 UDG candidates, of which 31 fall within the PIPER survey fields. The RUDG catalog contains 148 identified UDG candidates but only 10 of them are within the PIPER fields.

\section{Methodology: Log-Gaussian Cox Process} \label{sec:method}

Our goal is to introduce and apply the tools and formalism of spatial statistics, specifically LGCP, to extract information about the spatial distribution of GCs. Subsequently, we attempt to detect UDGs through the clustering signals of their associated GCs.

Before we introduce the basics of point processes, it is noteworthy that many concepts within spatial statistics are in fact already widely used in astronomy. However, due to the isolated development between the two fields in the past decades, the terminologies of the same concepts have diverged. We could translate the terminologies in spatial statistics back to their corresponding ones in astronomy, but we believe it is best to keep the terminologies used in spatial statistics so that interested readers from astronomy can search the spatial statistics literature more easily for other novel methods. We will instead provide the corresponding astronomy concepts for terminologies in spatial statistics if we deem necessary.

We begin by specifying the basic terminology and notation for point process theory. A \emph{random point process} is denoted by $\mathbf{X}$, while $\mathbf{x}$ denotes a configuration/realization of $\mathbf{X}$. $S \subset \mathbb{R}^d$ is the region or observation window on which $\mathbf{X}$ is defined and $s \in S$ denotes a location in the region. Here, we only focus on the process on the plane of the sky, so $d = 2$. 

The most basic statistical model for point processes is the Poisson point process (PPP), which represents \textit{complete spatial randomness}. It is solely determined by its intensity function $\lambda(s) (> 0)$ which represents the expected number of points per unit area at location $s$. In astronomy, the intensity function is often the mean surface number density or the mean point count per unit area. We include a formal definition of PPP and some of its defining properties in Appendix \ref{sec:ppp}.

Depending on the intensity function, a PPP can be classified as a \textit{homogeneous} Poisson process (HPP) or an \textit{inhomogeneous} Poisson process (IPP). An HPP has a constant intensity function everywhere, i.e., $\lambda(s) \equiv \lambda > 0$. In astronomy, an HPP is essentially a point pattern with points drawn from a uniform distribution. An IPP, on the other hand, has an intensity function that depends on the location $s$: the mean surface number density is different at different locations, but the occurrence of a point is independent from any other point.

Building on a PPP, a more flexible class of models is called log-Gaussian Cox process (LGCP; \citealt{Moller1998}). A LGCP is a type of doubly-stochastic (Cox) process that possesses a hierarchical structure. On the first (progenitor) level of the hierarchy, we have a log-Gaussian process (log-Gaussian random field) which generates an intensity function, while at the second level, the point process is assumed to be an IPP with intensity function being the one previously generated by the log-Gaussian process. 

An example of the LGCP can be illustrated through the galaxy distribution discussed in \cite{Coles1991}, where at the first level of the hierarchy, the large-scale matter (baryonic and dark matter) distribution in the Universe is modelled as a log-Gaussian random field. After a realization of such a random field is obtained, the second level of the hierarchy is then an IPP with intensity function being the realized log-Gaussian random field and locations of galaxies will then ``spawn" according to the IPP. With a concrete depiction of the LGCP in mind, we now introduce the mathematical formalism of the LGCP.

\subsection{Model Specification}

We start from an IPP $\mathbf{X}$ with intensity function $\lambda(s) \geq 0$, where $s \in S \subset \mathbb{R}^2$. Let $\mathbf{x} = \{x_1, \dots, x_N\} \subset S$ denote a realization of $\mathbf{X}$, and $N$ denote the number of points in $\mathbf{x}$. Then the probability density function of $\mathbf{X}$ w.r.t. a unit rate homogeneous Poisson process, i.e., $\lambda_{\mathrm{unit}}(s) \equiv 1$, is
\begin{equation}
\label{ppdensity}
    \pi(\mathbf{x}|\lambda(s)) = \exp\left(\int_S1 - \lambda(s)ds\right)\prod_{i=1}^N\lambda(x_i).
\end{equation}
If we further assume that the intensity function $\lambda(s)$ arises from a Gaussian process $\mathcal{U}(s)$ such that $\lambda(s) = \exp(\mathcal{U}(s))$, then $\mathbf{X}$ is called a log-Gaussian Cox process.

A Gaussian process (GP) or a Gaussian random field $\mathcal{U}(s)$ is a collection of random variables indexed by time or space such that any finite collection of these random variables is multivariate normally distributed. A GP is uniquely determined by its mean function $\mu(s)$ and its covariance function $\text{Cov}(\mathcal{U}(s), \mathcal{U}(t))$. If we further assume the process is isotropic and stationary, then the covariance function can be written as $\text{Cov}(\mathcal{U}(s), \mathcal{U}(t)) = \sigma^2\mathbf{C}(|s - t|)$, where $\sigma^2$ is the marginal variance and $\mathbf{C}(|s-t|)$ is the correlation between $\mathcal{U}(s)$ and $\mathcal{U}(t)$ as a function of the Euclidean distance $|s-t|$.

For a more succinct representation, the model can be written as
\begin{equation}
    \begin{aligned}\label{lgcp}
    \mathbf{X} &\sim \text{IPP}(\lambda(s)), \\
    \log(\lambda(s)) &= \boldsymbol{X}(s)'\boldsymbol{\beta} + \mathcal{U}(s),\\
    \mathcal{U}(s) &\sim \mathcal{GP}(\mathbf{0}, \sigma^2\mathbf{C}(\cdot)),\\
    \text{Cov}(\mathcal{U}(s), \mathcal{U}(t)) &= \sigma^2\mathbf{C}(|s-t|),\ s,t \in S.
\end{aligned}
\end{equation}
Note that $\boldsymbol{X}(s)$ is a row vector for some underlying covariates at location $s$, whose value is available at any location in the observation window. $\boldsymbol{\beta}$ is the vector of covariates' coefficients. $\boldsymbol{X}(s)'\boldsymbol{\beta}$ is the fixed effect that captures the large-scale spatial variation while $\mathcal{U}(s)$ is set to a zero mean GP to capture the spatial random (residual) effect. 

For the correlation function, the most widely used is the  Mat\'{e}rn correlation function: 
\begin{equation}
    \mathcal{M}_\nu(|s - t|) = \frac{|s-t|^\nu}{\Gamma(\nu)2^{\nu-1}h^\nu}K_\nu\left(\frac{|s-t|}{h}\right),
\end{equation}
where $h > 0$ is a range parameter that controls the scale of the correlation, $\nu > 0$ is the shape parameter that controls the smoothness of the GP, and $K_\nu$ is the modified Bessel function of the second kind. In most practical applications, $\nu$ is usually assumed to be known and excluded from inference. The parameters being inferred in the model are then $\boldsymbol{\beta}, \sigma,$ and $h$.

To facilitate inference, a traditional approach\footnote{Note that this is not the inference method we use, but due to its simplistic nature, we describe it here for illustrative purposes.} is to divide the spatial domain into an $n\times m$ grid with equally-sized cells, and then assume the intensity within each cell is constant. After the grid is constructed, the log-intensity within the $i$-th cell $y(c_i) = y_i$ is constant and indexed by the centroid $c_i, i \in \{1, \dots, nm\}$. Denoting $\mathbf{y} = (y_1, \dots, y_{nm})'$, by the definition of a GP, $\mathbf{y}$ then follows a multivariate normal distribution $\mathbf{y} \sim \mathcal{N}(\boldsymbol{X}(s)'\boldsymbol{\beta}, \sigma^2\mathbf{C}_{nm})$ where $\mathbf{C}_{nm}$ is the $nm\times nm$ correlation matrix with entry $\mathbf{C}(|c_i - c_j|)$. Let $\boldsymbol{\theta} = (\boldsymbol{\beta}, \sigma^2, h)$, $n_i$ the number of points in the $i$-th cell, and let $A$ denote the area of each individual cell. Under a Bayesian setting, the joint log-posterior density given the discretization rule is then
\begin{equation}\label{jposterior}
\begin{aligned}
    \log \pi(\boldsymbol{\theta}, \boldsymbol{y}|\mathbf{x}) =& \sum_{i}[y_in_i - A\exp(y_i)]\\ &- \frac{1}{2\sigma^2}(\mathbf{y} - \boldsymbol{X}'\boldsymbol{\beta})'\mathbf{C}_{nm}^{-1}(\mathbf{y} - \boldsymbol{X}'\boldsymbol{\beta})\\
    &- \frac{nm}{2}\log(\sigma^2) - \frac{1}{2}\log(\det(\mathbf{C}_{nm})) + \log \pi(\boldsymbol{\theta}),
\end{aligned}
\end{equation}
where $\pi(\boldsymbol{\theta})$ is the prior distribution of the parameters. If we are interested in inferring the parameters $\boldsymbol{\theta}$, then we will focus on the posterior marginal distribution $\pi(\boldsymbol{\theta}|\mathbf{x})$. If we are interested in the latent intensity function, then we will look at the posterior marginal distribution $\pi(\boldsymbol{y}|\mathbf{x})$ (i.e., the posterior predictive distribution).

For our purposes, GCs are divided into the following categories: (a) GCs within bright normal galaxies (including normal dwarfs), (b) GCs in the intergalactic medium (IGM), and (c) GCs within UDGs. In our modelling assumption, the spatial residual effect $\mathcal{U}(s)$ contains the latter two categories of GCs. Therefore, by studying the posterior predictive distribution of the intensity surface of the residuals $f(\mathcal{U}|\mathbf{x})$,\textit{ we can identify over-dense regions of GCs with some user-defined level of confidence, which in turn provides us with positions of UDG candidates.} 

One could argue that the LGCP method is essentially an over-elaborated variant of the kernel intensity estimator (KIE, \citealt{Diggle1985}). However, LGCP has distinct advantages over KIE for our problem such as theoretical guarantee of performance as well as a full probabilistic prediction of the latent intensity function instead of the point-wise prediction provided by KIE \citep{Diggle2013}.

Most importantly, KIE cannot incorporate covariate information. In our problem, a clear covariate effect is the presence of various bright, normal galaxies, where 11 out of the 20 pointings contain at least one of them. Since GCs are generally strongly clustered and abundant in normal galaxies, the clustering signals of GCs from these galaxies will dominate or contaminate the signals from smaller GC systems of UDGs, making UDGs undetectable. Using KIE cannot deal with the confounding signals from these normal galaxies. But for LGCP, we can easily construct covariate models, e.g. S\'{e}rsic models \citep{Sersic1963}, to eliminate the effect of GC systems from normal galaxies. Afterwards, the signals from UDGs are then captured in the residual effect.

For inference, LGCP can be fitted using traditional methods such as Markov chain Monte-Carlo (MCMC). However, due to the high-dimensional nature of the underlying GP, inference through MCMC is computationally prohibitive. Here, we adopt the state-of-the-art method called the integrated nested Laplace approximation (INLA; \citealt{Rue2009}) to conduct Bayesian inference. INLA is an alternative method to MCMC that directly approximates the posterior marginal distribution using Laplace approximation, which is orders of magnitude faster. Note that in most cases, we are only interested in the marginal distribution of the posterior instead of the usual high-dimensional full posterior. For a detailed overview of INLA, readers can refer to \cite{Rue2009}. 

The main computational tool for model inference will be the \href{https://www.r-inla.org/home}{\texttt{R-INLA}} package as well as \texttt{inlabru} \citep{inlabru} in \texttt{R} \citep{Rcore}, where \texttt{inlabru} is a much more integrated and streamlined version of \texttt{R-INLA}. In the spirit of reproducible research, we need to note that due to the automatic parallelization employed in the \texttt{R-INLA} package, different machines will produce slightly different results as the initialization values for inference are machine-dependent and out of user control. For the record, the results for the UDG detection in the Perseus cluster presented in Section \ref{sec:result} are obtained on a 64-bit Windows 10 desktop with an Intel i7-6800K 3.40GHz CPU. For the simulation analysis done in Section \ref{sec:simulation}, they are carried out using a high performance computing cluster with a Linux CentOS 7 operating system and an Intel Xeon E5-2683 v4 2.1GHz CPU. Although the results from \texttt{R-INLA} can differ for different machines, in general the differences are insignificant unless the data are ill-conditioned, which is not of concern for our problem.

Aside from employing INLA, we also utilize another state-of-the-art method for fitting LGCP models, namely approximating the latent GP as the solution to a stochastic partial differential equation (SPDE; \citealt{Lindgren2011, Simpson2015}). Due to the highly technical nature of the method, we do not over-elaborate the details here. It suffices to know that the SPDE approach is used for obtaining fast and accurate approximation of the unknown integral in Equation~\ref{ppdensity} and it is computationally superior to the aforementioned method where the spatial domain is divided into equally spaced cells. The solution of the SPDE is instead a continuous representation of the latent GP and it requires sufficiently fewer ``cells"\footnote{``Cell" here is only a nomenclature, where in fact SPDE does not construct cells but a Delaunay triangulation of the observation window based on the point pattern. The latent GP and its associated integral are then approximated based on the vertices of the triangulation using basis-function expansion.} to achieve a similar accuracy compared to when equally spaced cells are used, thereby drastically reducing the computational time. The SPDE approach is also implemented in the \texttt{R-INLA} and \texttt{inlabru}.

Note that there exists other point process model aside from LGCP that could potentially be suitable for our problem, e.g. Poisson cluster process \citep{Neyman1958, moller_2003}. LGCP is chosen here among others due to computational reasons. In general, fitting point process models is difficult since the likelihood of any point process model requires integrating the unknown intensity function in Equation~\ref{ppdensity}. The SPDE approach combined with LGCP can render the computation of this integral extremely fast. Moreover, within the context of point process models, LGCP is the only model framework that can enjoy the benefit of INLA as LGCP belongs to the class of latent Gaussian model \citep{Rue2009} for which INLA is specially designed. For our data set, LGCP combined with INLA only requires maximum of 10 minutes to run for one point pattern. For other point process models, however, the inference has to rely on MCMC and the computation can take days or hours. Since we also need to conduct a large-scale simulation in Section \ref{sec:simulation}, LGCP is the only feasible option.

\subsection{Detection through Excursion Sets and Excursion Function}\label{sec:excursion}

To detect UDGs using LGCP, we require a further method proposed by \cite{Bolin2015} to identify the so-called \textit{excursion sets} of a stochastic process, i.e., regions in the posterior intensity surface which exceed some threshold values with high probability. These excursion sets will then provide us with potential locations of UDGs. 

Formally, an excursion set with threshold $C$ and confidence level $1-\alpha$ is given as follows:
\begin{equation}
    E_{C, \alpha}(\mathcal{U}) = \arg\max_D\{|D|: P(D \subset A_C(\mathcal{U})|\mathbf{x}) \geq 1-\alpha\},
\end{equation}
where
\begin{equation}
    A_C(\mathcal{U}) = \{s: \mathcal{U}(s) > \log(C)\}.
\end{equation}
The interpretation is that $E_{C, \alpha}(\mathcal{U})$ is the set where \textit{all the points} within it exceed level $\log(C)$ with probability of at least $1-\alpha$. Note that a smaller value of $\alpha$ would indicate a higher significance level. As noted by \cite{Bolin2015}, however, $E_{C, \alpha}(\mathcal{U})$ only provides us with regions that surpass $C$ with probability $1-\alpha$. There can be many more hidden structures within $E_{C, \alpha}(\mathcal{U})$ that are left out. For example, we can set $\alpha = 0.5$, but $E_{C, 0.5}(\mathcal{U})$ can potentially contain regions such as $E_{C, 0.05}(\mathcal{U})$. Only reporting $E_{C, 0.5}(\mathcal{U})$ will not provide us information about these hidden regions. Therefore, the goal here is to obtain a description of the excursion set by something similar to that of $p$-value for each location. To address this issue, \cite{Bolin2015} proposed the following excursion function:
\begin{equation}
    F_C(s) = \sup\{1 - \alpha: s \in E_{C,\alpha}(\mathcal{U})\}.
\end{equation}
The excursion function takes a value between zero and one. A value close to zero at a given location $s$ indicates that $\mathcal{U}(s)$ will only exceed $\log(C)$ for large values of $\alpha$, hence the process is unlikely to exceed $\log(C)$ at $s$. On the other hand, a value close to one indicates that the process is very likely to exceed $\log(C)$ at $s$. We will not elaborate on the computational details of excursion sets and functions here, and instead refer the reader to \cite{Bolin2015}. The actual computations of excursion sets and excursion functions are facilitated using the \texttt{excursion} package \citep{Bolin2018} in \texttt{R}.

The last part of our methodology that needs to be addressed is the excursion threshold $C$. Intuitively, we would expect that regions where $\exp(\mathcal{U})$ exceeds high values of $C$ with high probability are likely to indicate strong clustering signals and hence potential locations of UDGs. However, the notion of high excursion threshold is not well-defined here. In contrast, in many disciplines such as environmental science and climate science, the threshold $C$ is often pre-determined. For example, one might be interested in regions where an air pollutant exceeds a certain preset threshold, or regions where the CO$_2$ level exceeds, say, $300$ PPM. These thresholds are generally determined before any study is carried out, and they possess clear, problem-specific motivation. But this is not the case for our problem since what is considered a strong clustering signal of GCs will vary for different GC point patterns. Thus, we utilize the quantile, $Q(p)$, of the posterior marginal distribution of $\mathcal{U}(s)$ as initially suggested by \cite{Diggle2005}. Using the quantiles will provide us with a principled procedure for computing excursion functions that are comparable across different GC point patterns. However, the original suggestion by \cite{Diggle2005} to compute the quantiles is rather simplistic and not suited for a Bayesian setting. The mathematical details for obtaining the posterior marginal quantiles are given in Appendix \ref{sec:quantile}. For our problem, we consider $p = 0.5, 0.75, 0.9, 0.95$ to compute the excursion functions, i.e., we consider $\log(C)$ to be the median, $75$-th, $90$-th, and $95$-th quantiles of the posterior marginal distribution of $\mathcal{U}(s)$.

\subsection{LGCP Model Form for UDG Detection}
Here we describe our LGCP models for detecting UDGs in the Perseus cluster. We denote the point pattern of GCs in each pointing by $\mathbf{X}$ and the field of view of each pointing by $S$, and we specify the following model structure based on Equation \ref{lgcp}:
\begin{equation}
\begingroup
\allowdisplaybreaks
    \begin{aligned}\label{lgcp:model}
    \mathbf{X} &\sim \text{IPP}(\lambda(s)), \\
    \log(\lambda(s)) &= \beta_0 + \sum_{k=1}^{N_G}\beta_k g(s; R_{k}, \alpha_k)1_{\{N_G \neq 0\}} + \mathcal{U}(s),\\
    \mathcal{U} &\sim \mathcal{GP}(\mathbf{0}, \sigma^2\mathcal{M}_1(\cdot; h)),\\
     \text{Cov}(\mathcal{U}(s), \mathcal{U}(t)) &= \mathcal{M}_1(|s - t|;h), \ s, t \in S.
\end{aligned}
\endgroup
\end{equation}

The model in Equation~\ref{lgcp:model} consists of two main components. First, the fixed effects comprise the intercept term $\beta_0$ to model the baseline mean intensity, and the summation term to model the covariate effects from normal galaxies within the pointing. Specifically, $N_G$ represents the number of normal galaxies in the pointing. Note that these galaxies do not need to be completely within the pointing. If a pointing includes part of these galaxies, they will be included. The coefficients $\beta_k$ represent the log-relative central surface intensity of GCs within these galaxies. $1_{\{\}}$ denotes the indicator function and if there is no normal galaxy in the pointing, then the fixed effect only contains the intercept term.

The function $g$ in Equation~\ref{lgcp:model} follows the functional form of the S\'{e}rsic model:
\begin{equation}\label{sersic}
    g(s; R_k, \alpha_k) = \exp\left(-\left(\frac{d(s, c_k)}{R_k}\right)^{\alpha_k}\right),
\end{equation}
where $c_k$ is the center of the $k$-th galaxy which is known, and $d(s, c_k)$ is the distance from $s$ to $c_k$. $R_k$ is the scale parameter for the GC system of the $k$-th galaxy and $\alpha_k$ is the inverse of the S\'{e}rsic index. For simplicity, we only follow the functional form of the S\'{e}rsic model instead of adhering to the traditional parameterization where the half-number radius and the S\'{e}rsic index are used. Moreover, we found that our parameterization and the traditional parameterization generally lead to similar results but the traditional parameterization of the S\'{e}rsic model led to higher computational cost ($20\%$ more computational time in some cases) and sometimes numerical instability. 

On another note, the model in Equation~\ref{sersic} is a simplified version of what we actually employ --- namely, we have dropped the information on position angle and ellipticity of the galaxies. However, since these two parameters can be easily measured with high accuracy, we treat them as known and do not explicitly include them in Equation~ \ref{sersic}. The ellipticities, position angles, as well as the half-light radii of normal galaxies are obtained using the same archival CFHT/MegaCam imaging of the entire Perseus cluster for the RUDG catalog. These quantities are obtained using SExtractor \citep{Bertin1996}. 

We need to point out that the position angle and the ellipticity can have a significant impact on the distribution of GC systems depending on the galaxy type. Although the exact impact varies from one galaxy to another, the general understanding is that the GC systems in ETGs tend to follow the galaxy isophotes, while GC systems are spherically distributed for disk type galaxies \citep{Harris1991,Wang+2013}. Thus, for simplicity, we adopt the previous observation by \cite{Harris1991} and \cite{Wang+2013} when constructing the GC profile structure of galaxies. 

The second component of the model in Equation~\ref{lgcp:model} is the residual effect modelled by the latent GP. Here we adopt a Mat\'{e}rn correlation function for the latent GP with the shape parameter $\nu = 1$. The choice of $\nu$ here is rather arbitrary (although under the SPDE approach, $\nu$ is required to be less than one for technical reasons), and we found that different values of $\nu$ do not seem to have any significant impact on the results.  

It needs to be mentioned that there exists another model formulation instead of Eq.~\ref{lgcp:model}: the GC point pattern can be regarded as the superposition of GCs from normal galaxies and GCs from the IGM and UDGs. This model formulation is probably more natural and used more often in an astrophysical context, e.g., when calculating background-subtracted GC counts in galaxies. Under such a formulation, the intensity function can be written as:
\begin{equation}\label{sup_PPP}
    \lambda(s) = \sum_{k=1}^{N_G}\beta_k g(s; R_{k}, \alpha_k)1_{\{N_G \neq 0\}} + \beta_0\exp(\mathcal{U}(s)),
\end{equation}
i.e., the intensity function of all GCs is the sum of intensity functions of GCs in all normal galaxies and the intensity function of GCs in IGM and UDGs. However, inference is now impossible with INLA as the model is no longer a LGCP. The best available solution is to incorporate INLA within MCMC combined with data augmentation \citep{Gomez-Rubio2017}, which is much more difficult to implement and extremely computationally expensive \footnote{A mere $5000$-iteration of INLA within MCMC will take roughly at least $17$ hours to run for one point pattern in our data. It only takes $\sim 10$ minutes under Eq.~\ref{lgcp:model}.} without significant gains in UDG detection performance. The difference between the two model structures lies in the physical interpretation of the data generating process of the GC point patterns: while both interpretations are acceptable, they are irrelevant for the purpose of our problem. Thus, Eq.~\ref{lgcp:model} is the one we adopt for the sake of feasibility.

\subsection{Prior Specification}

After model construction, we specify the prior distributions for our models. We denote all parameters by $\boldsymbol{\theta} = (\boldsymbol{\beta}, \boldsymbol{\alpha}, \mathcal{R}, \sigma, h)$ where $\boldsymbol{\beta} = (\beta_0, \dots, \beta_{N_G})$, $\boldsymbol{\alpha} = (\alpha_1, \dots, \alpha_{N_G})$, and $\mathcal{R} = (R_1, \dots, R_{N_G})$. For the prior of $\boldsymbol{\beta}$, we assign a $(N_G+1)$-dimensional standard normal distribution, i.e.,
\[
\boldsymbol{\beta} \sim \mathcal{N}(\boldsymbol{0}, 1000\times\mathbf{I}_{N_G+1}),
\]
where $\mathbf{I}_{N_G+1}$ is the $(N_G+1)$-dimensional identity matrix. For the prior distribution of the S\'{e}rsic indices $\boldsymbol{\alpha}$, since they are required to be positive, we assign an exponential distribution with rate parameter one to each individual component $\alpha_k$ of $\boldsymbol{\alpha}$ and assume that the individual components are independent. Note that the prior distributions chosen here are quite uninformative as we do not have a very good idea about the values of $\boldsymbol{\beta}$ and $\boldsymbol{\alpha}$.

For the scale parameters $\mathcal{R}$, we construct the prior distribution for individual components $R_k$ loosely based on the results from \cite{Forbes2017}. The author studied the scale ratio between the half-light radii and the half-number radii of GC systems for a sample of ETGs. It was found that on average, the half-number radii of GC system of a galaxy is roughly four times its half-light radii, and there exists a positive linear relationship between the stellar mass and the scale ratio, but the amount of scatter is quite significant. It was also noted that for low-mass ETGs, the scale ratio is roughly three times while for more massive galaxies, the scale ratio increases to an average of five times.

Due to the significant scatter in the scale ratio, we obtain a crude guess of $R_k$ by eye based on the galaxy's light profile and GC system with the observation of \cite{Forbes2017} as a rough guide. The contaminating normal galaxies in the PIPER survey can be divided into two distinctive sub-populations of giant ETGs and dwarf ellipticals (dEs). Thus, for giant ETGs, we construct a prior distribution for $R_k$ where it follows a log-normal distribution with mean equal to five times the galaxy's half-light radius with standard deviation equal to the half-light radius. The choice of a log-normal distribution is to ensure that $R_k$'s are positive. For dEs, we supply the prior distribution of $R_k$ with the same log-normal distribution but with the mean value being three times the half-light radius. For completeness, the prior distributions for $R_k$'s of all normal galaxies are provided in Appendix \ref{sec:Rk}.

We also need to mention that $R_k$ is not the half-number radius of the GC system, as the parameterization in our modelling is different from the standard S\'{e}rsic model. However, we found that the S\'{e}rsic indices of the GC systems for most of the galaxies in our data are relatively low ($0.5 \lesssim n \lesssim 1$). For a S\'{e}rsic index in the range of $0.5 \lesssim n \lesssim 1$, it can be shown that $R_k$ and the half-number radius differ less than a factor of two. Hence, to reduce the complexity of our model, we assume here that $R_k$ and the half-number radius are the same.

Lastly, for the hyperparameters $\sigma$ and $h$, we assign the penalized complexity (PC) priors based on the results from \cite{Simpson2017, Fuglstad2018}. The PC priors are currently the standard prior choice for modelling the latent GP within a LGCP. As the name suggests, PC priors penalize the complexity of the fitted model so that the resulting model possesses more predictive power. It does so by forcing the range parameter $h$ to be as large as possible while pushing the marginal standard deviation $\sigma$ to be as small as possible. Such a scenario corresponds to a Gaussian random field that is inherently smooth with almost no local fluctuation. Therefore, the model will not be heavily influenced by local spurious noise structures. Furthermore, PC priors are scale-invariant, meaning that changing the physical unit of distance measurement does not change the underlying model structures, which is highly desirable. Specifically, the hyperparameters are first reparameterized where:
\[
h = \rho/\sqrt{8\nu}, 
\]
Here, $\rho$ is the distance at which the correlation function for the GP is roughly 0.1 as per convention by \cite{Lindgren2011}. Then the PC priors for $\sigma$ and $\rho$ are specified jointly as
\begin{equation}\label{pcprior}
    f(\sigma, \rho) = \lambda_1\lambda_2\rho^{-2}\exp(-\lambda_1\rho^{-1}-\lambda_2\sigma), \ \sigma,\rho > 0.
\end{equation}
Furthermore, the parameters for the PC prior are specified implicitly by the constraints where $P(\rho < \rho_0) = a_1$ and $P(\sigma > \sigma_0) = a_2$ with $\lambda_1$ and $\lambda_2$ satisfying
\[
\lambda_1 = -\log(a_1)\rho_0,
\]
and
\[
\lambda_2 = -\frac{\log(a_2)}{\sigma_0}.
\]
The parameters $\rho_0$, $\sigma_0$, $a_1$ and $a_2$ are supplied by users.

For the prior of the range parameter $\rho$, we specify the prior parameters based on the sizes of identified UDGs within the Coma cluster obtained by \cite{VanDokkum2015}. The median of the half-light radius of the 47 UDGs they discovered within the Coma cluster are roughly 3 kpc. \cite{Forbes2017} also studied the half-number radii of the GC system of UDGs and found that the sizes of GC systems are roughly over twice the half-light radii, although their sample size is rather small (only three). Based on this information, since $\rho$ represents the distance at which the latent GP has correlation of less than 0.1, we assign $\rho_0 \approx 7$ kpc and $a_1 = 0.5$. The motivation for the choice of these parameters is that the majority of the correlation of the latent GP in the model should be caused by the presence of UDGs; at distances greater than the half-number radius of the GC systems of UDGs ($\sim 7$ kpc), the correlation of the latent GP governing the GC point patterns should be rather small. Hence, $\rho_0$ is set to $\sim 7$ kpc while this value is the median of the distribution. 

As for the variance parameter $\sigma$, there is not much prior information available to us. Hence, we set $\sigma_0 = 1.5$ and $a_2 = 0.5$ to represent a rather uninformative prior.

\section{Results}\label{sec:result}

\begin{figure*}
\centering
    \gridline{\fig{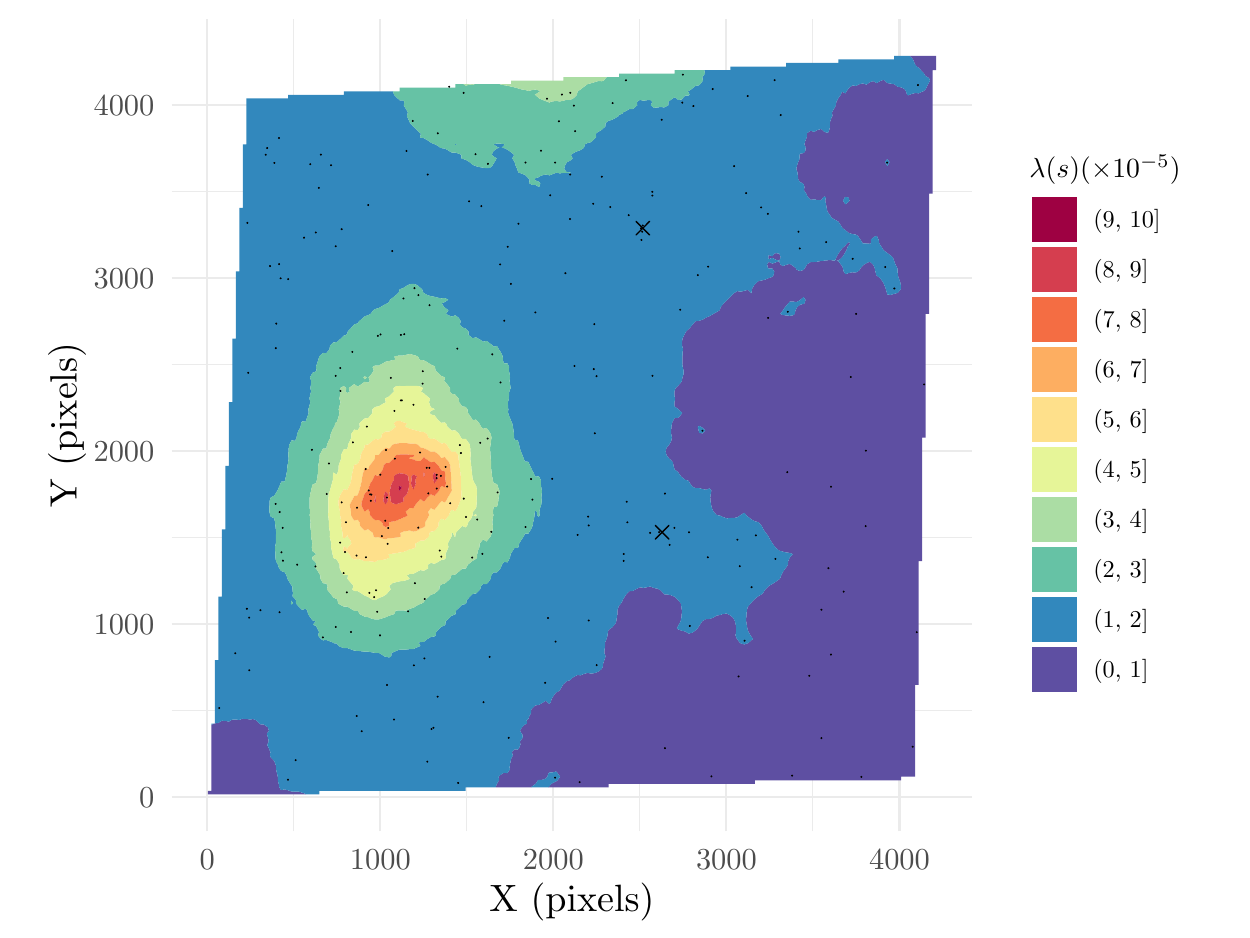}{0.51\textwidth}{(a)}}
    \gridline{\fig{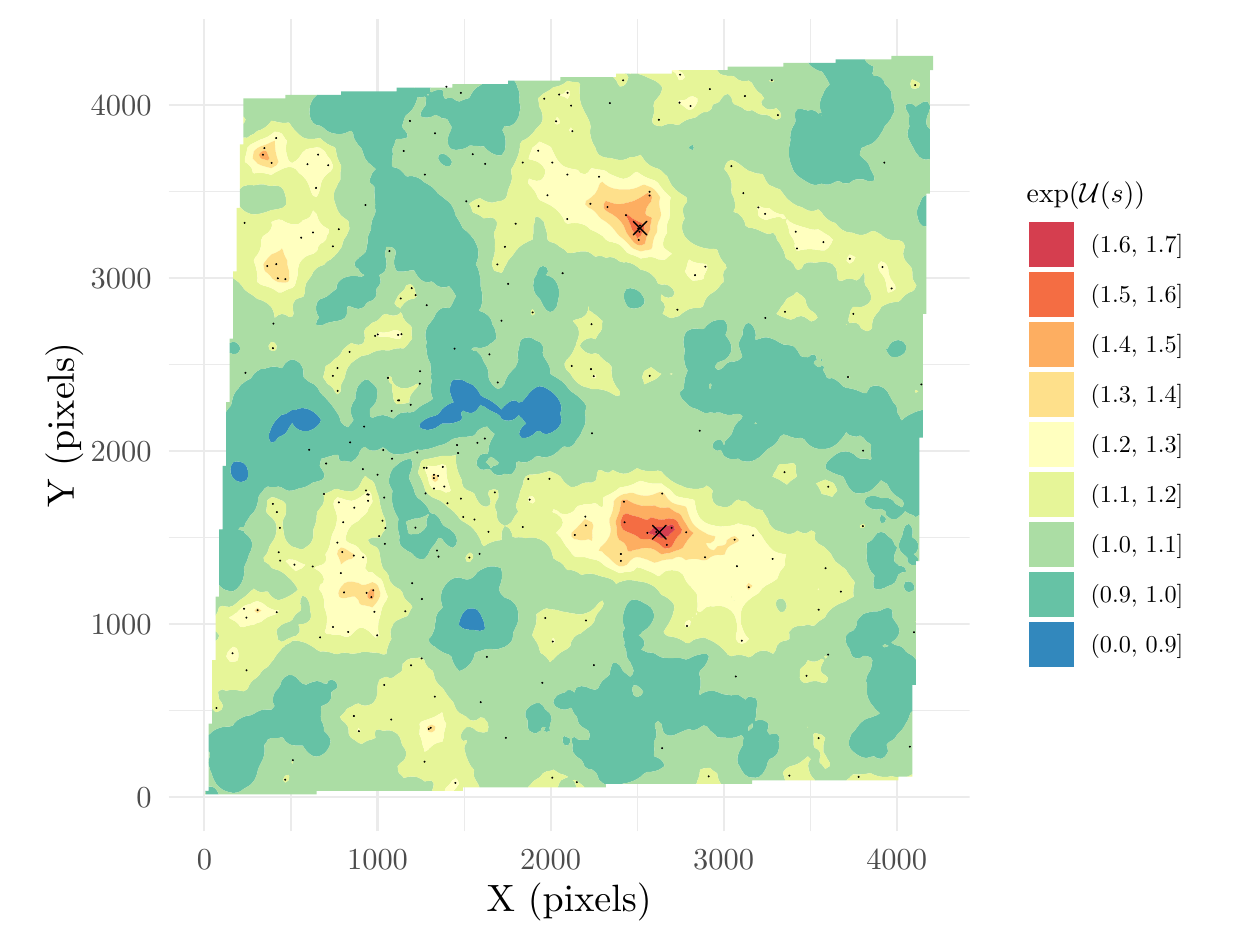}{0.51\textwidth}{(b)}
    \fig{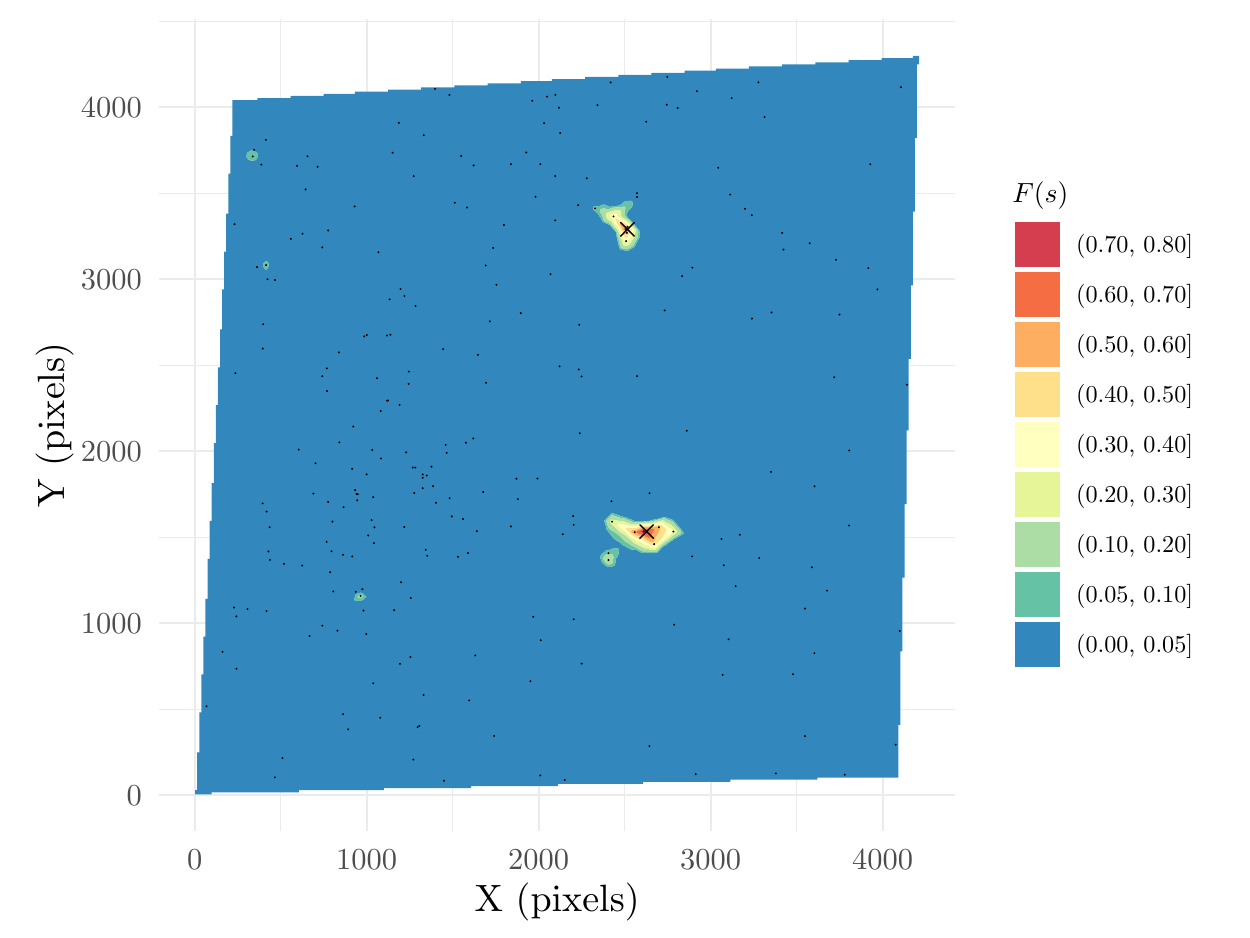}{0.51\textwidth}{(c)}}
    \caption{Posterior results for V11-ACS pointing. 
    Black points are locations of GC candidates. Black crosses are locations of previously detected UDG candidates
    (WUDG-89 at top, WUDG-88 at bottom).
    North and East are approximately downward and rightward, respectively.
    (a) Posterior mean intensity $\lambda(s)$.
    The signals from UDGs are dominated by large GC systems associated with two giant ETGs (PGC 012437 at left center, and UGC~02673 just beyond the top of the field of view).
    (b) Posterior mean spatial random effect $\exp(\mathcal{U}(s))$, shown as  $\exp(\mathcal{U}(s))$. After eliminating the signals from the GC systems from the two ETGs, the signals of two UDGs clearly show up within the spatial random effect and they are the strongest signals in the field of view. (c) The excursion function, $F_C(s)$, of $\mathcal{U}(s)$, with $C=1$, i.e., the median of the posterior marginal of $\mathcal{U}(s)$. The excursion function picks out the two UDGs in the field with almost no false positive detection.}
    \label{fig:v11acs_results}
\end{figure*}

\begin{figure*}
    \gridline{\fig{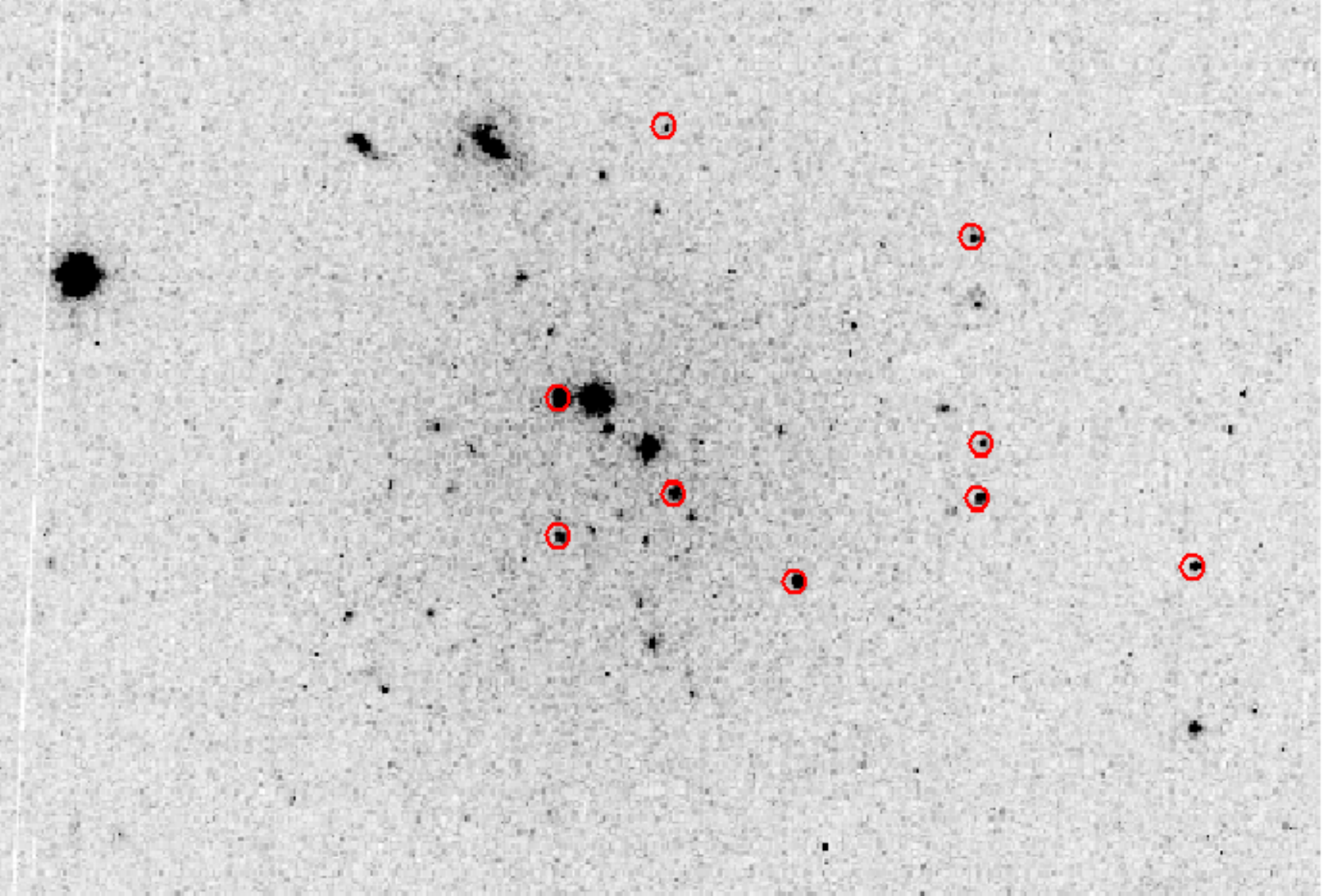}{0.22\textwidth}{RUDG-84}
              \fig{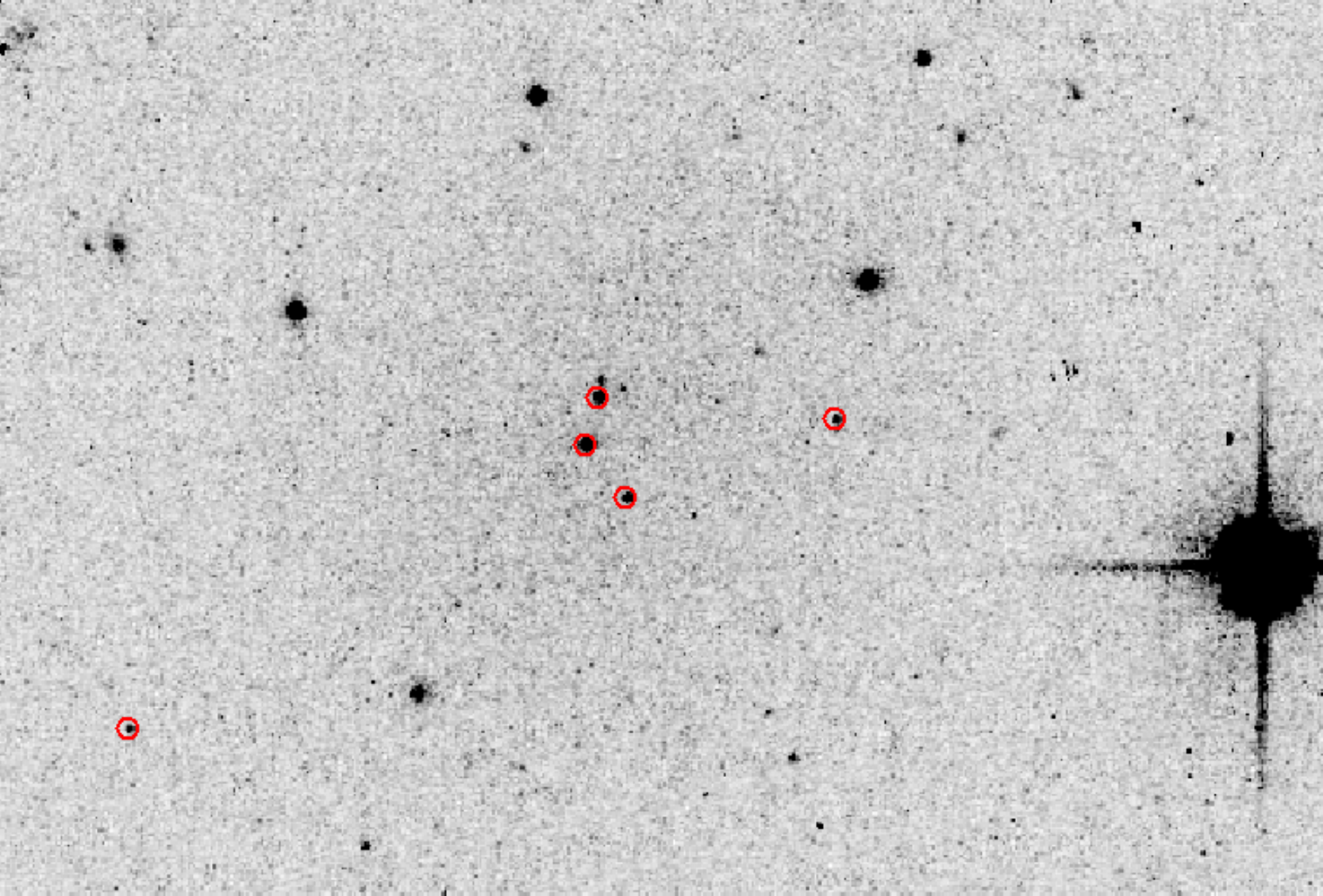}{0.22\textwidth}{RUDG-5}
              \fig{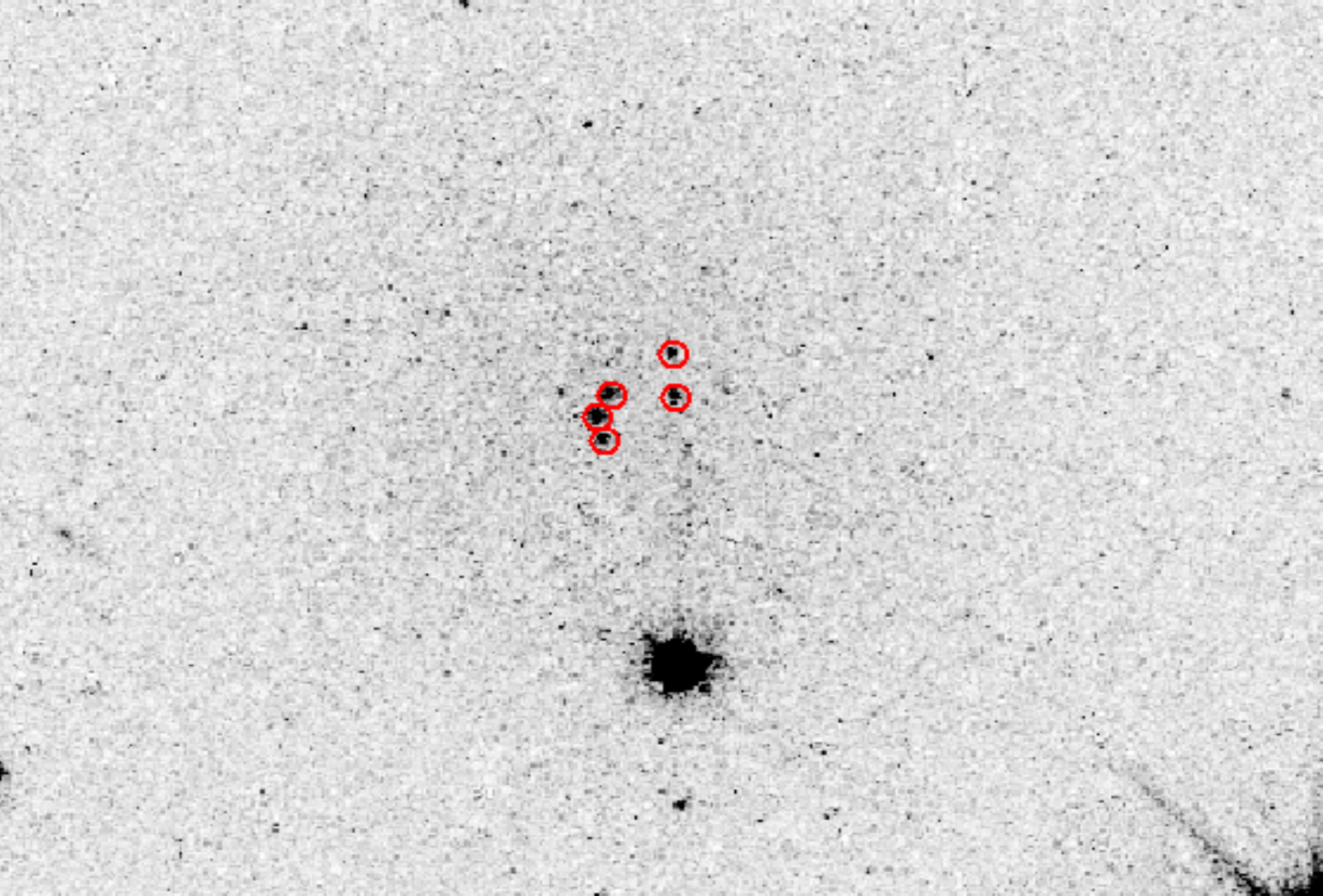}{0.22\textwidth}{RUDG-6}
              \fig{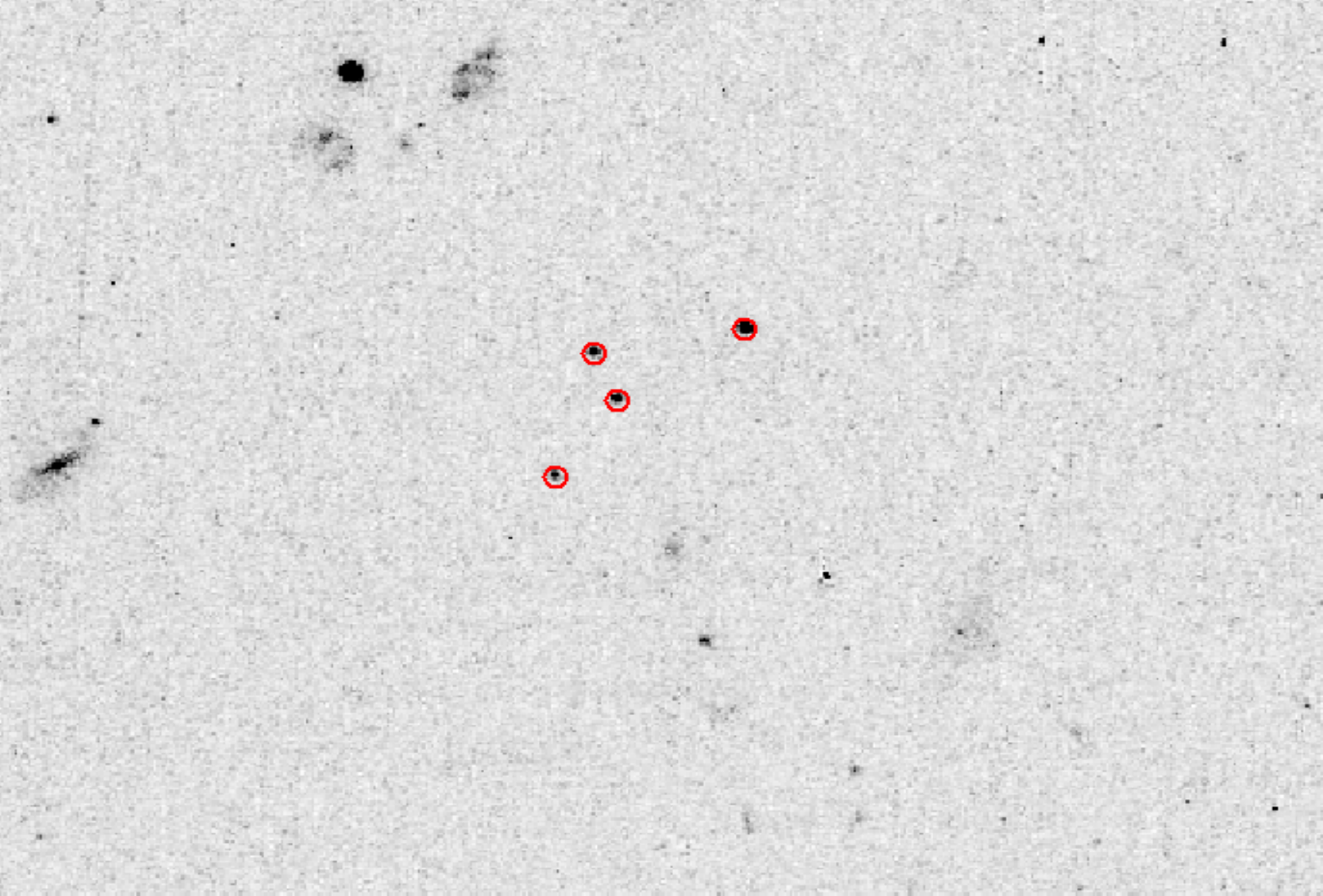}{0.22\textwidth}{\mbox{CDG-1} \textasteriskcentered}}
    \gridline{\fig{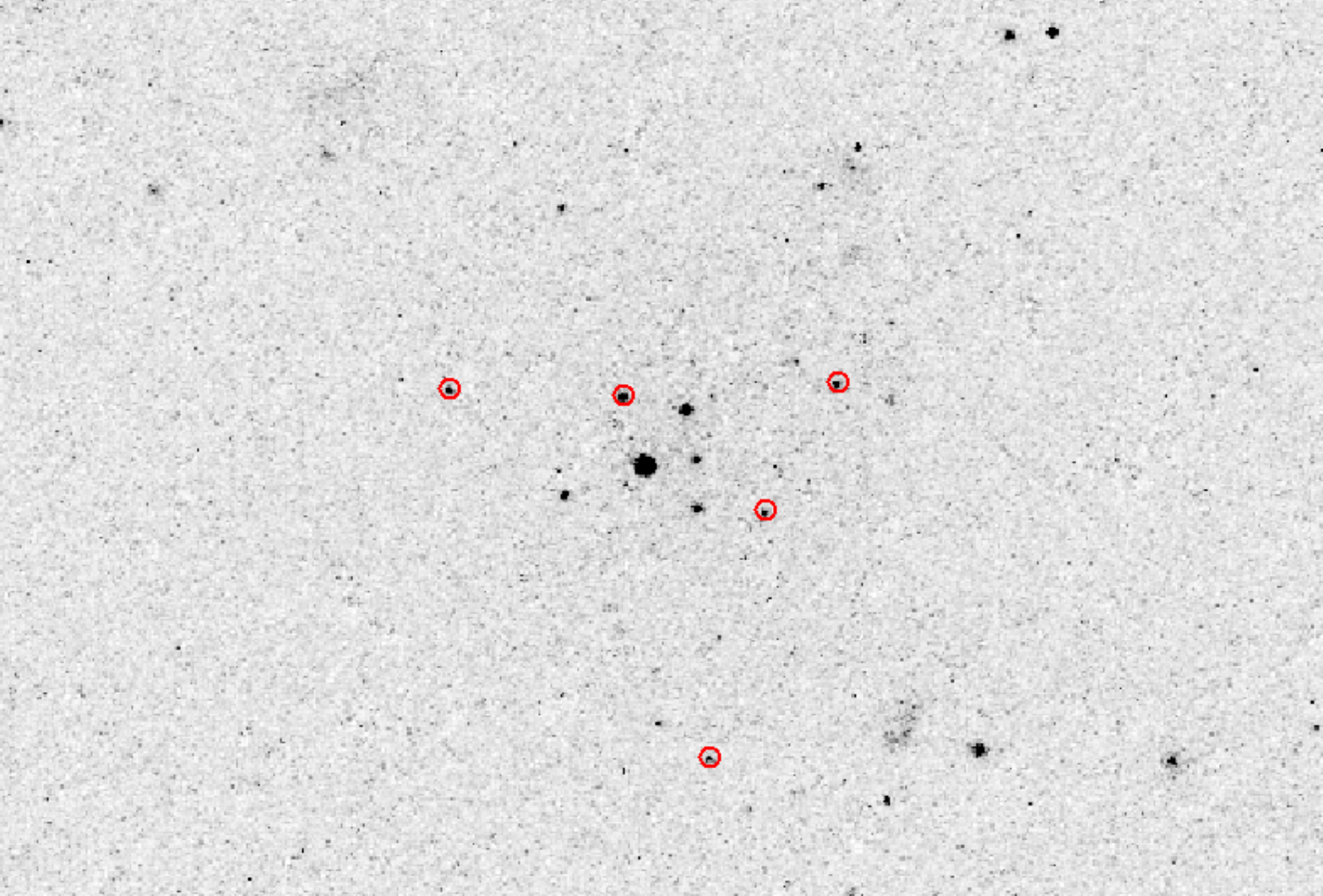}{0.22\textwidth}{WUDG-33}
              \fig{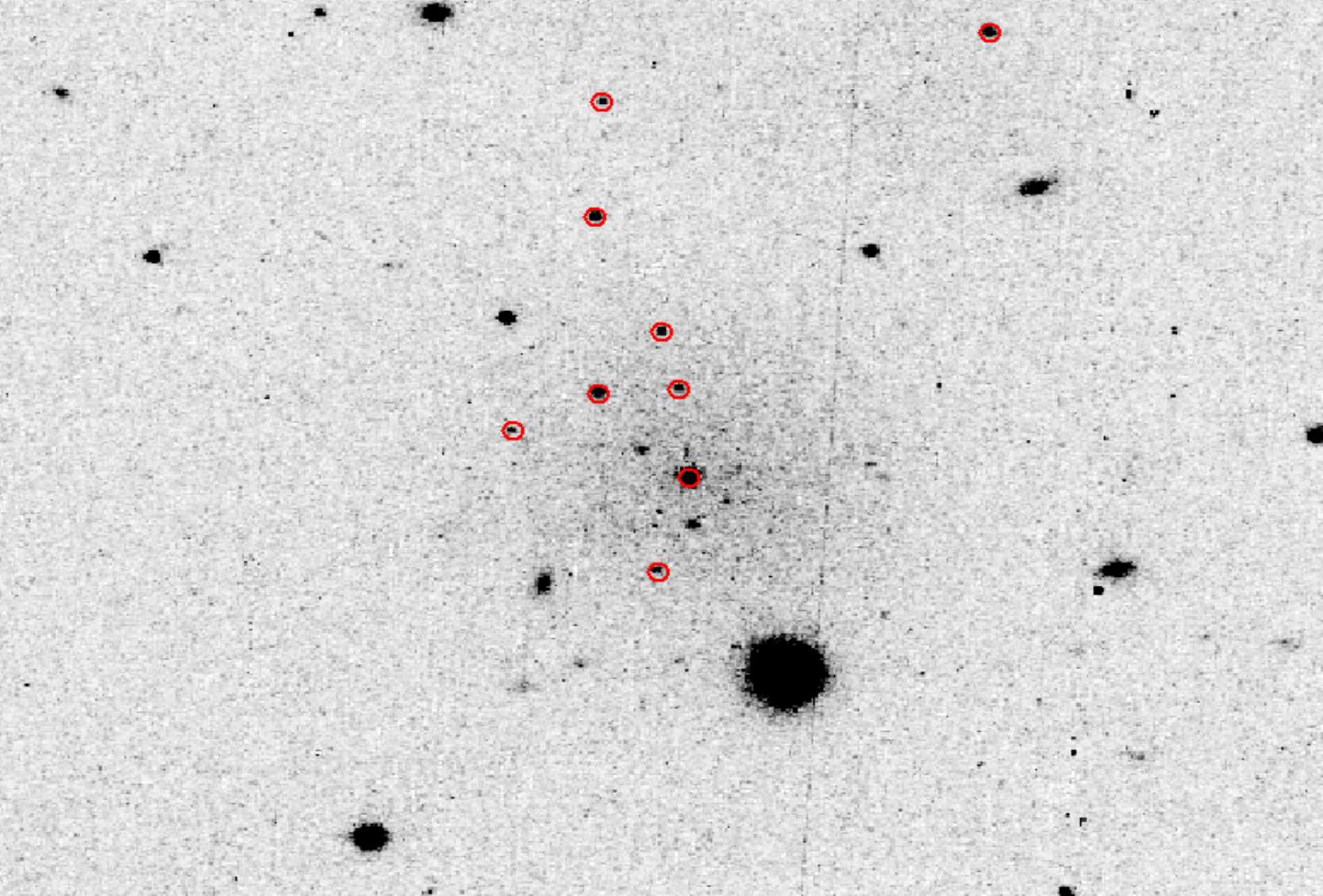}{0.22\textwidth}{RUDG-60}
              \fig{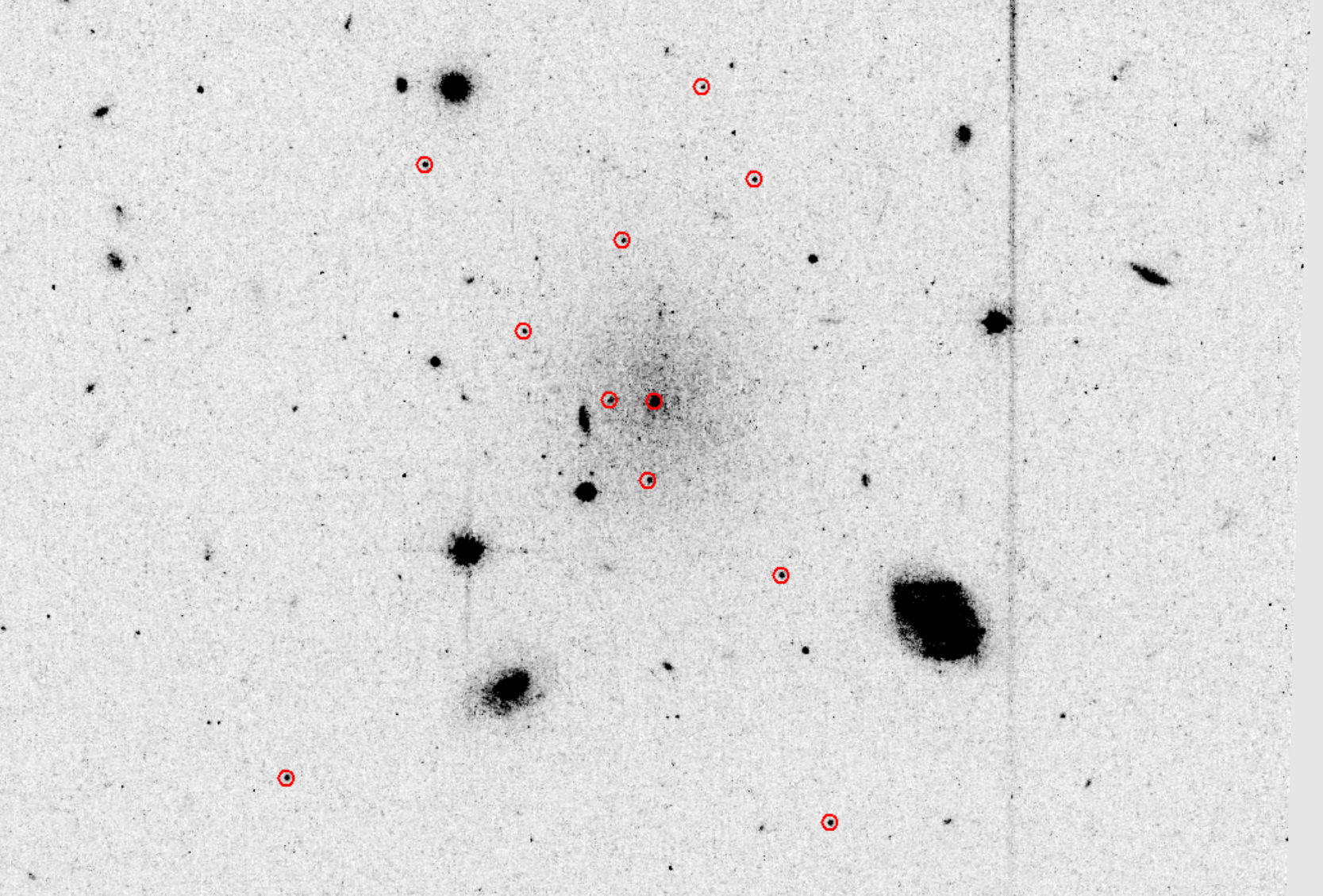}{0.22\textwidth}{RUDG-21}
              \fig{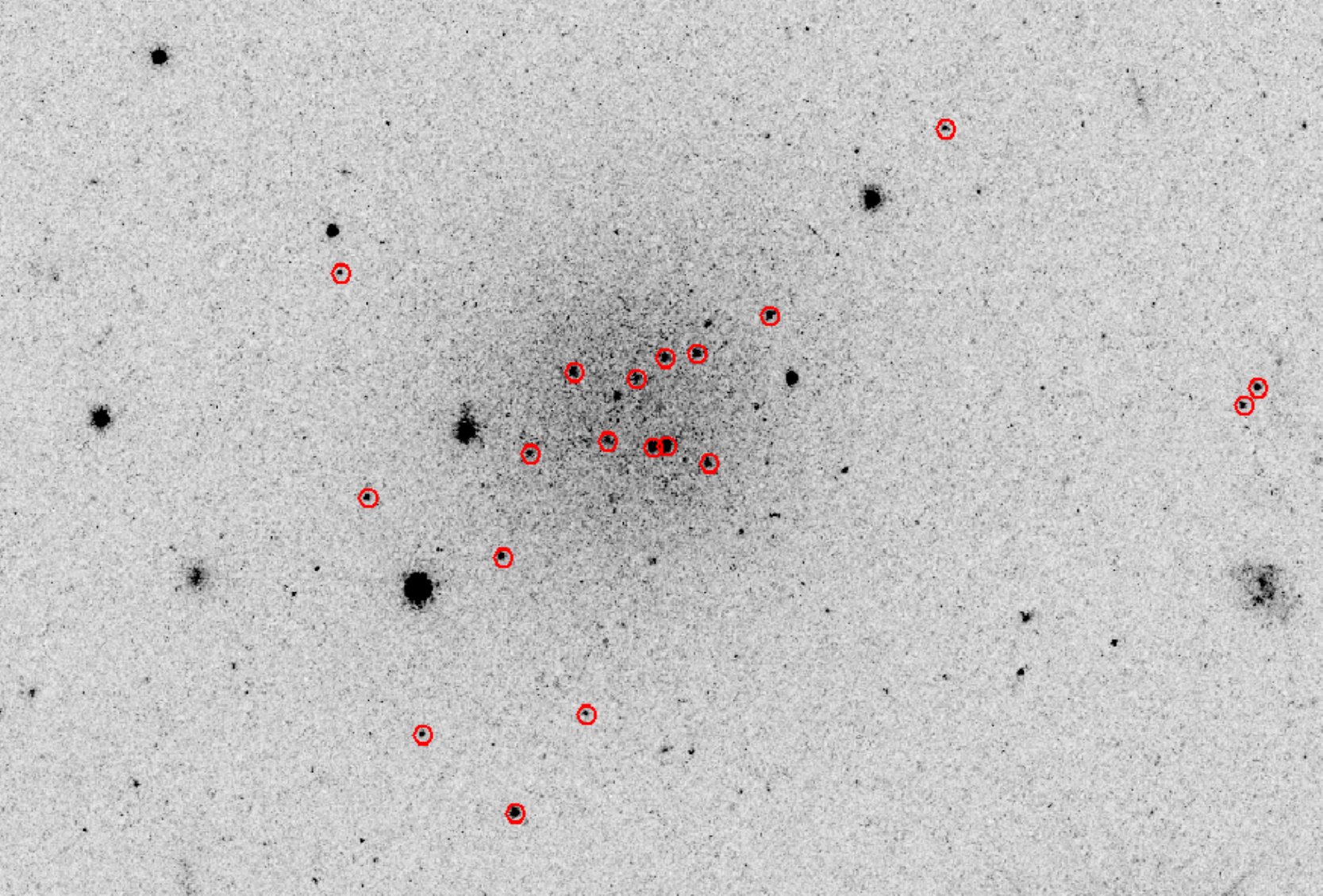}{0.22\textwidth}{RUDG-27}}
    \gridline{\fig{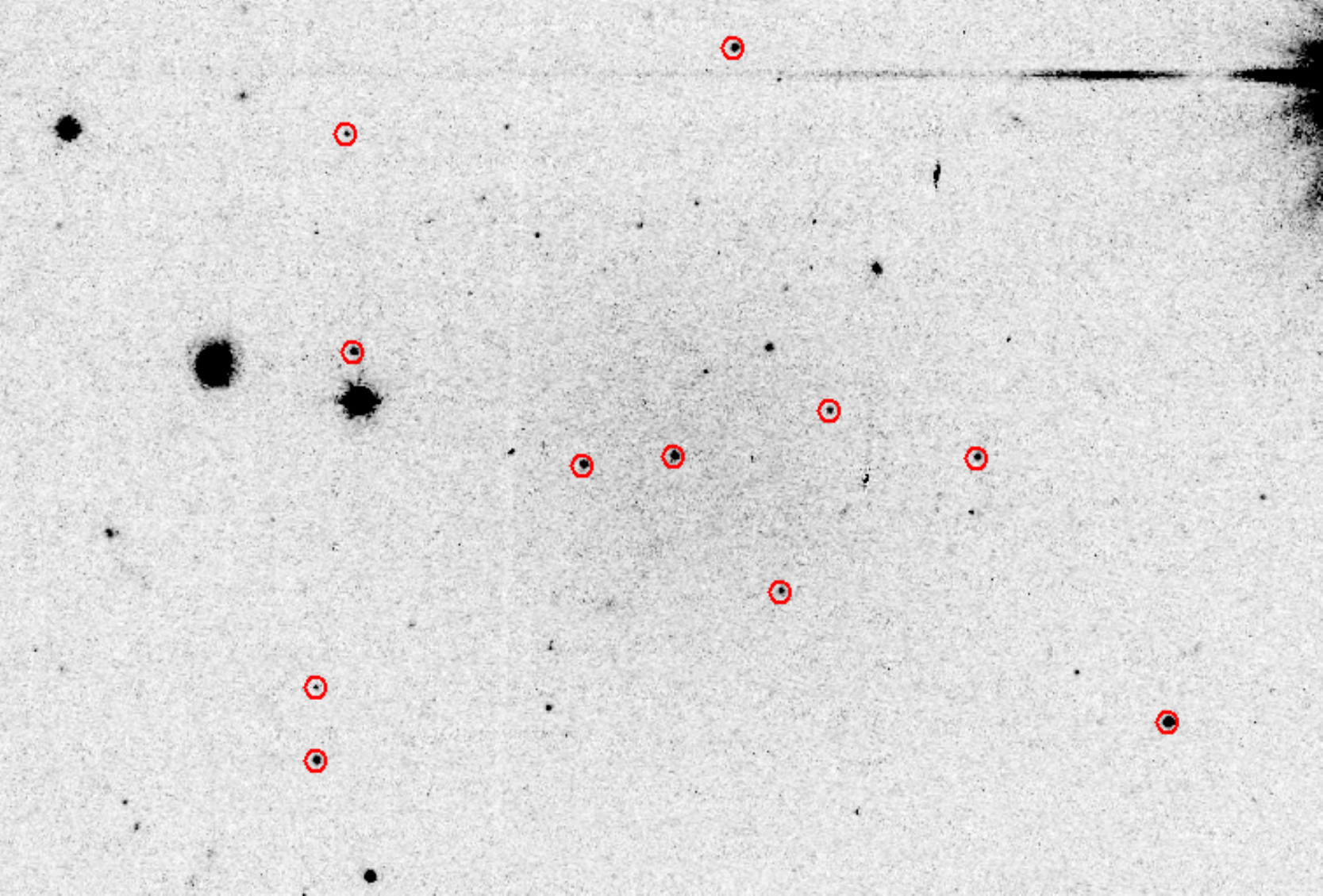}{0.22\textwidth}{WUDG-88}
              \fig{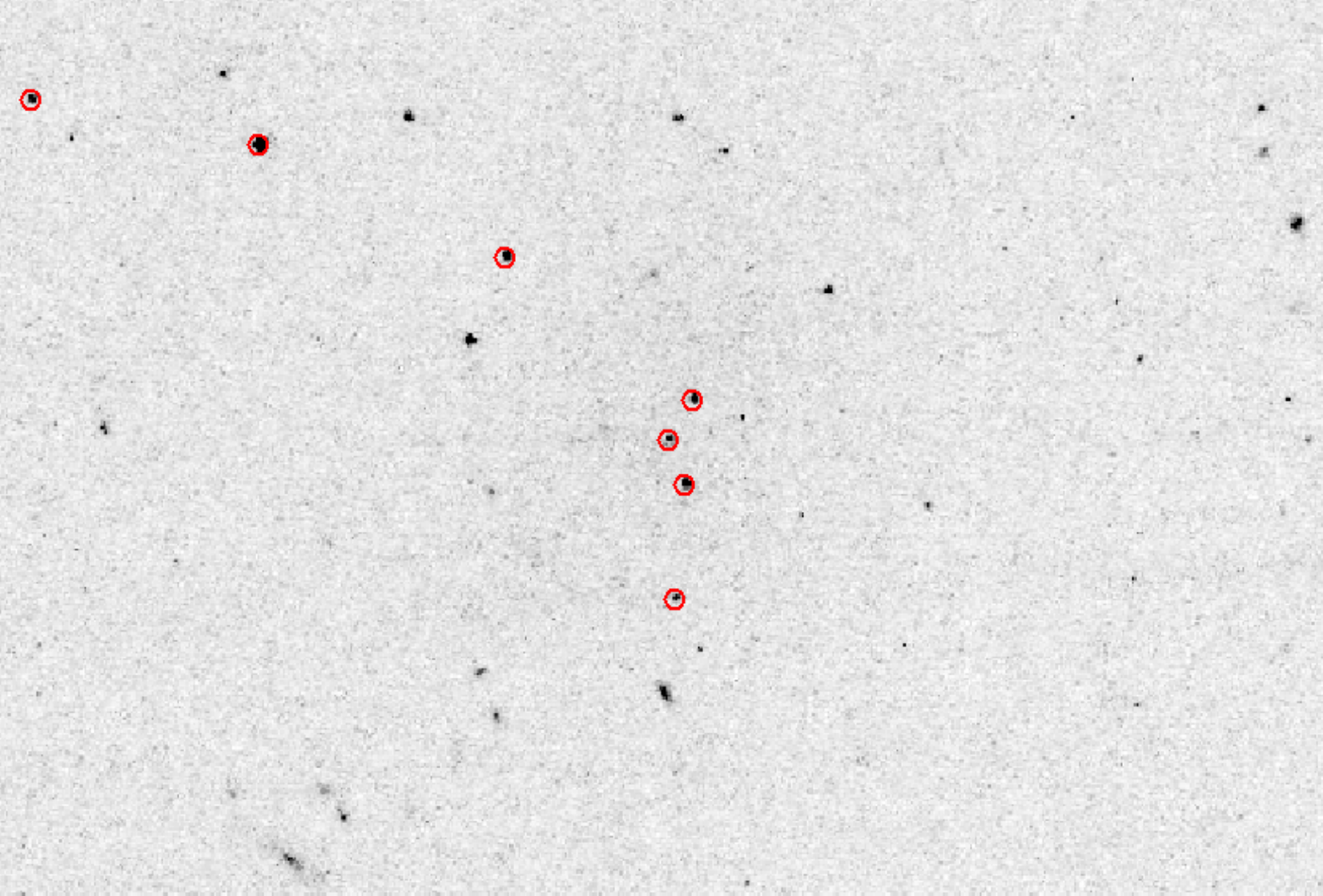}{0.22\textwidth}{WUDG-89}
              \fig{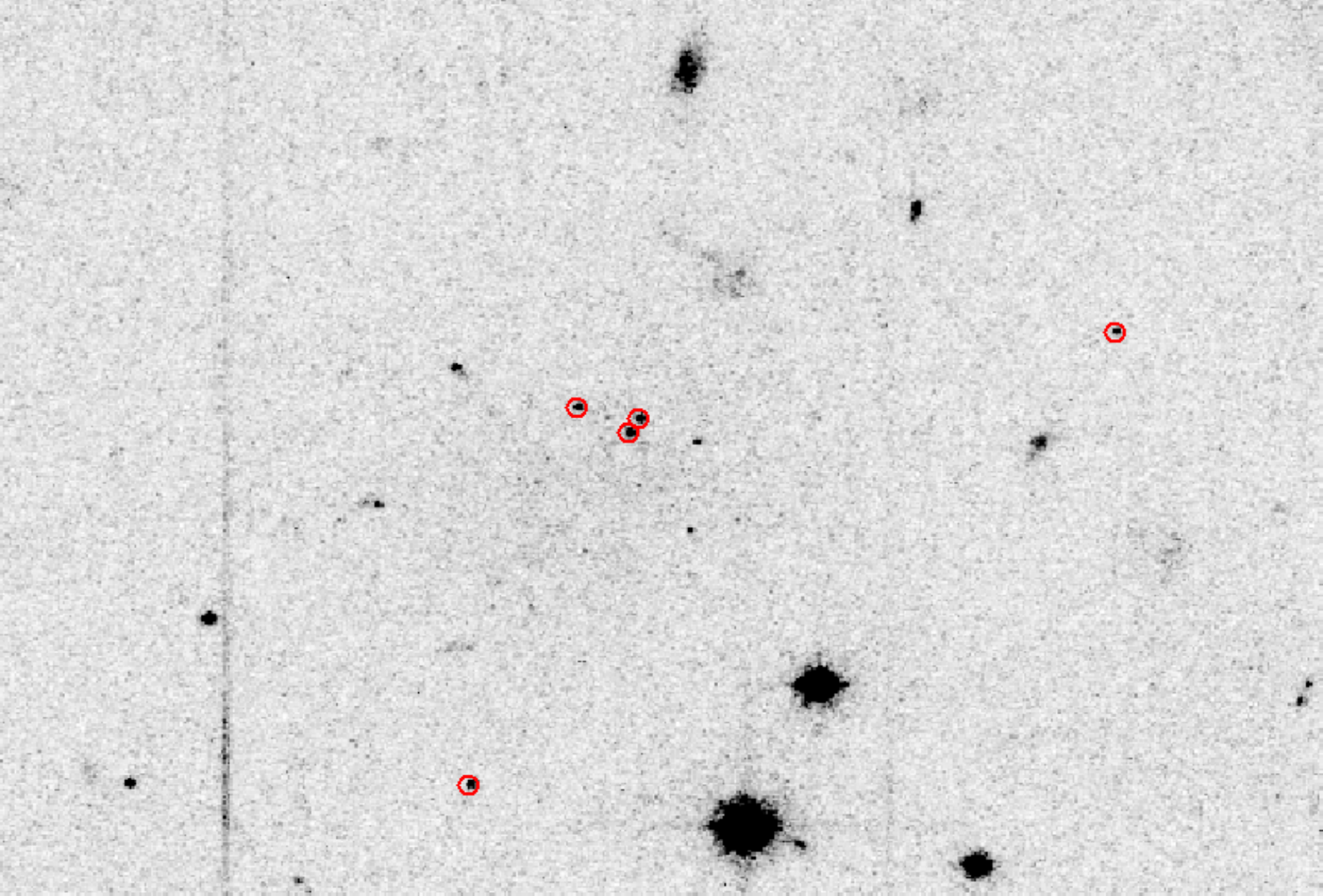}{0.22\textwidth}{WUDG-56}}
\caption{All UDG candidates detected within the PIPER survey data using LGCP. \textasteriskcentered \ represents the potential dark galaxy we found that has no detectable diffuse light. WUDG represents UDGs in the catalog of \cite{Wittmann2017} and RUDG represents UDGs in the catalog of Romanowsky et al. Red circles in the images are the selected GC candidates from the PIPER survey.}
\label{UDG_candid}
\end{figure*} 

\begin{deluxetable*}{lcclrcccccc}[ht]

\tablecaption{Object ID, coordinates, visit pointings, observed number of associated GCs ($N_{\rm GC}$), half-number radius of GC system ($R_{\rm GC}$) from Janssens et al.\ (in progress), and the maximum excursion function values for the detected UDGs. The coordinates for WUDGs are from \cite{Wittmann2017}, the coordinates for RUDGs are obtained from the RUDG catalog, the coordinate for \mbox{CDG-1} is taken as the average of the coordinates of its constituent GCs. The excursion function values are presented for each of the quantile excursion thresholds (see details in the text). The excursion function here can be loosely interpreted as the probability that a UDG signal exceeds the excursion thresholds. The last column represents whether or not there are bright normal galaxies within or near the pointing.\label{tab:UDG_F_function}}
\tablenum{1}

\tablehead{\colhead{ID} & \colhead{R.A.} & \colhead{Dec.} & \colhead{Pointing} & \colhead{$N_{\rm GC}$} & \colhead{$R_{\rm GC}$} & \colhead{$F_{Q(0.5)}$} & \colhead{$F_{Q(0.75)}$} & \colhead{$F_{Q(0.9)}$} & \colhead{$F_{Q(0.95)}$} & \colhead{Galaxies?}\\ 
\colhead{} & \colhead{(J2000)} & \colhead{(J2000)} & \colhead{} & \colhead{} & \colhead{(kpc)} & \colhead{} & \colhead{} & \colhead{} & \colhead{} & \colhead{}} 
\startdata
RUDG-84 & 03 17 24.80 & +41 44 20.84 & V7-ACS & 9 & 1.24 & 0.999 & 0.999 & 0.996 & 0.987 & N \\
RUDG-5 & 03 17 34.50 & +41 45 21.97 & V7-ACS & 4 & 0.61 & 0.998 & 0.984 & 0.859 & 0.697 & N \\
RUDG-6 & 03 16 58.90 & +41 42 09.64 & V7-WFC3 & 5 & 0.53 & 0.999 & 0.999 & 0.997 & 0.984 & N \\
\mbox{CDG-1} & 03 18 12.27 & +41 45 57.96 & V8-ACS & 4 & 0.67 & 0.870 & 0.700 & 0.518 & 0.385 & Y \\
WUDG-33 & 03 18 25.86 & +41 41 06.90 & V8-WFC3 & 5 & 0.86 & 0.732 & 0.520 & 0.350 & 0.262 & Y \\
RUDG-60 & 03 19 36.10 & +41 57 25.67 & V9-ACS & 8 & 1.45 & 0.999 & 0.999 & 0.999 & 0.997 & N \\
RUDG-21 & 03 20 29.40 & +41 44 50.70 & V10-ACS & 10 & 2.78 & 0.983 & 0.946 & 0.831 & 0.693 & Y \\
RUDG-27 & 03 19 43.40 & +41 42 47.72 & V10-WFC3 & 16 & 1.55 & 0.999 & 0.999 & 0.999 & 0.999 & N \\
WUDG-88 & 03 19 59.10 & +41 18 33.10 & V11-ACS & 7 & 1.41 & 0.800 & 0.607 & 0.421 & 0.322 & Y \\
WUDG-89 & 03 20 00.20 & +41 17 05.10 & V11-ACS & 6 & 1.19 & 0.639 & 0.363 & 0.171 & 0.100 & Y \\
WUDG-56 & 03 18 48.02 & +41 14 02.40 & V15-ACS & 3 & 0.60 & 0.943 & 0.882 & 0.603 & 0.469 & Y \\
\enddata

\end{deluxetable*}
\subsection{Detection of Known UDGs}

We now present the results for UDG detection in the Perseus cluster using our LGCP model and GC candidates from the PIPER survey data. To validate our results, we compare our detections to the WUDG \citep{Wittmann2017} and RUDG catalogs. Recall that 31 of the WUDGs and 10 of the RUDGs are within the PIPER fields. We note that some of these UDGs contain bright nuclei and the nuclei are included in the GC data that we use for detection.

For illustrative purposes, we present the results obtained from our LGCP model for one of the pointings, V11-ACS, in Figure \ref{fig:v11acs_results}. The figures for all other detected UDGs are presented in Appendix \ref{sec:detect_figs}. V11-ACS is the most contaminated pointing of all; a large lenticular (S0) galaxy (PGC~012437) takes up a significant portion of the pointing and another elliptical galaxy (UGC~02673) just outside the pointing has a GC system that clearly extends into the field of view. In Figure \ref{fig:v11acs_results}, previously detected UDGs from other catalogs are denoted by black crosses. Figure \ref{fig:v11acs_results}(a) shows the posterior mean intensity surface $\lambda(s)$ obtained from LGCP. Figure \ref{fig:v11acs_results}(b) presents the posterior spatial random effect $\mathcal{U}(s)$, shown on the exponential scale. Figure \ref{fig:v11acs_results}(c) presents the excursion function of $\mathcal{U}(s)$ with the excursion threshold being the median of the posterior marginal of $\mathcal{U}(s)$ or $C = 1$. For simplicity and to save space, we only show the results for the median here and omit the results for other quantile excursion thresholds.

We can see in Figure \ref{fig:v11acs_results}(a) the intensity is dominated by the elliptical galaxy at the left-center region of the pointing. The high intensity region near the top is the other elliptical galaxy that falls partially in the field. It is clear from Figure \ref{fig:v11acs_results}(a) that the signals from known UDGs are drowned out, and not picked up by $\lambda(s)$. However, if we strip away the fixed effects from the elliptical galaxies, the spatial random effect $\mathcal{U}(s)$ detects the UDG signals (Figure \ref{fig:v11acs_results}(b)). In fact, the locations of the two identified and previously known UDGs have the highest intensity in $\mathcal{U}(s)$. The excursion function in Figure \ref{fig:v11acs_results}(c) successfully picks out the two UDGs with almost no false positive noise. The selection of UDG candidates is then based on the results as demonstrated in Figure \ref{fig:v11acs_results}(c), where regions with the highest excursion function values are chosen as potential locations for UDGs. Our results demonstrate the extraordinary power of LGCP for detecting UDGs.

Based on the excursion functions obtained for all 20 PIPER survey pointings, LGCP picked up $11$ UDG candidates within nine different pointings, which are shown in Figure \ref{UDG_candid}. Among these, ten UDGs were already reported in WUDG and RUDG catalogs, with four in the WUDG catalog and six in the RUDG catalog. Five of these UDGs have cross-IDs from the PCC catalog \citep{Wittmann2019}: PCC~2251 (WUDG-33), PCC~3023 (WUDG-56), PCC~5374 (WUDG-88), PCC~5402 (WUDG-89), and PCC~4867 (RUDG-27). It turns out that these ten UDGs are all the previously confirmed UDGs within the PIPER survey that have an observed GC counts greater than two. For other confirmed UDGs, their observed GC counts are all fewer than two and thus undetectable.

Note that the observed GC count we consider here is the observed number of GCs within three times the half-number radius of the UDG GC system where the half-number radii are from Janssens et al. \ (in prep). This GC count is our observable: it contains the majority of GCs in UDGs that allow for detection, but it also  contains background contamination. GCs outside this region will not be clustered enough to have a significant effect on detection. Although one would be more interested in knowing how the background-subtracted GC count affects the UDG detection, its true value is never known in real data. Furthermore, it is traditionally obtained by fitting a S\'{e}rsic model to the GC system with background contamination subtracted, such an estimation can be subject to high level of uncertainty especially for systems with small GC populations.

For an independent comparison based on GC counts, we also consider the traditional background-subtracted GC counts for all 41 known UDGs. The background-subtracted GC counts are obtained by Janssens et al. \ (in prep). Out of the 41 known UDGs, 14 have background-subtracted GC counts greater than two. Of these, five were not detected using our method, namely WUDG-5, WUDG-8, WUDG-13, WUDG-22, and WUDG-29. However, these UDGs all have background-subtracted GC counts just above two and the estimates are highly uncertain. Moreover, none of these UDGs has previously defined observed GC counts greater than two. On the other hand, there is one UDG we detected --- WUDG-56 --- that has a background-subtracted GC count lower than two. But the PIPER GC data shows that there are in fact three GC candidates tightly clumped at the center of this UDG and our method is able to detect it.

\subsection{Detection of \mbox{CDG-1}}

In addition to detecting known UDGs, the most interesting object we have detected is \mbox{CDG-1} --- \textit{Candidate Dark Galaxy-1}.  \mbox{CDG-1} consists of four densely clumped observable GC candidates but has no detectable diffuse stellar content at that location.
Since we are not yet certain of the nature of \mbox{CDG-1}, we will not classify it as a potential UDG.

CDG-1 is in the V8-ACS pointing, and it is the only over-density that stands out. We include it here since, in theory, there is no condition that says a galaxy needs to have any diffuse stellar content. It is entirely possible to have galaxies that host GCs but are devoid of field stars. A more detailed detection analysis of this object will be covered in Section \ref{sec:CDG1}.

\subsection{Effects of Neighboring Galaxies}

Table \ref{tab:UDG_F_function} lists the detailed detection results for the 11 UDGs shown in Figure \ref{UDG_candid}. The main results shown in Table \ref{tab:UDG_F_function} are the excursion function values for each detected UDG. Note that the computed excursion function values are presented for the four quantile excursion thresholds as mentioned in section \ref{sec:excursion}. Furthermore, the presented excursion function values are obtained by considering a region within the half-number radius from the center of each UDG \footnote{Such a region is guaranteed to contain the maximum excursion function value associated with each UDG.}, and the maximum value of the excursion function in this region is chosen. We also present some of the characteristics of the UDGs that may have a significant impact on their detections, namely the observed associated GC counts, estimated half-number radii of the GC systems, and whether or not there are bright normal galaxies within the pointing. The half-number radius for \mbox{CDG-1} is obtained by fitting a separate S\'{e}rsic model to its observed GC system. We can see that many UDGs have extremely high excursion function values which indicates that there is no issue of detecting them. However, there are some UDGs with excursion function values dropping to a very low level when the quantile thresholds are increased. Unsurprisingly, bright normal galaxies within the field of observations do seem to have a large impact on the excursion function values of the UDGs. It is worth noting that the excursion function values of \mbox{CDG-1} are not the weakest in our list; in fact, it has a stronger detection signal than three out of the ten known UDGs.

To confirm the bright-galaxy effect, we conducted a preliminary analysis on the detection results using a simple linear regression with the excursion function values transformed into log-odds scale versus $N_{\rm GC}$, $R_{\rm GC}$ and whether or not there are bright normal galaxies within the pointing. We found that for all four quantiles, pointings with normal galaxies indeed have significantly lower excursion function values. This is somewhat expected since for pointings with no normal galaxies, as we did not include any covariate information in the LGCP models, i.e., there is only the intercept term. In other words, the presence of normal galaxies increases the uncertainty of the signal, which decreases the values of the excursion functions. This is especially clear with WUDG-89, which is extremely close ($\sim 40$ kpc) to an elliptical galaxy (UGC~02673). This proximity causes the LGCP model to be uncertain about whether the GC over-density belongs to the elliptical or is a system of its own. Moreover, there is a larger UDG (WUDG-88) with a stronger signal in the same pointing, whose presence also negatively affects the detection probability of WUDG-89: its signal is the weakest amongst UDGs with similar $N_{\rm GC}$ and $R_{\rm GC}$. Nevertheless, we are still able to pick out the signal from WUDG-89. 

The results from the linear regression model shows that $N_{\rm GC}$ and $R_{\rm GC}$ have no significant impact on the detection probability. This is likely due to the small sample size, such that the actual effects of $N_{\rm GC}$ and $R_{\rm GC}$ are undetectable. Furthermore, the numerical ranges of $N_{\rm GC}$ and $R_{\rm GC}$ are too restricted to obtain any meaningful relationship.

\subsection{ROC Analysis of UDG Detection}\label{sec:ROC_detect}

The detection results presented previously only address LGCP's performance on providing true positive detection. We still require analysis on the overall performance of LGCP as a binary classifier which demands assessing LGCP's performance with respect to false positive, false negative, and true negative detection. To this end, we provide an overall detection analysis using the spatial receiver operating characteristic. Note that in what follows, the UDGs we are interested in are restricted to GC-rich UDGs.

Receiver operating characteristic (ROC; \citealt{Fawcett2006,Mas2013}) is a widely used performance analysis tool in machine learning for assessing the overall performance of any binary classifier. In our problem, for any given excursion threshold $C$, the resulting excursion function obtained from LGCP can be viewed as a binary classifier. By choosing a probability threshold $q$, the excursion function will produce a corresponding Boolean map: pixels (regions) with excursion function values greater than $q$ will be assigned the true value, i.e., part of a UDG, while pixels with excursion function values less than $q$ will be assigned the false value, i.e., not part of a UDG. We can then compare the obtained Boolean map to the ``truth" map: a reference map indicating which pixels are truly part of a UDG and which are not. The comparison will result in the predicted Boolean value of each pixel to fall into one of the four categories: true positive (TP), false positive (FP), true negative (TN), and false negative (FN). To assess the performance of a classifier, the true positive rate (TPR) and the false positive rate (FPR) defined below are used jointly as performance measures:
\[
\text{TPR} = \frac{\text{TP}}{\text{TP} + \text{FN}},
\]
\[
\text{FPR} = \frac{\text{FP}}{\text{FP} + \text{TN}}.
\]
If we are able find a value of $q$, so that the resulting TPR is one and FPR is zero, it then means the classifier has perfect performance. 

\begin{figure}[h]
    \centering
    \includegraphics[width = 0.5\textwidth]{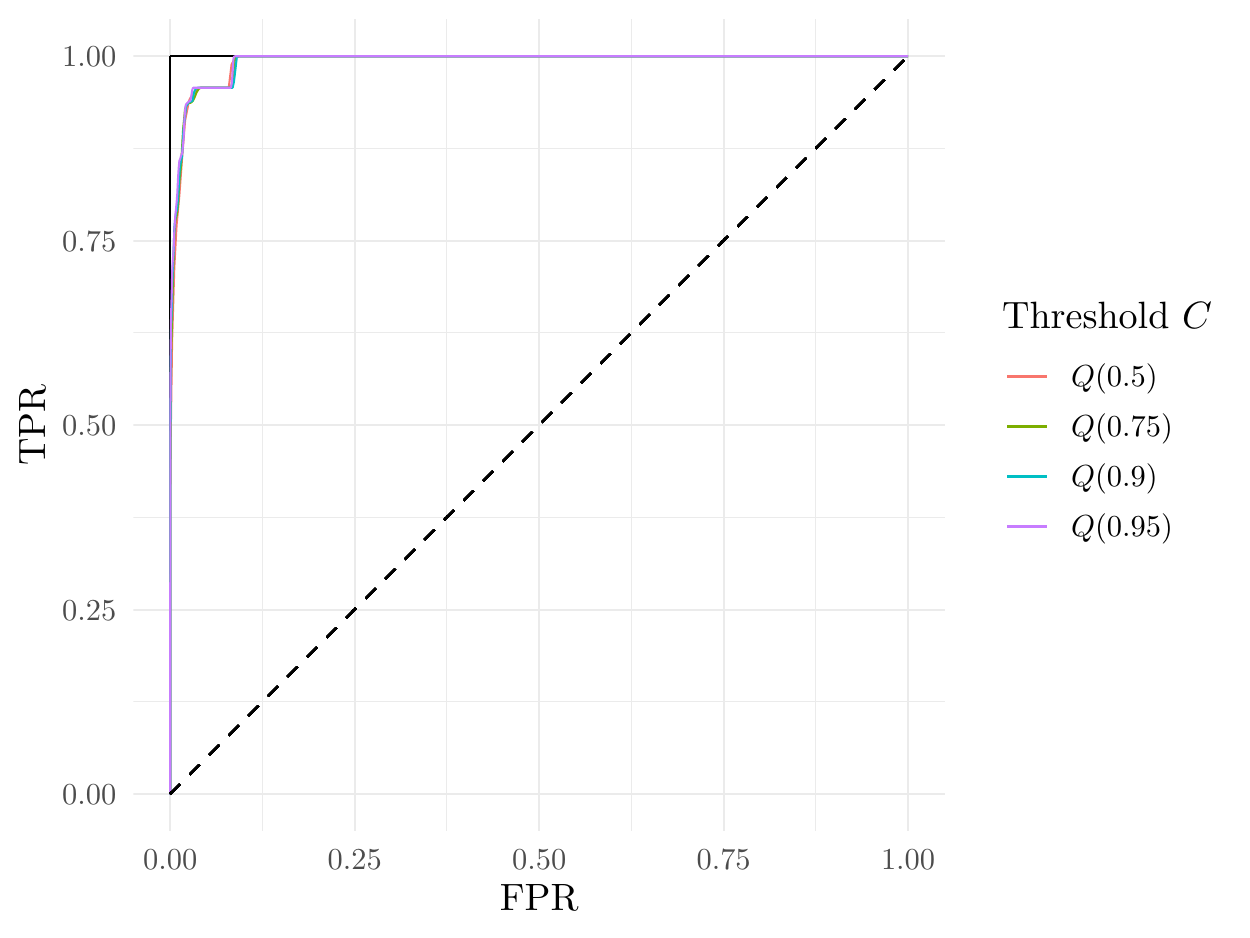}
    \caption{ROC curves of the four quantile excursion thresholds obtained for the pointing V11-ACS. Dashed black curve represents a random classifier; Black solid curve represents a perfect classifier.}
    \label{fig:ROC_v11}
\end{figure}

\begin{deluxetable*}{lcccc}

\tablecaption{AUC computed from the ROC curve of each of the quantile excursion thresholds for all pointings with confirmed UDGs.\label{tab:AUC_table}}
\tablenum{2}

\tablehead{\colhead{Pointing} & \colhead{AUC$_{Q(0.5)}$} & \colhead{AUC$_{Q(0.75)}$} & \colhead{AUC$_{Q(0.9)}$} & \colhead{AUC$_{Q(0.95)}$}} 
\startdata
V7-ACS & 0.99721 & 0.99777 & 0.99795 & 0.99806 \\
V7-WFC3 & 0.99974 & 0.99984 & 0.99981 & 0.99981 \\
V8-WFC3 & 0.99917 & 0.99928 & 0.99952 & 0.99950 \\
V9-ACS & 0.99592 & 0.99709 & 0.99756 & 0.99763 \\
V10-ACS & 0.99517 & 0.99812 & 0.99833 & 0.99808 \\
V10-WFC3 & 0.99667 & 0.99812 & 0.99871 & 0.99915 \\
V11-ACS & 0.99222 & 0.99254 & 0.99270 & 0.99311 \\
V15-ACS & 0.99942 & 0.99961 & 0.99942 & 0.99937 \\
\enddata

\end{deluxetable*}

For almost all binary classifiers, perfect performance is rare. Furthermore, the performance of the classifier depends on the choice of $q$: a low value of $q$ will produce high TPR while a high value of $q$ will produce low FPR. Therefore, an overall performance assessment of the classifier is obtained by plotting the TPR against the FPR with the corresponding value of $q$ ranging from $q = 0$ to $q=1$. This procedure will result in a curve called the receiver operating characteristic (ROC) curve. For illustration purposes, we present in Figure \ref{fig:ROC_v11} the ROC curves of the four quantile excursion thresholds for the pointing V11-ACS. For the truth map of the V11-ACS pointing, true values are given to regions within twice the GC system half-number radius (shown in Table \ref{tab:UDG_F_function}) from the center of the UDGs in the pointing, while the rest of the regions are assigned false values. If we can find $q$ such that the classifier has perfect performance, then its ROC curve will correspond to the black solid curve in Figure \ref{fig:ROC_v11} since it will have to hit the point $(0,1)$ (TPR $=1$, FPR $=0$). If a classifier is a random classifier, i.e., the classifier's performance is the same as random guessing, its ROC curve then corresponds to the dashed line in Figure \ref{fig:ROC_v11}. Hence, when comparing two classifiers, the better classifier will effectively have a ROC curve that approaches closer to the point $(0,1)$. We can see from Figure \ref{fig:ROC_v11} that all the ROC curves for the four thresholds are essentially the same and they are all extremely close to the point $(0,1)$, indicating that the performance of our method is very good.

Intuitively, we might expect the performance of our classifier to be exactly perfect since the LGCP method can pick out the two UDGs in the V11-ACS pointing perfectly without any noise (see Figure \ref{fig:v11acs_results}). However, we do not have a perfect classifier as shown in Figure \ref{fig:ROC_v11} because the truth map used in our ROC analysis has a mismatch with the intuitive notion of truth in our problem. The intuitive truth for our problem is the true locations of the UDGs instead of the projected regions occupied by the UDGs. However, to conduct the ROC analysis, we require the projected regions occupied by UDGs. It is impossible to obtain a perfect recovery of said regions since a UDG occupies a region with a poorly defined boundary. Furthermore, even if the region occupied by a UDG is the same as the ones we have set to, it is still not possible to obtain a perfect performance since the detected regions returned by LGCP will unlikely coincide exactly with the ones in the truth map. Nevertheless, we only need the ROC analysis for a bulk part assessment of the overall performance of LGCP, and it is completely adequate for our purpose here.

As shown in Figure \ref{fig:ROC_v11}, the overall performance of LGCP is quite spectacular for V11-ACS. However, it is highly inconvenient to present the performance results in the format of Figure \ref{fig:ROC_v11} for all the pointings. A simple method to extract a one-number summary from a ROC curve is by the area under the curve (AUC), since the total area under the ROC curve will generally be greater for a better classifier. A perfect classifier will then have an AUC of one while a random classifier has AUC being a half.

Table \ref{tab:AUC_table} provides the computed AUCs obtained for the four quantile excursion thresholds for all the pointings that contain detected UDGs. Note that the pointing V8-ACS, which contains \mbox{CDG-1}, is excluded as the nature of \mbox{CDG-1} is not yet confirmed. Furthermore, we only consider the performance on detecting UDGs with sufficient GC populations. The truth map for each of the pointings are obtained in the same way as the one we have set for obtaining the ROC curves in Figure \ref{fig:ROC_v11}. As shown in Table \ref{tab:AUC_table}, the AUC values for all excursion thresholds and pointings are almost one, indicating that the LGCP's performance as a binary classifier for identifying UDGs with significant GC populations is essentially perfect. 

The detection results of the known UDGs and \mbox{CDG-1} show a promising future for our proposed method, and the potential confirmation of \mbox{CDG-1} as a ``dark" galaxy would have profound implications for the understanding of galaxy evolution as well as fundamental physics. Thus, in the next section, we present some preliminary analysis to determine the nature of \mbox{CDG-1} using observational data at hand. 

For the sake of rigor and a better understanding of the performance and applicability in a more general setting, we require a calibrated assessment of our method using simulation. For interested readers, the simulation results are presented in Section \ref{sec:simulation}. Otherwise, one can skip to Section \ref{sec:conclusion} for a brief summary and the future plan of our work.

\section{\mbox{CDG-1}: A Pure GC Galaxy?}\label{sec:CDG1}

\subsection{Detection Results}
\begin{figure*}
    \centering
    \includegraphics[width = 0.7\textwidth]{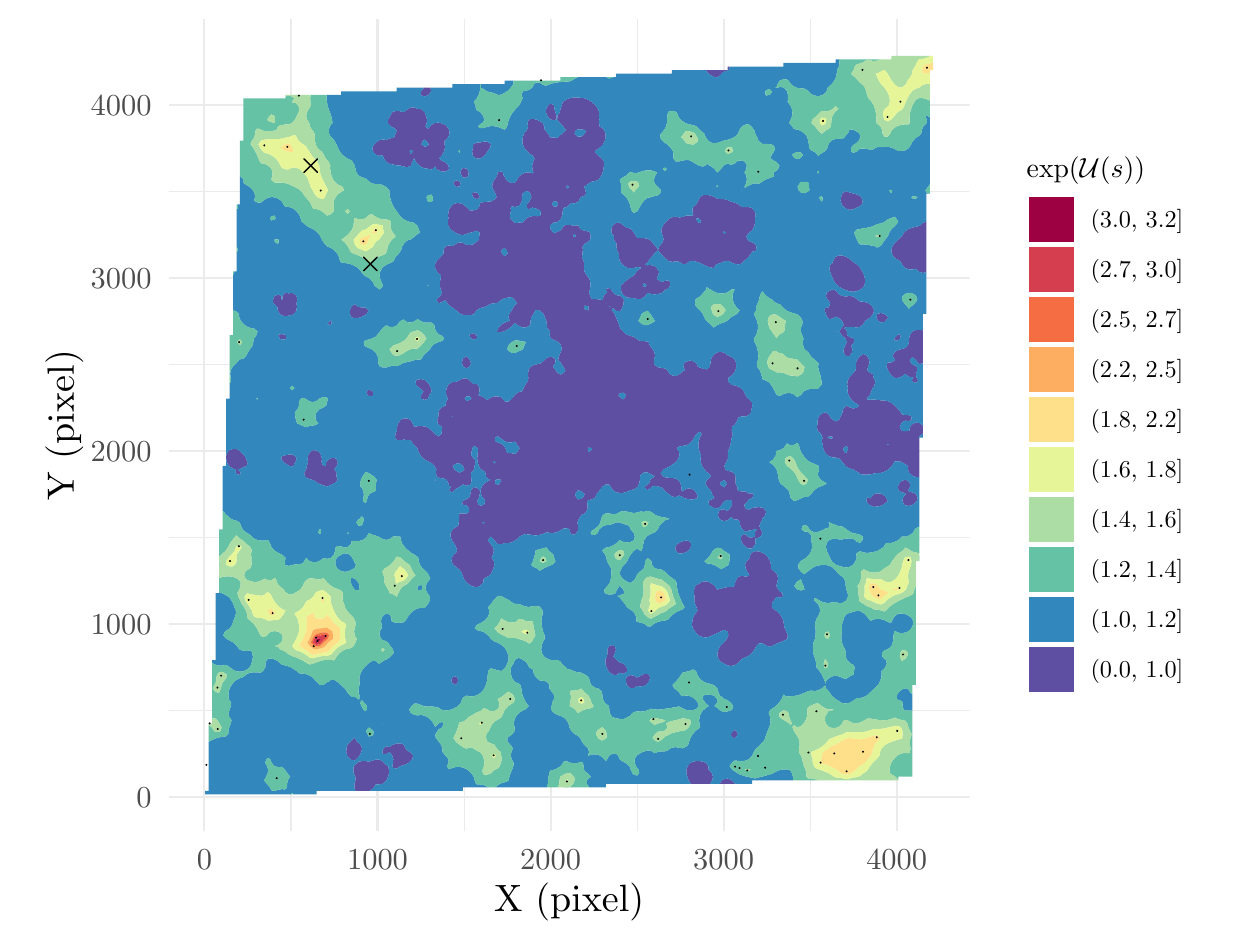}
    \caption{Posterior spatial random effect $\mathcal{U}(s)$ for V8-ACS, shown in exponential scale. Black points are GC candidates from the PIPER survey. Two known UDGs (WUDG-28 and WUDG-29) are in the upper left corner shown as black crosses. The object \mbox{CDG-1} is in the lower left corner, with the strongest signal in the entire pointing.}
    \label{fig:v8_Us}
\end{figure*}

\begin{figure*}
 \gridline{\fig{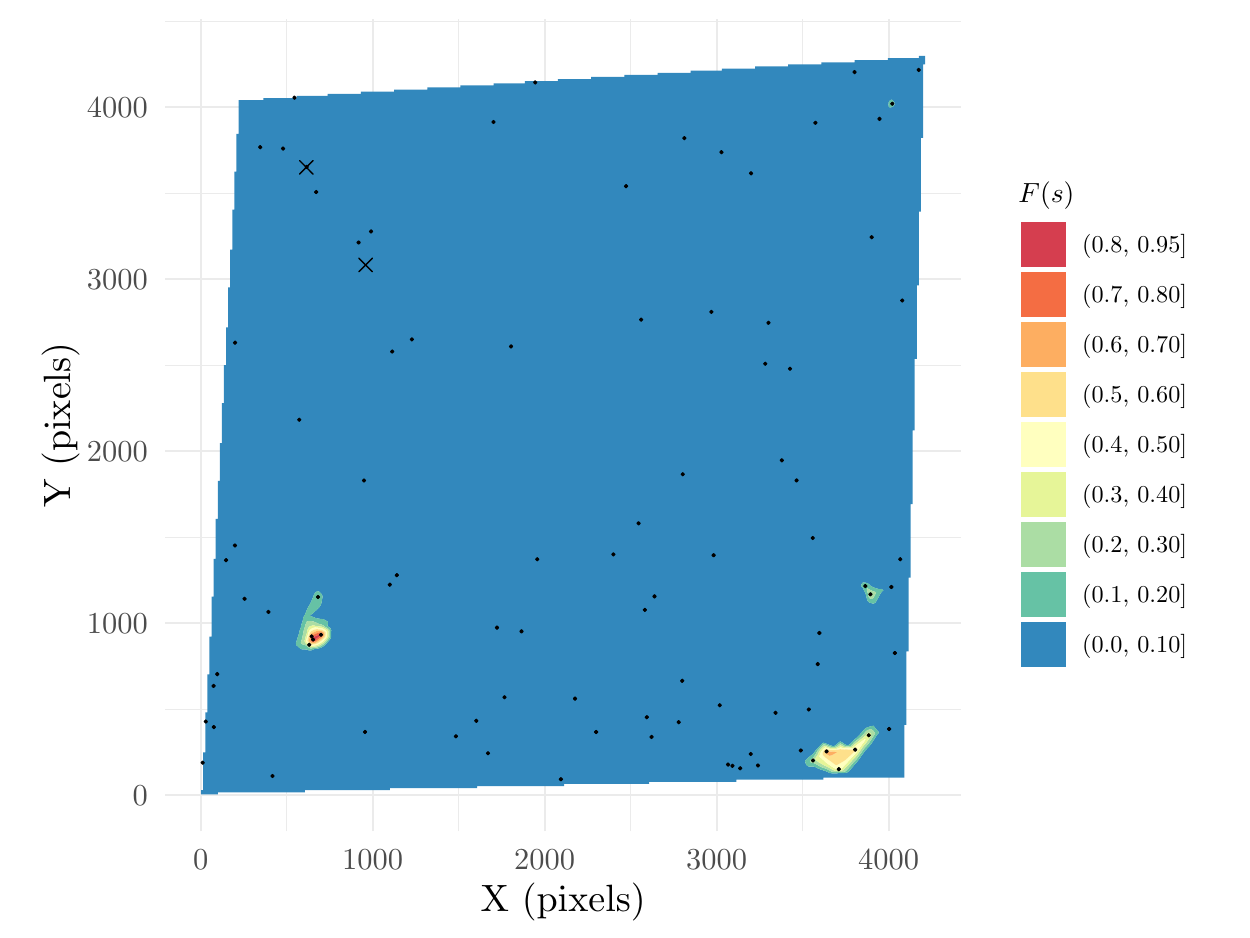}{0.5\textwidth}{(a) $F_{Q(0.5)}(s)$}
           \fig{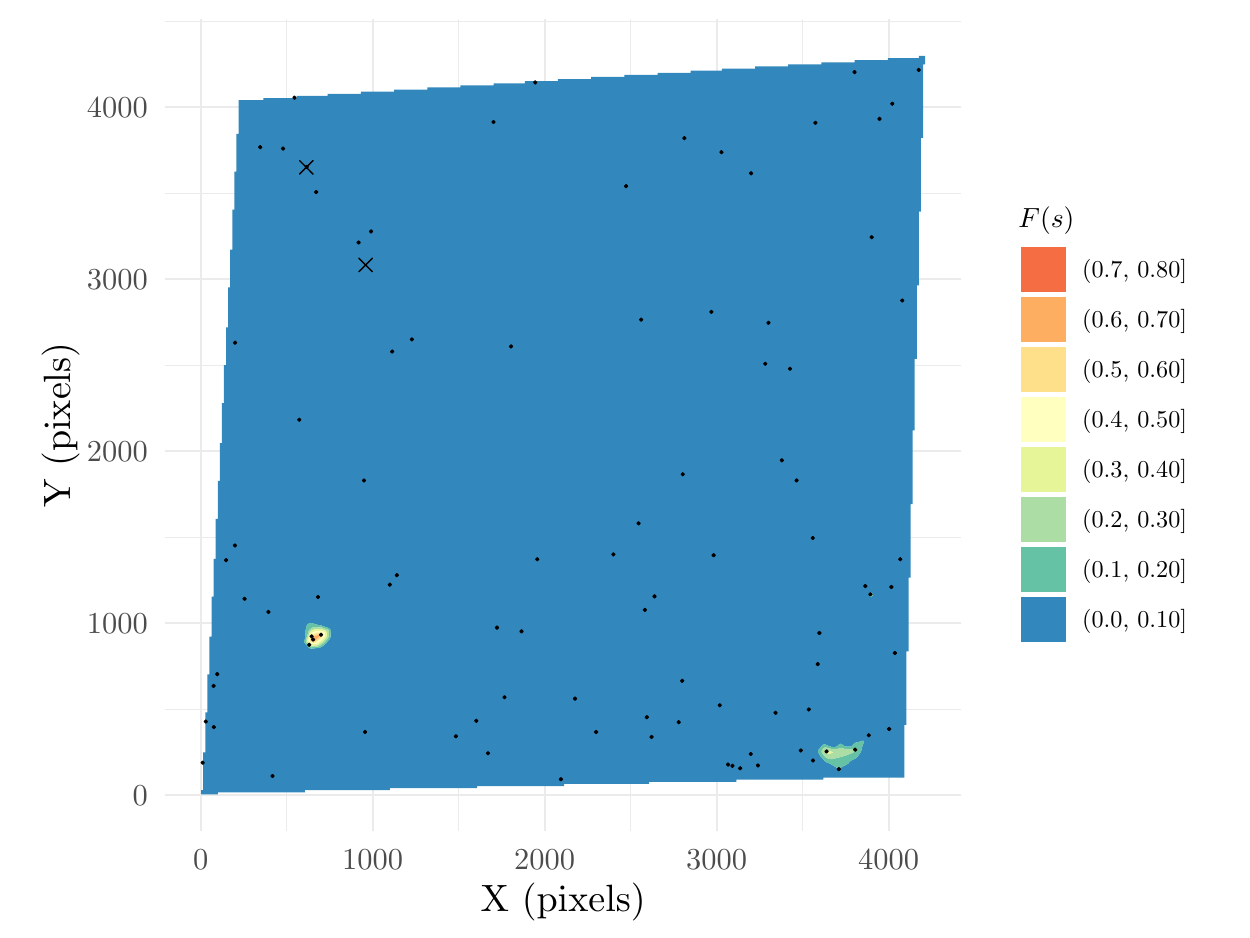}{0.5\textwidth}{(b) $F_{Q(0.75)}(s)$}}
 \gridline{\fig{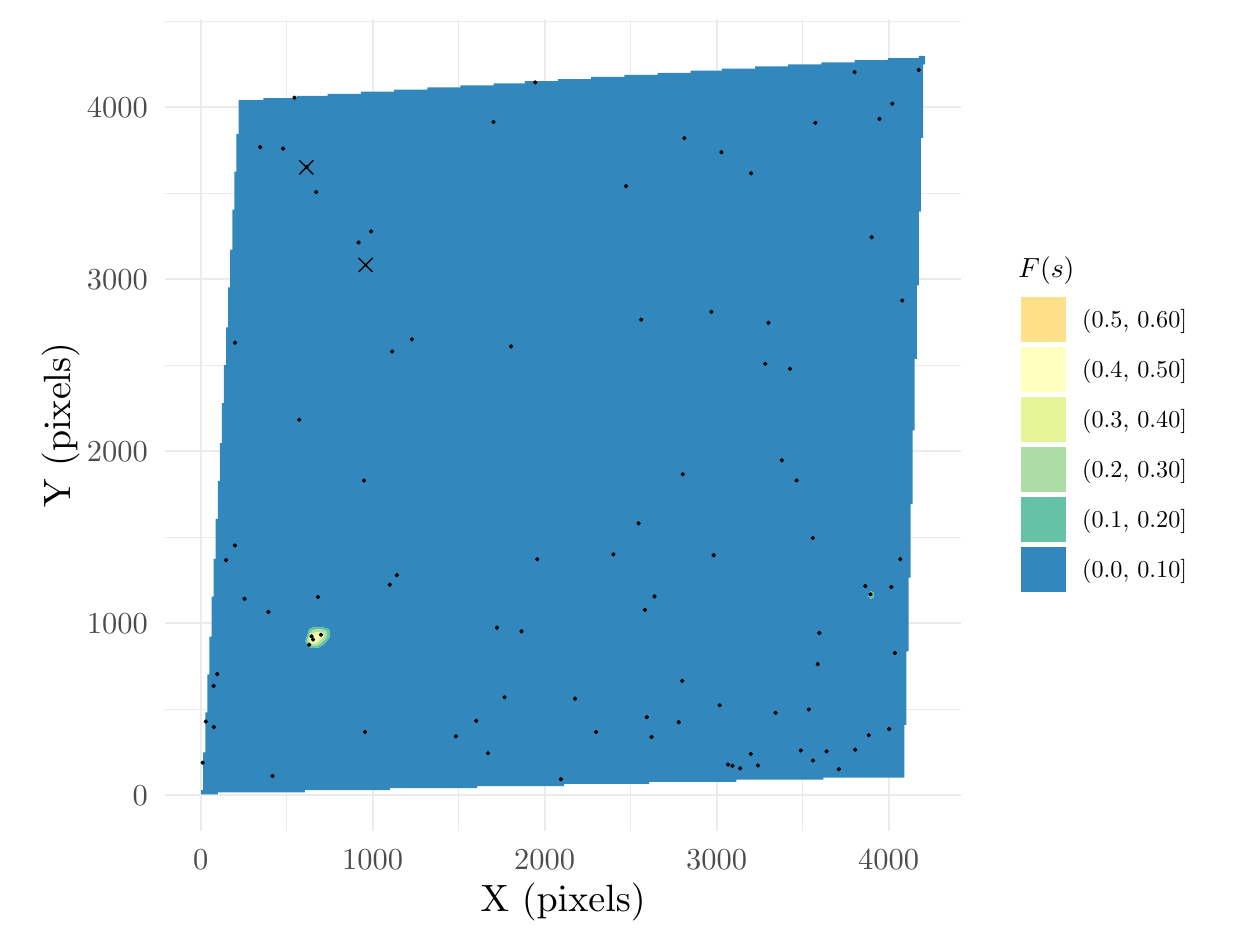}{0.5\textwidth}{(c) $F_{Q(0.9)}(s)$}
           \fig{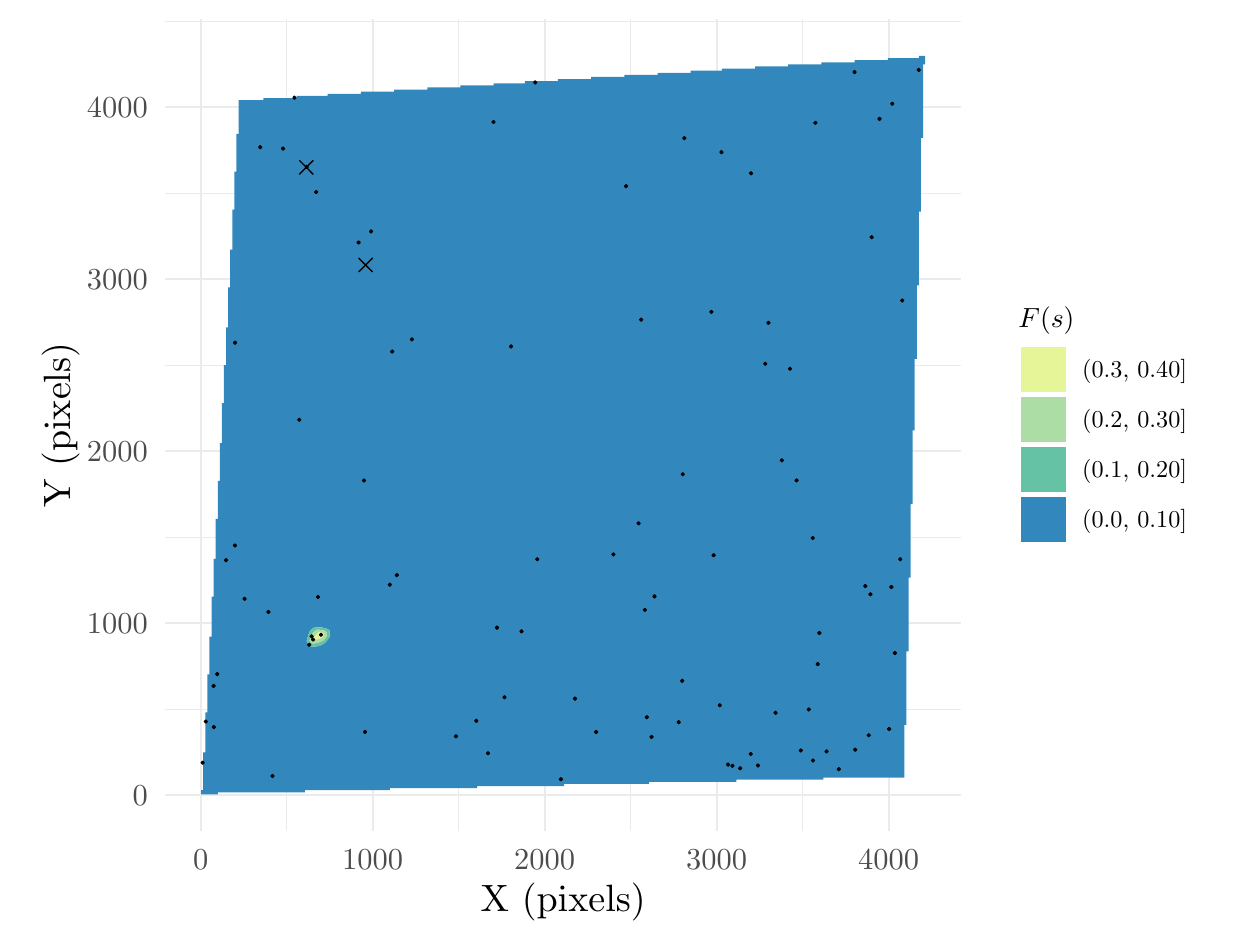}{0.5\textwidth}{(d) $F_{Q(0.95)}(s)$}}
\caption{Excursion functions for V8-ACS pointing. 
Black points are the locations of GC candidates from the PIPER survey. (a)-(d) are excursion functions for $\mathcal{U}(s)$ with four quantile excursion thresholds.
Two known UDGs (WUDG-28 and WUDG-29) are in the upper left corner but not detected significantly through GC counts. The weaker signals in the lower-right corner in (a) and (b) are most likely contaminating GCs from the neighboring giant radio galaxy NGC 1265 (see text for details). The ``dark galaxy'' candidate \mbox{CDG-1} is clearly shown in the lower-left corner of all the panels.}
\label{fig:v8_results}
\end{figure*}

Here we present the detection results for \mbox{CDG-1}. Figure \ref{fig:v8_Us} and \ref{fig:v8_results} show the results for the V8-ACS pointing. Figure \ref{fig:v8_Us} shows the posterior mean spatial random effect $\exp(\mathcal{U}(s))$ and Figure \ref{fig:v8_results} presents the excursion functions for $\mathcal{U}(s)$ computed using the four quantile excursion thresholds. The highest intensity region in the lower-left corner of Figure \ref{fig:v8_Us} is the location of \mbox{CDG-1}. Figure \ref{fig:v8_results} confirms with high probability across all thresholds that \mbox{CDG-1} is indeed something worth our attention. Although there are some other signals (lower-right corner) in the excursion function when the excursion threshold is low, these signals are rather weak in comparison with \mbox{CDG-1}. They also completely disappear ($< 0.1$) for higher thresholds and only the signal from \mbox{CDG-1} persists. 

The weaker signals in Figure \ref{fig:v8_results} seem to be arising from the remote halo region of the massive ETG NGC~1265.  This giant galaxy is $\simeq 120$ kpc away, well outside the field of view of the ACS, but the globular cluster systems of giant ellipticals often extend more than 100 kpc outward.\footnote{A giant galaxy with the luminosity of NGC 1265 would normally have a population of $\sim 5000$ GCs for a GC specific frequency $S_N \simeq 4$ \citep{Harris+2013}.  Though there are individual differences from galaxy to galaxy, some dozens of GCs could easily extend beyond 100 kpc galactocentric radius, a segment of which seems visible in the V8-ACS pointing.}   This weak signal can be safely regarded as contaminating noise. We did not include NGC~1265 as a covariate because there is also a bright foreground star at the lower-right corner in the same direction. Thus, modelling the GC system of NGC 1265 would be extremely difficult and add additional unwanted uncertainty. In this sense, the actual signal strength of \mbox{CDG-1} is quite strong as we did not remove the contaminating noise from NGC~1265.

A similar but lower-amplitude issue is that a smaller ETG, the lenticular IC 312, lies just off the
lower left corner of V8-ACS and generates a contaminating signal of its own. Due to its closeness to \mbox{CDG-1}, IC~312 is included as a covariate effect in our model. Further comments about IC 312 and NGC 1265 are given below.

\subsection{Is \mbox{CDG-1} Real?}

Despite the strong signal from \mbox{CDG-1}, it is possible the four GCCs in \mbox{CDG-1} are a random clustering signal arising from a PPP, i.e., they are a random association of intergalactic GCs belonging to the Perseus Intracluster Medium (ICM). In V8-ACS and the adjacent V8-WFC3, the mean intensity of GCs in the magnitude and color range we are using ($22.0 <$ F814W$_0$ $< 25.5$, $1.0 <$ (F475W$-$F814W)$_0$ $< 2.4$ as defined above) is $\sim 13.7$ per arcmin$^2$. If we make the simplest assumption that the GC distribution in V8-ACS follows a HPP, then the probability that four random GCs would fall within a similar area as in \mbox{CDG-1} is $\sim 3\times 10^{-6}$, just a few parts per million.\footnote{Here we assume the area is the smallest circle encompassing all GCCs in \mbox{CDG-1}, which leads to an area of $0.0074$ arcmin$^2$. Following condition 1 in Appendix \ref{sec:ppp} with $\lambda = 13.7$ per arcmin$^2$ and $|B| = 0.0074$ arcmin$^2$ would give the estimated probability by computing $\mathbb{P}(N(B) = 4)$.} However, this probability is an overestimate since V8-ACS and V8-WFC3 are contaminated by multiple ETGs; the actual GC mean intensity in the ICM should be much lower than $13.7$ per arcmin$^2$ and so should the previously computed probability. After a crude removal of regions with contaminating ETGs, the previous probability drops to $\sim 1 \times 10^{-6}$.

Another alternative is that the \mbox{CDG-1} clump may be some sub-structure that belongs to the halo of either IC~312 or NGC 1265, the bigger galaxies that are in the neighborhood. 
We need the complete GC catalogs of both galaxies to test  the association of \mbox{CDG-1} further, but these are not yet available.

Nevertheless, based on the observed features and location of \mbox{CDG-1}, it is unlikely that \mbox{CDG-1} is a sub-structure in the halo of either galaxy; If it is, then it is either an infalling satellite or a clump of GCs within the larger galaxy. An infalling satellite, however, should still exhibit diffuse light.

\mbox{CDG-1} is also unlikely to be a clump of GCs belonging to either galaxy. For NGC~1265, \mbox{CDG-1} is too far ($120$ kpc) from the center of the giant ETG for its GC system to have such an over-dense clump of GCs. For IC~312, \mbox{CDG-1} has a projected distance of $\sim 26$ kpc to its center (assuming distance to IC~312 is $75$ Mpc) while IC~312 has a half-light radius of $\sim 2$ kpc. If we assume the ``five times" scale ratio between the half-number radius of the GC system and the half-light radius of IC~312, the half-number radius of its GC system is then $\sim 10$ kpc. From the Hyperleda catalog, the total absolute magnitude of IC~312 is $M_B = -20.32$ or $M_V \simeq -21.22$.  For a specific frequency $S_N = 2$ typical for galaxies in that range \citep{Harris+2013}, the estimated total GC population of IC~312 would be $N_{\rm GC} \sim 600$. Of these, about 160 will be brighter than our adopted limit of F814W at 25.5 by assuming a normal Gaussian-like luminosity distribution for the GCs in IC~312. Moreover, \mbox{CDG-1} lies in the direction of the minor axis of IC~312 while IC~312 has an ellipticity of $\sim 0.6$. If the result in \cite{Wang+2013} is assumed where the GC distribution follows the galaxy isophotes, then a very generous estimate of the number of GCs lying further than \mbox{CDG-1} would be at most six. This suggests it is statistically improbable for IC~312 to have four GCs lying as far and clumping as close as in \mbox{CDG-1}.

One last possibility can also be considered: the four GCs defining \mbox{CDG-1} are a group of faint background galaxies whose near-stellar morphologies and colors happen to match those of GCs seen at the Perseus distance.  Close inspection of the objects in the region adjacent to \mbox{CDG-1} shows that other similarly faint and obviously non-stellar objects are present.  However, similarly faint background galaxies are scattered everywhere across the field and the region close to \mbox{CDG-1} does not particularly stand out.  The majority of these faint background galaxies are also much redder or much bluer than the color range of true GCs \citep{Harris2020}. Any Milky Way foreground stars are almost all distinctly redder than the GC range, so the same argument applies for them \citep[see][for more extensive discussion of the background population]{Harris2020}. Though this option cannot be definitively ruled out at present, it must be viewed as unlikely. Nevertheless, we carefully inspect the properties of the GCs in the next section.

\subsection{Properties of the GCs in \mbox{CDG-1}}

\begin{deluxetable*}{lccccccc}[ht]

\tablecaption{Data for the globular cluster candidates in \mbox{CDG-1}.\label{GC_candidates}}
\tablenum{3}

\tablehead{\colhead{ID} & \colhead{R.A.} & \colhead{Dec.} & \colhead{F814W$_0$} & \colhead{(F475W$-$F814W)$_0$} & \colhead{$r_h$} & \colhead{$\log(L/L_{\odot})$} & \colhead{[Fe/H]} \\
\colhead{} & \colhead{(J2000)} & \colhead{(J2000)} & \colhead{(Vegamag)} & \colhead{} &  \colhead{(pc)} & \colhead{} & \colhead{(dex)} }
\startdata
GCC-1 & 3 18 12.23 & +41 45 58.03 &	24.709 $\pm$ 0.035 & 1.445 $\pm$ 0.088 & 4.3 & 5.49 & $-1.19 \pm 0.28$ \\ 	
GCC-2 & 3 18 12.43 & +41 45 59.55 & 	23.535 $\pm$ 0.046 & 1.606 $\pm$ 0.066 & 6.7 & 5.97 & $-0.68 \pm 0.21$ \\ 	
GCC-3 & 3 18 12.06 & +41 45 57.60 & 	24.926 $\pm$ 0.050 & 1.446 $\pm$ 0.098 & 6.7 & 5.41 & $-1.19 \pm 0.31$\\ 	
GCC-4 & 3 18 12.29 & +41 45 57.23 &	24.773 $\pm$ 0.039 & 1.684 $\pm$ 0.076 & 6.5 & 5.48 & $-0.44 \pm 0.24$\\
\enddata

\end{deluxetable*}

The detailed aspects of the four globular cluster candidates (GCCs) in \mbox{CDG-1} are provided in Table \ref{GC_candidates}. Specifically, we provide their coordinates, magnitudes and colors, luminosities, half-light radii, and metallicities.

The four GCCs in \mbox{CDG-1} have magnitudes and color indices appropriate for genuine globular clusters.  Very close inspection of their morphologies on the images with \emph{iraf/imexamine} shows that all of them have profiles with full-width-at-half-maximum (FWHM) $\simeq 2.2$ pixels, whereas an ideal star-like object would have FWHM $= 2.0$ px.  The GCCs may then be marginally resolved.  
To quantify their intrinsic scale sizes, we applied ISHAPE \citep{Larsen99}, which convolves a \citet{King62} profile with the PSF to produce a best-fit profile for each object and hence a half-light radius.  The results for the four targets are presented in Table \ref{GC_candidates} under the column $r_h$.  These scale radii are about twice as large as the most common GCs in the Milky Way, for example, but fall well within the range of observed GCs considered more widely, particularly in dwarf galaxies \citep[e.g.,][]{Georgiev10}.  A more extended
discussion of the $r_h$ distribution for the intergalactic GCs in Perseus is given by \citet{Harris2020}, for which $r_h \sim 5-7$ pc is also a very typical value.

The four candidates, as GCs, are moderately bright:  converting their F814W magnitude to luminosity gives the $\log(L/L_{\odot})$ values in Table \ref{GC_candidates}. These are similar to the bright Messier-type GCs in the Milky Way.  If these four candidates are the brighter members of a GC population that follows anything like a normal luminosity distribution, then deeper exposures could be expected to reveal many less luminous ones and to add to the known population. Such data would enable a strong test of our interpretation.  Conversely, they may turn out to resemble the very top-heavy GCLF seen in NGC1052-DF2 and DF4 \citep{Shen2021}, where clusters fainter than the classic turnover point are present but at surprisingly low numbers.

Since GC integrated color is primarily a function of their heavy-element abundance, the de-reddened color indices (F475W$-$F814W)$_0$ can be converted to give rough estimates of metallicity [Fe/H] \citep{Harris+2016}.  These are listed in the last column of Table \ref{GC_candidates} and again fall in the typical range for luminous GCs, though on average they are somewhat more metal-rich (redder) than most of the ones found in previously known dwarfs in the Local Group, Virgo, and nearby groups \citep[e.g.][]{Georgiev10,Peng+2006}.  However, the photometric data from \citet{Peng+2006} show that a significant range of colors (metallicities) can be found for the GCs in any one dwarf.

In brief, the available data and inferred properties are consistent with the interpretation that the four objects within \mbox{CDG-1} are genuine GCs.  A clinching argument for their identities, though observationally quite challenging, would be direct measurement of their radial velocities.

\subsection{Comparison to GCs in Other Systems}

Aside from the spatial features, the colors of GCCs in \mbox{CDG-1} should also satisfy certain properties if \mbox{CDG-1} is indeed a dark UDG or dark galaxy. Based on the observation in \cite{Peng+2006}, GC color dispersion can exhibit quite a range of variation within dwarfs. Hence, we should expect the GC color dispersion within \mbox{CDG-1} to be less than the GC color dispersion of the whole UDG population. However, since there are only four GCs in \mbox{CDG-1}, any statistical tests conducted to determine color dispersion differences has little meaning. Especially since it is much more likely to produce a false negative result than a false positive result for small sample size. Nevertheless, a simple numerical comparison shows that the GC color dispersion of \mbox{CDG-1} is not out of the ordinary with a mean color of $1.546$ and a dispersion of $0.119$. The average GC color for the ten known UDGs detected in this paper is $1.503$ and the average color dispersion is $0.143$. Compared to the ten known UDGs, the GC color dispersion in \mbox{CDG-1} is lower than eight of them. Hence, one should not be surprised if \mbox{CDG-1} is indeed some sort of UDG or dwarf. 

We also compare the color distribution of \mbox{CDG-1} GCs and that of GCs within IC~312. This requires the complete GC data of IC~312, but there are only five GCs within V8-ACS that should belong to IC~312. The colors of these five GCs are $1.813, 1.85, 1.845, 1.502$, and $1.384$. Again, statistical tests for differences between the color distribution is not meaningful due to small sample sizes. Nevertheless, the average color of the five GCs in IC~312 is $1.678$ with standard deviation of $0.219$. Comparing this with the four GCs in \mbox{CDG-1}, we see that, at least with the available data, the \mbox{CDG-1} GCs are bluer with a much smaller color dispersion than IC~312 GCs. 

In brief, observational evidence in hand is entirely consistent with the interpretation that \mbox{CDG-1} is a dark -- or, at least, extremely faint -- galaxy consisting of at least four GCs and (as yet) no other stellar population to the current limit of detection. Deeper imaging should be capable of reinforcing (or disproving) this hypothesis. Moreover, for a normal GC mass-to-light ratio $(M/L)_V \simeq 2$, the total mass of the four GCs in \mbox{CDG-1} adds up to $M_{\rm GCS} = 3.3 \times 10^6 M_{\odot}$.  The normal relation between GC system mass and host galaxy stellar mass, for dwarf galaxies that contain GCs, is 
\begin{equation}
    {\rm log} M_{\rm GCS} = -0.725 + 0.788 {\rm log} M_*
\end{equation}
(Eadie et al. 2022 in progress). This relation would predict that \mbox{CDG-1} should show up as a dwarf with $M_* \simeq 2.7 \times 10^9 M_{\odot}$, which would be easily visible on the HST images.  Similarly, if the standard near-uniform ratio between GCS mass and host galaxy halo mass 
of $M_{\rm GCS}/M_h = 2.9 \times 10^{-5}$ \citep{Harris+2017}
applies in the dwarf regime \citep{Forbes+2018,Burkert_Forbes2020}, then \mbox{CDG-1} should have had $M_h \sim 10^{11} M{\odot}$ at time of formation.  The fact that nothing is visible of any dwarf now may then be a signal that considerable stripping of its stellar mass has happened since.

\section{Simulation}\label{sec:simulation}

In this section, we analyze the performance of LGCP for UDG detection in more general scenarios by simulating the GC systems of UDGs, assuming the GC system is dictated by a S\'{e}rsic model. The question of interests here is: if there is a UDG in a certain region (in our case this would be the observational pointing made by \textit{HST}), how well can LGCP detect the UDG given its physical structure? In other words, given the S\'{e}rsic model, we seek to understand how UDG detection is affected by (i) the S\'{e}rsic index ($n$), (ii) the half-number radius ($R_e$), and (iii) the number of GCs within the GC systems ($N_{\rm GC}$). 

Another variable of interest is environment. By environment, we mean the spatial distribution of GCs that do not belong to the UDGs we want to detect, e.g., intergalactic GCs and GCs within normal galaxies. We shall call these GCs noise GCs. Intuitively, one would expect the more noise GCs there are in a region, the harder it is to detect the UDGs. However, generating realistic simulations that encompass the broad range of environments where UDGs exist is quite challenging. Therefore, we simply place simulated UDGs inside the pointings obtained by the PIPER survey. This ensures that the environment variable we consider is realistic while also being computationally manageable. 

There are a total of 20 pointings surrounding the center of the Perseus cluster in the PIPER survey, but for the simulation conducted here, we only consider placing simulated UDGs in the eight pointings listed in Table \ref{tab:AUC_table} (i.e, those containing identified UDGs with sufficient GC populations).

Below we introduce some mathematical notations for our simulation procedures:
\begin{enumerate}
    \item Denote the eight point patterns of GCs obtained by the PIPER survey as $X_k = \{x_{k,1},\dots, x_{k,N_k}\} \subset S_k, \ k = 1,\dots, 8$. $N_k$ is the number of GCs in $X_k$ and $S_k$ is the observational window (field of view) for the $k$-th pointing.
    \item For $S_k$, we select ten locations $p_{k,j} \in S_k, \ j = 1,\dots,10$ to place the simulated UDG. These ten locations are chosen to cover most of $S_k$. Note that regions where there are normal galaxies and confirmed UDGs are avoided.
    \item For location $p_{k,j}$, generate a simulated UDG based on the S\'{e}rsic model with parameters $\boldsymbol{\theta}_i = (n, R_e, N_{\rm GC})_i$. Denote the point pattern of GCs within such a simulated UDG as $U_{kji}$. We then obtain a simulated point pattern $X_{kji} = X_k \cup U_{kji}$. The parameter values are set to: $n=\{0.5, 1, 1.5\}$; $R_e \ (\text{pixel}) = \{80, 110, 140, 170, 200\}$; $N_{\rm GC} = \{8, 12,16, 20, 24\}$. Hence, there is a total of $75$ configurations for $\boldsymbol{\theta}_i$ with $i = 1,\dots, 75$.
    \item Fit the LGCP model to the point pattern $X_{kji}$ and determine the detection measure, $D_{kji}$, of the simulated UDG $U_{kji}$.
    \item Carry out $15$ iterations of step 3 and 4 to obtain a sample of detection measures of the simulated UDG for pointing $k$, location $j$, and parameter configuration $i$. Denote this sample as $\boldsymbol{D}_{kji} = \{D_{kji}^l\}_{l=1}^{15}$. We then integrate out the variations in the locations $j$ and the iterations $l$ by computing the average detection measure of the simulated UDG for the $k$-th pointing and the $i$-th parameter configuration:
    \[
    \bar{D}_{ki} = \frac{\sum_{l=1}^{15}\sum_{j=1}^{10}D_{kji}^l}{150}.
    \]
\end{enumerate}
We then investigate how the environment and the structural parameters of UDGs affect the performance of our method.

Note that the unit of $R_e$ specified in step 3 is set to pixels instead of the physical unit of kpc. The reason for doing this is that the pixel widths in the ACS and WFC3 pointings have different physical scalings, but the dimensions of the observational windows for the two pointings are the same in pixels. The difference in physical scaling causes the same UDG to have different apparent sizes in the ACS and WFC3 pointings. Apparent size here means the size of the UDG with respect to the size of the observation window. The difference in the apparent size will then artificially introduce differences on the effect of $R_e$ in the two camera pointings if the unit of $R_e$ is kpc. Therefore, the solution is to let the simulated UDGs have different physical sizes but the same apparent sizes in different pointings. The values of $R_e$ given in step 3 roughly corresponds to a physical sizes in kpc of $\{1.45, 2, 2.5, 3.1, 3.6\}$ in the ACS pointing and $\{1.2, 1.6, 2, 2.5, 3\}$ in the WFC3 pointing.

Indeed, the variable that ultimately affects the UDG detection is not the actual physical sizes of the UDGs but their apparent sizes. To see this, we can consider placing the same UDG at two different distances (assuming the same numbers of GCs are observed) but keeping the size of the observation window unchanged. It is easier to detect the UDG \emph{strictly through its GC} when it is placed at a farther distance than at a closer distance. If the UDG is too close, it can take up the entire observation window, rendering itself undetectable. Although the root of the issue in our problem here is not difference in distance but difference in the pixel scaling, the underlying logic is the same. Note that we do not consider distance as a physical parameter that affects the UDG detection since in this work all UDGs have the same distances in the Perseus cluster.

To ensure the UDG simulation is as realistic as possible, we also consider the \emph{HST} limit for detecting GCs. Specifically, for each of the simulated GCs, we also simulate a luminosity in $I$ band based on the GC luminosity function derived in \cite{Harris2020}. The resulting simulated apparent magnitude in F814W is then obtained by assuming $M_I=$ F814W. The apparent magnitude in F475W is obtained by assuming $M_B = $ F475W and $M_B - M_I = 1.5$ with a Gaussian scatter of $0.25$ \citep{Harris2020}. We then discard the simulated GCs whose brightness is below the \emph{HST} detection limit. The detection limits are given in \cite{Harris2020} where F475W $> 27.67$ and F814W $>27.07$. The remaining simulated GCs will be the GC system that we utilize for analysis. 

\begin{figure*}
    \centering
    \includegraphics[width = \textwidth]{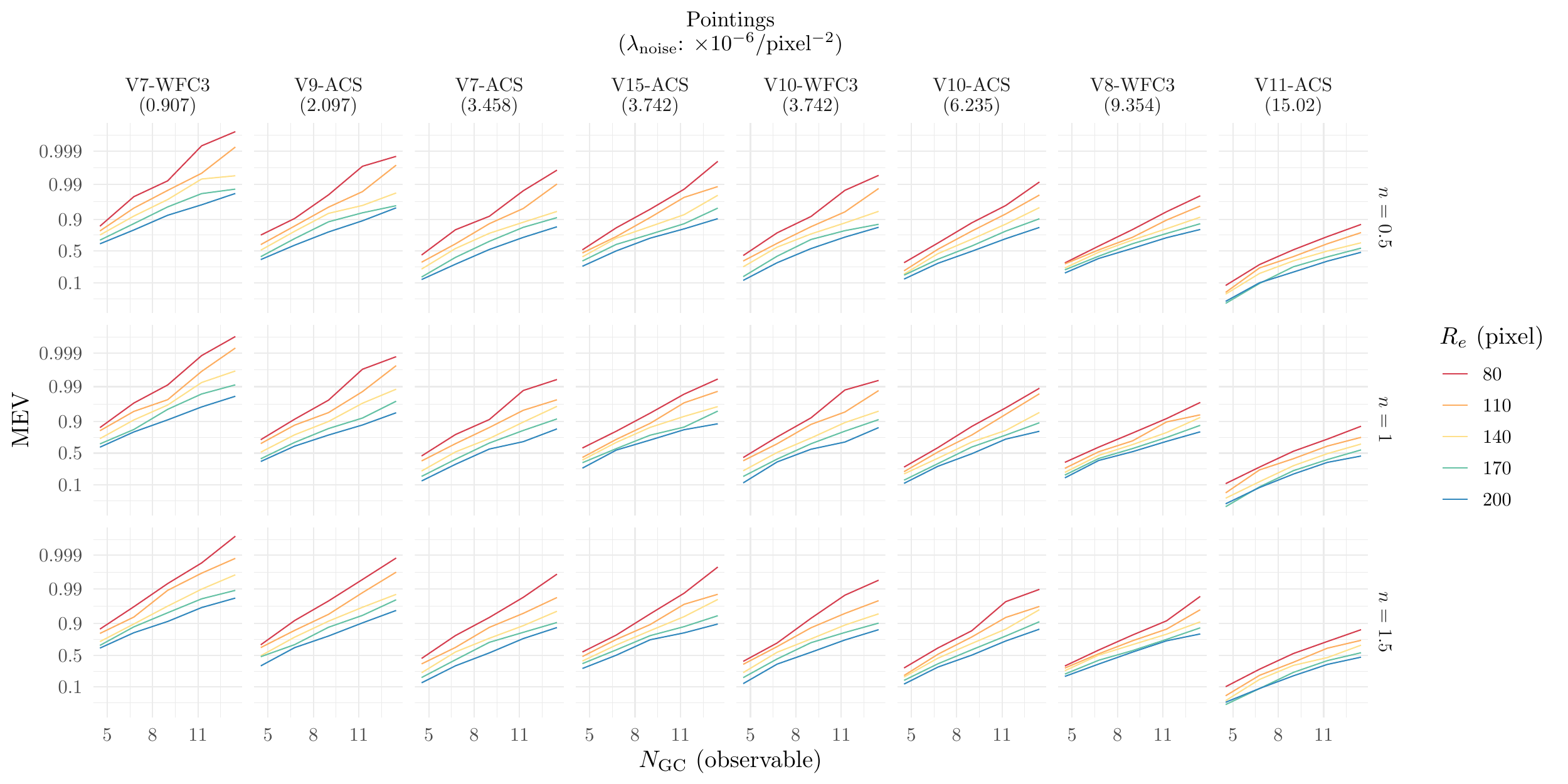}
    \caption{Maximum excursion values (MEV) for the $95$-th quantile threshold against the ``observable" number of GCs ($N_{\rm GC}$), the S\'{e}rsic index ($n$), the GC system half-number radius ($R_e$), and the noise GCs intensity level ($\lambda_{\text{noise}}$). Every row has the same S\'{e}rsic index, every column shows the observational pointing and the corresponding $\lambda_{\text{noise}}$. The MEVs are shown in log-odds scale.}
    \label{fig:sim_MEV}
\end{figure*}

\begin{figure}
    \centering
    \includegraphics[width = 0.5\textwidth]{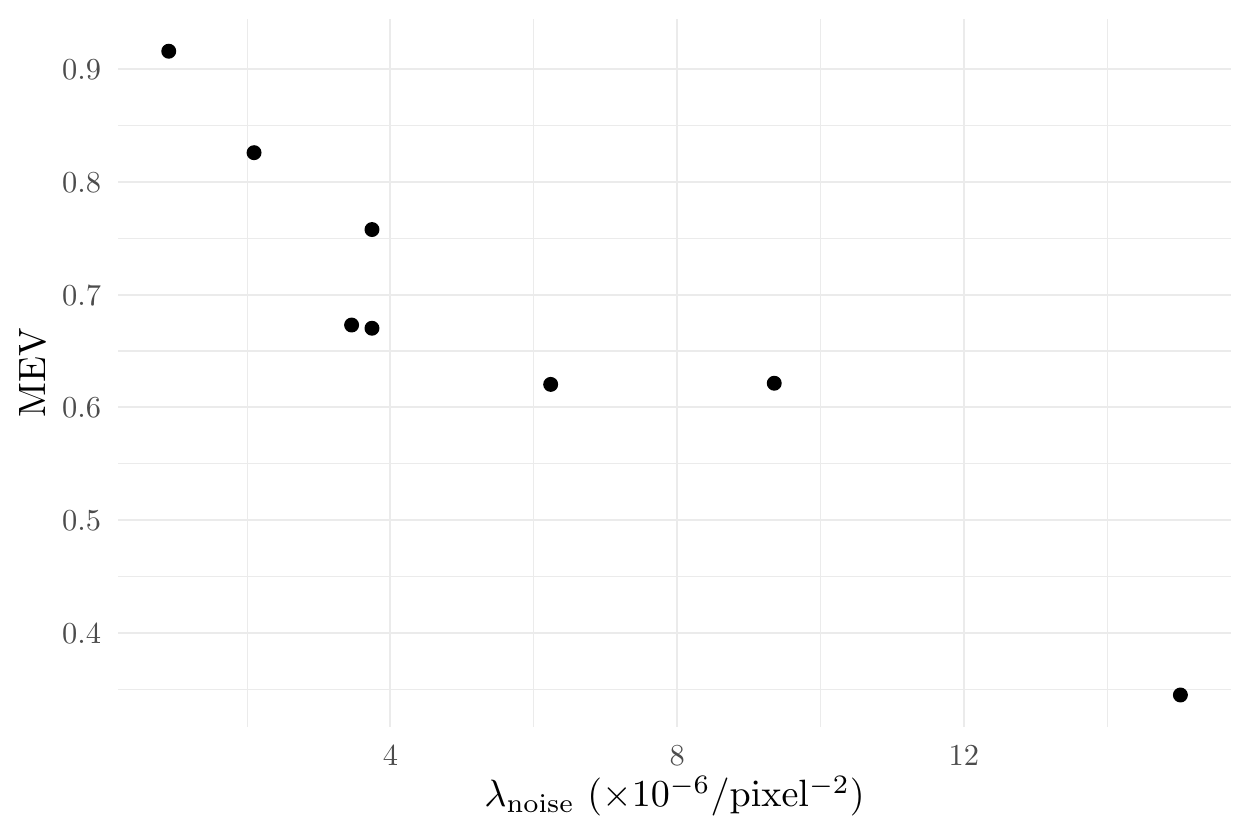}
    \caption{Average MEV for the $95$-th quantile threshold vs. the noise GC intensity level of each pointing.}
    \label{fig:noise_MEV}
\end{figure}

When we remove the GCs that are below the detection limit, the number of GCs, $N_{\rm GC}$, provided in step 3 above is reduced, and the actual ``observable" number of GCs in the simulated GC system is now random. Since the ``observable" number of GCs potentially affects the UDG detection and it is the actual variable we want to investigate, we consider the average number of ``observable" GCs of all simulated UDGs by averaging out all other variables. The ``observable" number of GCs here does not include the background contamination since we know the exact value of observable number of GCs due to simulation.

Before presenting the simulation results, we also require the quantification of the environment (noise GCs) for each $S_k$. Since there can be numerous features characterizing the environment, e.g., number of noise GCs and the presence of normal galaxies, it is difficult to consider all such features. For simplicity, we consider the noise GC intensity level, $\lambda_{\text{noise}}$, as a general characterization of the environment. $\lambda_{\text{noise}}$ here is computed for each $S_k$  and defined as the total number of noise GCs in $S_k$ divided by the area of $S_k$. 

We present the simulation results and analysis next. Similar to the detection measures employed in Table \ref{tab:UDG_F_function} and \ref{tab:AUC_table}, we will examine how the $\lambda_{\text{noise}}$ and the structural parameters affect both the detection probability and the overall performance measured by AUC.

\subsection{Detection Probability}

The detection probability considered here is only restricted to the simulated UDGs instead of real UDGs already in the pointings. Hence, when we compute $\lambda_{\text{noise}}$, the GCs from the real UDGs within a pointing is also considered as noise GCs since these GCs are confounding noise signal with respect to the simulated UDG we try to detect.

The measure of the detection probability is similar to the one used in Table \ref{tab:UDG_F_function}, i.e., for each of the simulated UDG, we consider a circle centered at the location of the simulated UDG with radius equal to the half-number radius of the simulated GC system, and we find the maximum excursion function value within the circle for a given excursion threshold. For simplicity, we shall call this the maximum excursion value (MEV). 

Figure \ref{fig:sim_MEV} shows the results obtained from simulation. To save space, we only present the results for the MEVs computed with the excursion threshold being $Q(0.95)$. The results are more or less similar for other excursion thresholds. Moreover, we found that the MEVs for lower levels of excursion thresholds are subject to more noise. This is caused by the relatively small number of iterations ($15$ times). Higher levels of excursion thresholds are more resistant to such a noise since the range of MEVs for higher threshold is smaller, hence the variance of MEVs is smaller. 

As shown in Figure \ref{fig:sim_MEV}, the ``observable" number of GCs heavily influences the detection probability. As expected, the more GCs there are within a UDG, the easier it is to detect the UDG. Moreover, Figure \ref{fig:sim_MEV} shows that the MEV is above $0.5$ if the ``observable" number of GCs is $\gtrsim 13$. We focus on the value of $0.5$ as a reference point since a region with MEV greater than $0.5$ means that it exceeds the excursion threshold more often than not. If we exclude the simulation results for V11-ACS, we only need the ``observable" number of GCs to be $\gtrsim 9$ for the MEV to be above $0.5$. Hence, the performance of LGCP for detecting UDGs with larger GC populations is of no concern. Moreover, since the results we have shown here are for the highest excursion threshold --- $Q(0.95)$, the resulting MEVs are inevitably low. The results obtained for the excursion threshold $Q(0.5)$ indicate that the MEVs are above $0.5$ for all observed number of GCs except for V11-ACS. This means that in using the excursion threshold $Q(0.5)$, we are very likely to detect UDGs in most scenarios, as long as the ``observable" number of GCs is $\gtrsim 4$. This is an extremely promising result; the majority of the pointings within which we want to detect UDGs, they are unlikely to be as noisy as V11-ACS (at least within an environment similar to the Perseus cluster). Since our ultimate goal is to discover nearby UDGs with median to high number of GC population, the results from the simulation indicate we should have no issue detecting such a UDG using LGCP, if one does exist. One caveat of using a lower threshold such as $Q(0.5)$ is its tendency to produce false positive signals. Hence, our advice is to look at the excursion function for all thresholds and focus on signals that persists through all thresholds.

Figure \ref{fig:sim_MEV} shows that the half-number radius also has a strong negative effect on the UDG detection. This coincides with our intuition since the larger the half-number radius, the more dispersed the GCs within UDGs would appear, which reduces the strength of clustering of the GCs. To better detect UDGs with larger sizes, we can then look at the detection probability for lower excursion thresholds. Again, the MEVs obtained for the median ($Q(0.5)$) threshold are above $0.5$ for all half-number radii except for V11-ACS. Another method to mitigate the negative effect of larger sizes on UDG detection is to consider a larger observation window, since the relationship we have discovered here is regarding the apparent sizes of the UDGs in a given observation window. A UDG with a smaller apparent size in an observation window would be easier to detect. However, we certainly need to balance the noise increase introduced by increasing the size of the observation window.

\begin{figure*}
    \centering
    \includegraphics[width = \textwidth]{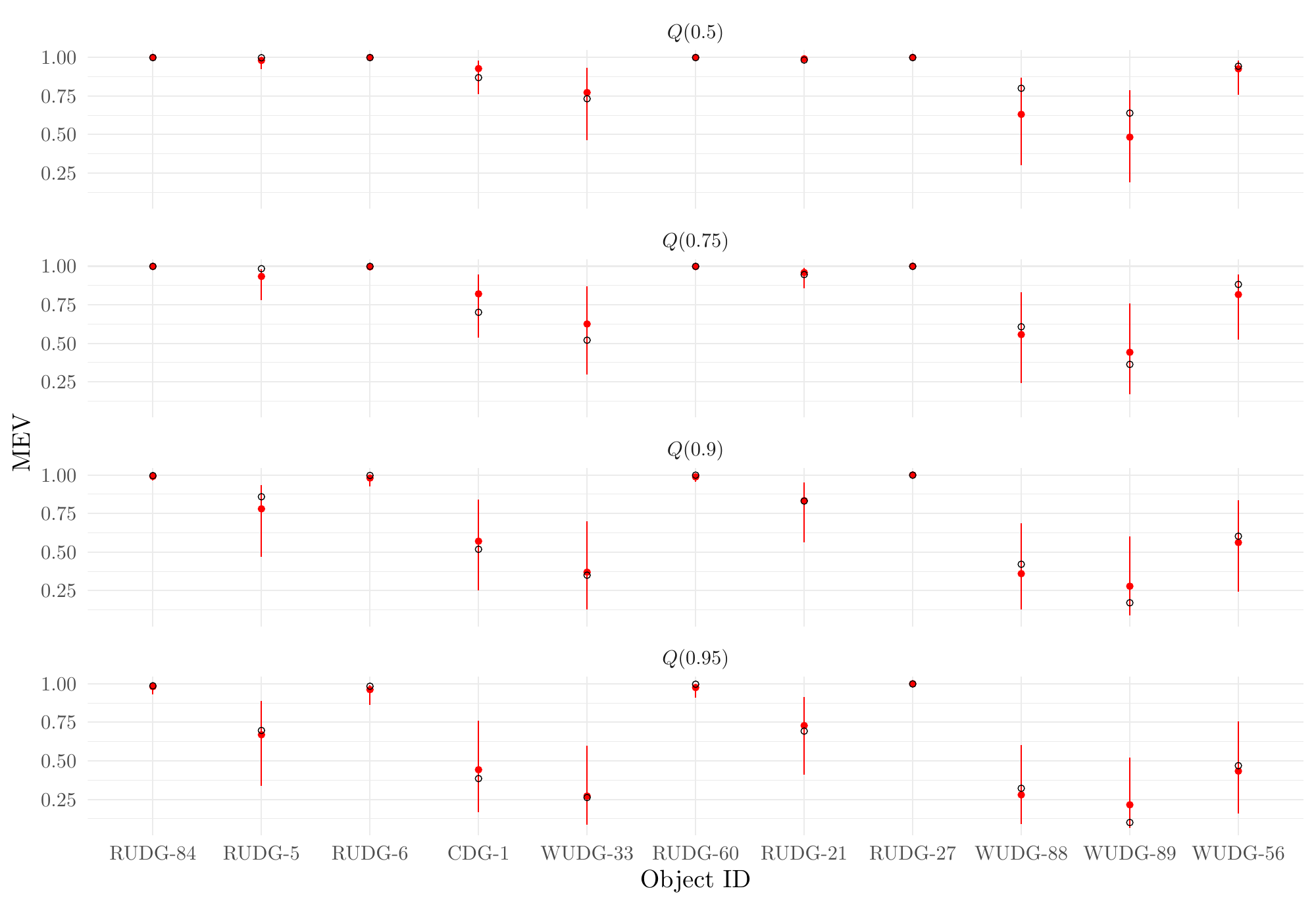}
    \caption{Predicted MEVs for all detected UDGs given in Table \ref{tab:UDG_F_function} using Model \ref{prediction_model} fitted with the simulated data. Red dots represent the predicted values while red vertical lines indicate the $95\%$ prediction intervals. Black circles show the true MEVs obtained in Table \ref{tab:UDG_F_function}.}
    \label{fig:F_pred}
\end{figure*}

Unlike the number of GCs and the half-number radius, the S\'{e}rsic index does not appear to have any significant impact on the detection probability at all. This result is somewhat surprising since higher S\'{e}rsic indices would lead to a more concentrated distribution of GCs within a UDG, hence increasing the clustering strength of GCs. We tested whether this is due to the range of S\'{e}rsic index being too restrictive. Hence, we performed another smaller set of simulations (results not shown here) with the S\'{e}rsic index ranging from $\{0.5, 1, 1.5, 2, 2.5, 3\}$. The result is similar to the one shown in Figure \ref{fig:sim_MEV}. We believe that the S\'{e}rsic index does have an impact on UDG detection, but the effect would only be apparent when we compare, say, $n = 0.5$ to $n = 8$. For the vast majority of currently detected UDGs, few have S\'{e}rsic indices exceeding $3$. Therefore, we conclude that the S\'{e}rsic index is not a strong factor in UDG detection using LGCP.

Lastly, in Figure \ref{fig:sim_MEV}, we have also ordered the results of each pointing in ascending order of $\lambda_{\text{noise}}$. We can see that as $\lambda_{\text{noise}}$ increases, the MEVs decreases, which is expected. UDG detection will be harder in a more contaminated environment. For a better visualization of the numerical relationship, Figure \ref{fig:noise_MEV} shows the average MEVs computed for each pointing (averaging out all other variables) against the corresponding $\lambda_{\text{noise}}$. We can see an almost perfect negative linear relationship exists between the MEVs and $\lambda_{\text{noise}}$. Certainly, there are fluctuations in the relationship, e.g., V15-ACS has a higher average MEV than V7-ACS, while V7-ACS is less ``noisy" than V15-ACS. This is an expected random fluctuation since $\lambda_{\text{noise}}$ is an extremely simple one-number summary of the environment. Upon detailed inspection, V15-ACS consists mostly of empty space where a relatively small region contains the majority of the noise GCs that belong to a dE within the pointing. In contrast, V7-ACS has GCs evenly distributed everywhere. Hence, $\lambda_{\text{noise}}$ is not a summary statistics that can sufficiently highlight the aforementioned difference between V15-ACS and V7-ACS but it is simple enough to provide us with useful insights. 

Aside from the effects of the related parameters on the UDG detection probability, we are also interested in whether the data from simulation can be used to provide us with any prediction capability. Essentially, we are aiming to answer the following interesting and important question: for a UDG with given physical parameters regarding its GC systems, what would be its detection probability under LGCP? Being able to answer this question would ensure our method is applicable in more general settings. We hence fit the following polynomial regression model using the simulated data:
\begin{equation}\label{prediction_model}
    \log\left(\frac{\text{MEV}}{\text{1-MEV}}\right) = P_2(N_{\rm GC}^o, R_e, \lambda_{\text{noise}}, p) + \epsilon.
\end{equation}
$P_2(\cdot)$ here is a second-order polynomial. The observed number of GCs ($N_{\rm GC}^o$), the half-number radius ($R_e$), $\lambda_{\text{noise}}$, and the value $p$ for the quantile excursion thresholds $Q(p)$ are used to construct the basis for the polynomial. We did not include the S\'{e}rsic index since it is demonstrated in Figure \ref{fig:sim_MEV} that the S\'{e}rsic index does not affect the MEVs. $\epsilon$ here is a mean-zero Gaussian error term. Note that $N_{\rm GC}^o$ here is obtained the same way as the $N_{\rm GC}$ values given in Table \ref{tab:UDG_F_function} since in real data we do not know the true number of observable GCs associated with a UDG. We have also fitted the following linear regression model:
\begin{equation}\label{linear_model}
    \log\left(\frac{\text{MEV}}{\text{1-MEV}}\right) = \beta_{N}N_{\rm GC}^o + \beta_R R_e + \beta_\lambda\lambda_{\text{noise}} + \beta_pp + \epsilon,
\end{equation}
but we found that Eq.~\ref{prediction_model} fits the data much better compared to Eq.~\ref{linear_model}. We then compute the predicted MEVs and the $95\%$ prediction intervals of MEVs for all the detected UDGs listed in Table \ref{tab:UDG_F_function} and compare the prediction to the true MEVs listed in Table \ref{tab:UDG_F_function}. The results are shown in Figure \ref{fig:F_pred}.

Figure \ref{fig:F_pred} shows that almost all the true MEVs for the detected UDGs are within the $95\%$ prediction intervals produced by Eq.~\ref{prediction_model}. The UDGs whose MEVs are outside of the prediction intervals are those with extremely high MEVs ($\gtrsim 0.98$), and their predicted MEVs are extremely high as well. Hence, the prediction error is in fact minuscule and negligible. For UDGs with lower true MEVs, the majority of the predictions from the model match with the true MEVs extremely well, except for WUDG-89 in the V11-ACS pointing. The larger prediction error for this UDG is somewhat expected since as mentioned, V11-ACS is a severely contaminated pointing and WUDG-89 is extremely close to the boundary of the giant ETG UGC~02673. Thus, we do not expect the prediction to match the true MEVs well in this case. 

The results in Figure \ref{fig:F_pred} suggest that the simulated data can be quite useful in determining the expected detection probability for any UDGs, at least within an environment similar to the Perseus cluster. The prediction model thus can be used as a screening mechanism for UDG detection: we will obtain the MEVs for UDG candidates in any pointing and then serve the corresponding parameters of the detected candidate into the prediction model and compute the predicted MEVs. If the true MEVs is close or much higher than the predicted MEVs, then it suggests the detection may be real and subsequent confirmation and verification will follow suit.

We also want to highlight the prediction for \mbox{CDG-1}. We can see that the predicted MEVs for \mbox{CDG-1} match quite well with the true MEVs for all thresholds, although there does seem to be a small degree of over-estimation from Eq.~\ref{prediction_model}. We suspect the over-estimation is likely due to the contaminating noise from NGC 1265, which may have driven down the MEVs of \mbox{CDG-1}. Since the pointing (V8-ACS) where \mbox{CDG-1} resides is not used in the simulation, in machine learning language, \mbox{CDG-1} is essentially a test data point for model \ref{prediction_model}. The matching between the predicted MEVs and the true MEVs for \mbox{CDG-1} suggests the prediction produced by model \ref{prediction_model} is quite generalizable.

\begin{figure*}
    \centering
    \includegraphics[width = 0.8\textwidth]{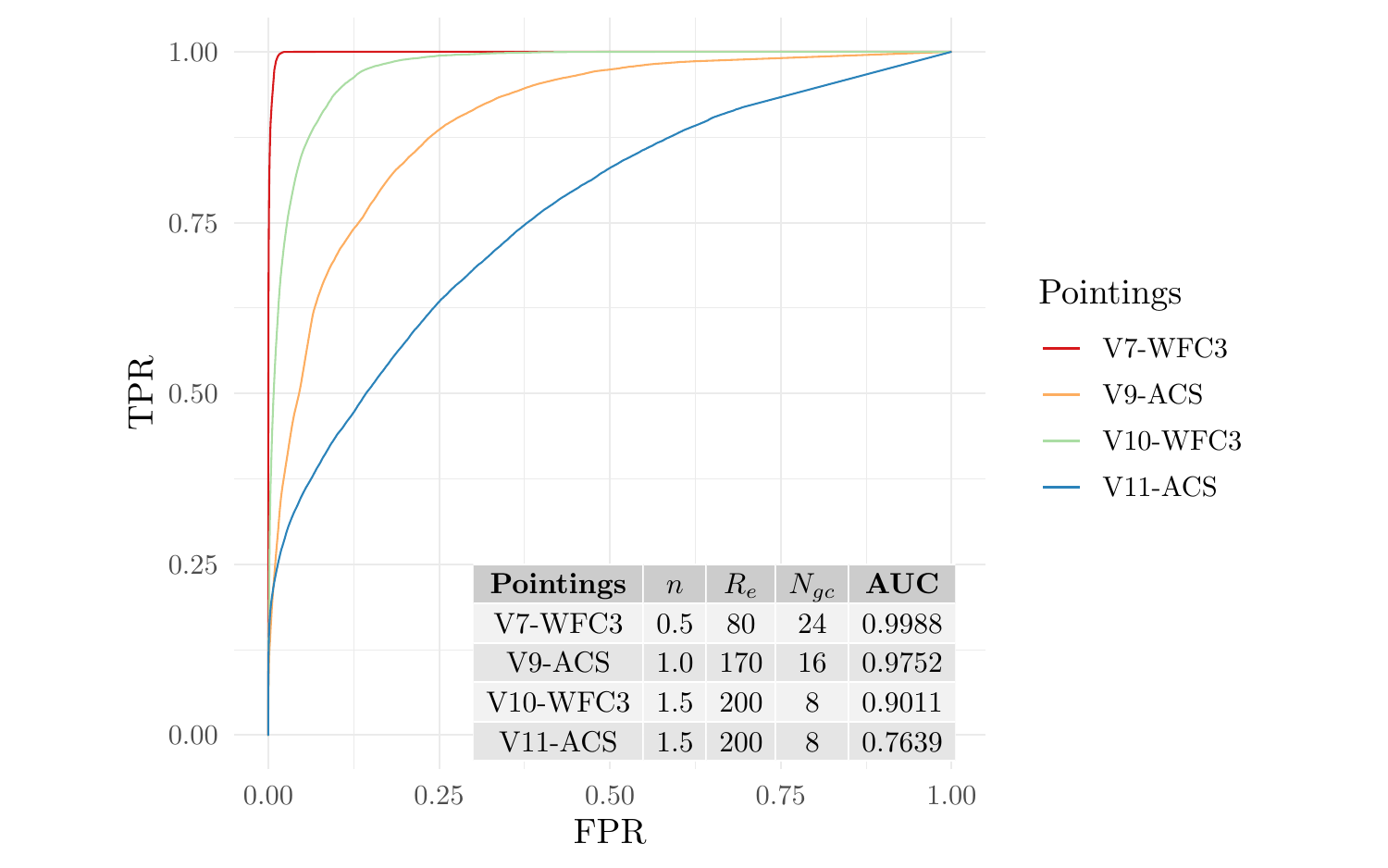}
    \caption{Example ROC curves computed for the $95$-th quantile threshold from four different sets of simulations. The sets of simulations chosen are to represent the complete range of performance of LGCP from best to worst in terms of AUC.}
    \label{fig:ROC_sim}
\end{figure*}

\subsection{ROC Analysis}

In this section, we analyze how the physical parameters affect the overall detection performance of LGCP using ROC analysis. Unlike the analysis for detection probability where we only focused on detecting the simulated UDGs, we consider the detection of both the already identified UDGs and simulated UDGs in each pointing for the overall performance. This is done to ensure we understand how the overall performance of LGCP behaves when there are multiple UDGs in a pointing. Moreover, if we only consider simulated UDGs, the real UDGs in each pointing will be considered as ``false positive" when we conduct ROC analysis, which is not desirable. Therefore, for the ROC analysis, the truth map of each pointing consists of regions with detected UDGs, simulated UDGs , and no UDGs. For regions where there are UDGs, we determine them the same way as in section \ref{sec:ROC_detect} where regions within twice the GC system half-number radius from the center of a UDG are assigned the true values while other regions are assigned false values. 

Since it is not possible to present the ROC curves for all the parameter configurations, we present in Figure \ref{fig:ROC_sim} example average ROC curves obtained for the $95$-th quantile excursion threshold from four different sets of simulations. These roughly encompass the best to worst performance of LGCP in terms of AUC in our simulation. Figure \ref{fig:ROC_sim} shows that even when LGCP has the worst performance, the corresponding AUC has a value of $0.76$. Thus, LGCP classification is still much better than random guessing.

\begin{figure*}
    \centering
    \includegraphics[width = \textwidth]{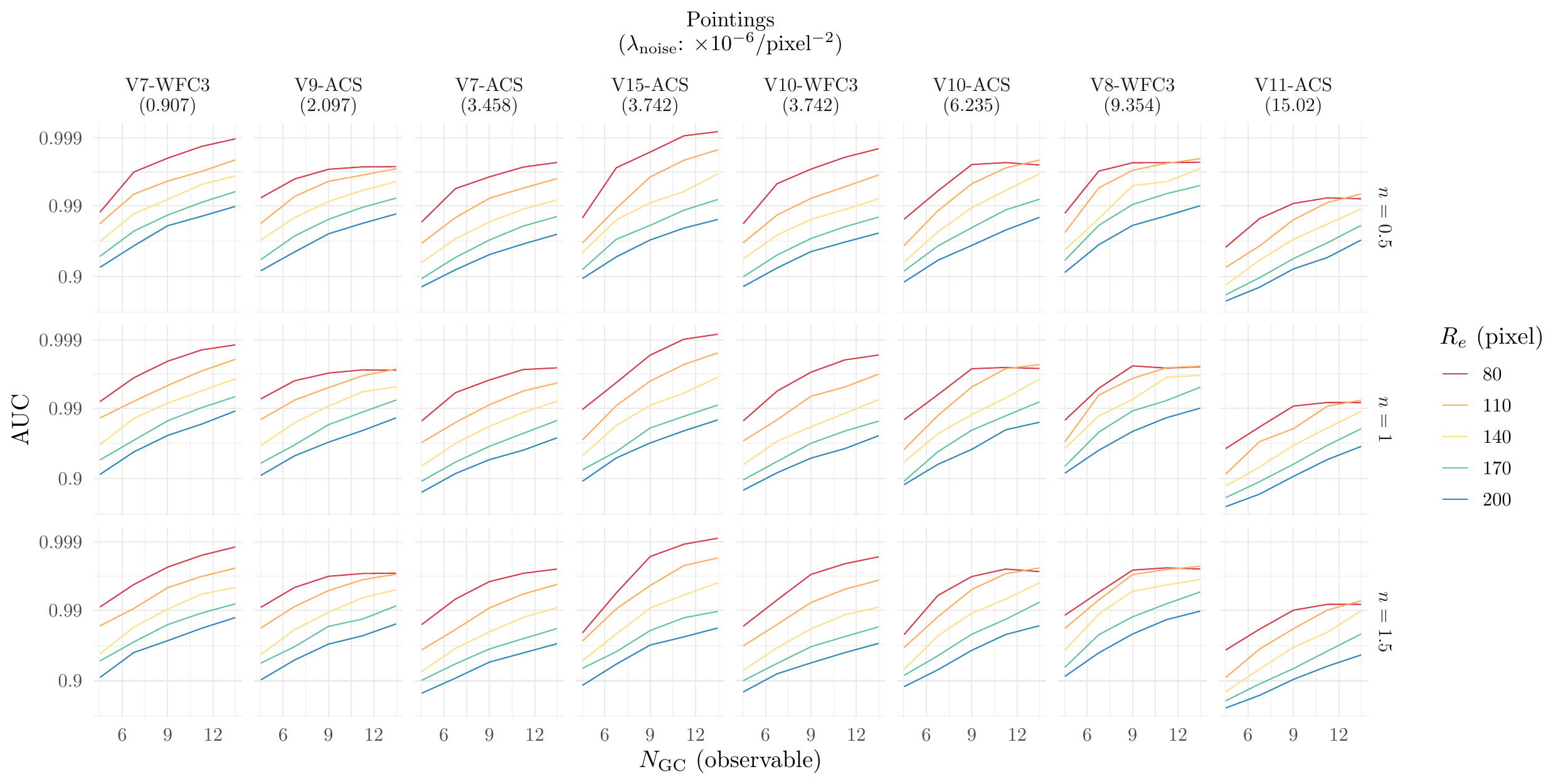}
    \caption{AUCs for the $95$-th quantile threshold against the ``observable" number of GCs ($N_{\rm GC}$), the S\'{e}rsic index ($n$), the GC system half-number radius ($R_e$), and the noise GCs intensity level ($\lambda_{\text{noise}}$). Every row has the same S\'{e}rsic index, every column shows the observational pointing and the corresponding $\lambda_{\text{noise}}$. The AUCs are shown in log-odds scale.}
    \label{fig:sim_AUC}
\end{figure*}

To fully investigate how the overall performance of LGCP is affected by the physical parameters, we consider the AUC values. The AUC values for each parameter configuration are computed based on the simulation procedure described in section \ref{sec:simulation} where AUC is now the detection measure $D$. The results are presented in Figure \ref{fig:sim_AUC}. Note that similar to the detection probability, we also only present the ROC analysis results for the $95\%$ quantile excursion threshold. The ROC analysis results are almost exactly the same across all excursion thresholds.

As shown in Figure \ref{fig:sim_AUC}, similar to the detection probability, the ``observable" number of GCs positively affects the AUC while the half-number radius of the GC system negatively impacts the AUC. The S\'{e}rsic index still does not seem to affect the AUC. The explanation of these observed effects are essentially the same as the ones previously illustrated for detection probability. Furthermore, for the majority of parameter configurations, the AUC values are above $0.9$, indicating that the overall performance of LGCP for UDG detection is quite remarkable. However, despite the similarity to the results for detection probability, the effects of the physical parameters on the AUCs also have a few notable differences. 

Firstly, the effect of the ``observable" number of GCs and the half-number radius on the AUCs are highly non-linear. We can see in Figure \ref{fig:sim_AUC} that the AUCs will reach a plateau as the number of observed GCs increases. We believe this is unlikely due to a physical reason. As mentioned in section \ref{sec:ROC_detect}, the Boolean map obtained by LGCP is highly unlikely to match exactly with the truth map we have employed. Therefore, once the AUC reaches a certain level, it cannot increase any further as there are always a very small number of pixels in the truth map that LGCP simply cannot correctly classify.

Secondly, the noise GCs intensity level ($\lambda_{\text{noise}}$) of each pointing does not seem to have any clear relationship with the AUCs. This is in contrast to the observed negative relationship between $\lambda_{\text{noise}}$ and the detection probability as demonstrated in Figure \ref{fig:sim_MEV} and \ref{fig:noise_MEV}. We suspect the environmental variables affecting the AUC cannot be properly characterized by $\lambda_{\text{noise}}$. We can easily see that one of the environmental variables that would affect the AUC is whether there are over-densities of noise GCs within a pointing since these can become false positive detection and severely affect the AUC. However, $\lambda_{\text{noise}}$ cannot characterize whether a pointing has over-densities of noise GCs or not, and there is no good summary statistics for characterizing such an environmental variable. 

Note that we are not suggesting the $\lambda_{\text{noise}}$ has absolutely no effect on the AUC since we can see from Figure \ref{fig:sim_AUC}, V11-ACS pointing still has the lowest AUC values out of all pointings. Since V11-ACS is extremely contaminated, the high level of noise does ultimately cause the AUC to be low. Another thing to note is that V7-WFC3 and V15-ACS have two of the highest AUC values. These two pointings have large areas where there are no GCs, and they also have no significant clumps of noise GCs. Thus, the resulting AUCs for these two fields are high.

In contrast, the detection probability is a rather locally defined quantity: having over-densities of noise GCs far away from a UDG is unlikely to affect the detection probability of the UDG. However, it will be strongly affected by $\lambda_{\text{noise}}$ of the pointing since, intuitively, raising the base level of noise inevitably reduces the strength of the signal.

Another conclusion we can draw from Figure \ref{fig:sim_AUC} is that we should have no trouble of detecting multiple UDGs within the same pointing. Looking at the highest AUC values in Figure \ref{fig:sim_AUC} for each pointing, they are very similar to the ones obtained in Table \ref{tab:AUC_table}. We focus on the highest AUC values since it ensures the simulated UDGs in the pointings are correctly identified. Furthermore, these cases correspond to the scenarios where the signals from simulated UDGs are the strongest, which makes the signals from real detected UDGs most likely to be drowned out. Nevertheless, the AUC values in Figure \ref{fig:sim_AUC} and Table \ref{tab:AUC_table} are nearly identical, suggesting that aside from the simulated UDGs, the real detected UDGs in each pointing are also detected. 

\section{Summary \& Future Works}\label{sec:conclusion}

In this paper, we have introduced the log-Gaussian Cox process --- a novel method for detecting ultra-diffuse galaxies by identifying the over-density of their associated globular clusters instead of first finding their associated diffuse light. We have demonstrated the usage and effectiveness of the method by applying it to data from the the PIPER Survey in the Perseus cluster. The LGCP was able to successfully detect all the confirmed UDGs with sufficient GC populations within the PIPER Survey field of view. ROC analysis has shown that the method also produced no false positive detection at least within the data used.  

Using our method, we have detected a potential ``dark" galaxy in the Perseus cluster --- \mbox{CDG-1} in our catalog --- which is not visible with the traditional detection method through diffuse light. Examination of the spatial features of its constituent GCs showed it is unlikely a random alignment of intracluster GCs in Perseus. However, it is near a normal lenticular galaxy IC~312 and currently the GC catalog of IC~312 is still proprietary, hence we cannot yet definitively confirm whether \mbox{CDG-1} is some sub-structure of IC~312.  The giant ETG NGC 1265 is also near enough in projection (120 kpc) that its large GC system may overlap the region of \mbox{CDG-1}.  Nevertheless, preliminary analysis on its spatial and color distribution reveal that the clump of objects defining \mbox{CDG-1} is also highly unlikely to be associated with these galaxies. In future work, we intend to obtain deeper imaging of this object and the neighboring galaxies using \textit{HST}. Combined with data from the Subaru Suprime camera once it is available, we aim to test and confirm the nature of \mbox{CDG-1}.

We also analyzed the performance of the LGCP method in terms of true detection and overall performance for UDG detection within the Perseus cluster using simulation. We found that under a S\'{e}rsic model for the simulated GC systems of UDGs, the S\'{e}rsic index does not have any significant impact on the detection probability of UDGs, while the number of GCs has a strong positive effect on the detection probability, and the half-number radius of the GC systems negatively affects the detection probability. We also concluded that the environment where a UDG resides strongly influences its detection probability: as the environment becomes more noisy, the detection probability decreases. Nevertheless, we found that for the majority of the environments considered, LGCP performs remarkably at detecting a UDG. 

Regarding the overall performance of LGCP measured by AUC, we have found that the physical parameters of the GC systems of UDGs have similar influences on the AUCs as they do on the detection probability. However, the effect of the environment on the overall performance of LGCP is rather unclear as it depends on a set of descriptive statistics that are hard to quantify. Moreover, the simulated results showed that the LGCP has no issue of detecting multiple UDGs within a pointing which is a characteristic that is highly desired.

Lastly, other applications of LGCP can also lead to fruitful results. Specifically, LGCP could be used to find ultra-faint satellite galaxies in the Local Group, as well as faint star clusters in the halos of nearby galaxies. In both of these cases, we will be looking for clumps of stars embedded in a background population. The significant advantage of LGCP at tackling these problems is that satellites and star clusters in the halo will be close to their host galaxies, thus the ability of LGCP to incorporate covariate information to exclude the signals from host galaxies can be extremely efficient at detecting the weaker signals from satellites and star clusters.

In conclusion, LGCP is quite a reliable method for detecting UDGs as demonstrated by its efficacy at detecting confirmed UDGs, as well as simulated UDGs. However, the results shown in this paper only serve as a preliminary first step, since the method was only applied to UDG detection in the Perseus cluster. In the future,  work is needed to calibrate the method in a much more general setting. In short, we intend to apply the method to other data sets to better understand the capability of LGCP for UDG detection in different environments other than the Perseus cluster.

Furthermore, more simulation studies would also assist in understanding and calibrating the detection procedure and criteria for UDG detection using LGCP. The ultimate goal here is to detect potential dark UDGs in the nearby Universe which only consist of observable GC populations and are devoid of individual stars. The discovery of such an object in the nearby Universe would be groundbreaking and significantly advance the fields of astronomy and fundamental physics.

\section*{ACKNOWLEDGEMENTS}
DL is supported by the Multidisciplinary Doctoral (MDoc) Trainee Program, funded by the Ontario Regional Centre of the Canadian Statistical Sciences Institute (CANSSI Ontario). GME is supported by a Discovery Grant from the Natural Sciences and Engineering Research Council of Canada (NSERC, RGPIN-2020-04554) and a University of Toronto Connaught New Researcher Award. WEH is supported by a Discovery Grant from NSERC.
AJR was supported as a Research Corporation for Science Advancement Cottrell Scholar.
Support for Program number HST-GO-15235 was provided through a grant from the STScI under NASA contract NAS5-26555.
SD is supported by NASA through Hubble Fellowship grant HST-HF2-51454.001-A awarded by the Space Telescope Science Institute, which is operated by the Association of Universities for Research in Astronomy, Incorporated, under NASA contract NAS5-26555. We thank Dr. Finn Lindgren (\href{mailto:finn.lindgren@ed.ac.uk}{finn.lindgren@ed.ac.uk}) for thoughtful discussion and recommendation of the method to compute the marginal quantile of the posterior spatial random effect of LGCP. The high performance computing cluster used in this work is SHARCNET (Shared Hierarchical Academic Research Computing Network).
\facility{HST (ACS, WFC3)}
\software{R Statistical Software Environment \citep{Rcore}; Main R packages include: \href{https://www.r-inla.org/home}{\texttt{R-INLA}}, \texttt{inlabru} \citep{inlabru}, \texttt{excursions} \citep{Bolin2018}; Secondary R packages: \texttt{tidyverse} \citep{tidyverse}, \texttt{spatial}\citep{spatial}}


\bibliography{reference}

\begin{thebibliography}{}
\expandafter\ifx\csname natexlab\endcsname\relax\def\natexlab#1{#1}\fi
\providecommand{\url}[1]{\href{#1}{#1}}
\providecommand{\dodoi}[1]{doi:~\href{http://doi.org/#1}{\nolinkurl{#1}}}
\providecommand{\doeprint}[1]{\href{http://ascl.net/#1}{\nolinkurl{http://ascl.net/#1}}}
\providecommand{\doarXiv}[1]{\href{https://arxiv.org/abs/#1}{\nolinkurl{https://arxiv.org/abs/#1}}}

\bibitem[{Abraham \& van Dokkum(2014)}]{Abraham2014}
Abraham, R.~G., \& van Dokkum, P.~G. 2014, \pasp, 126, 55,
  \dodoi{10.1086/674875}

\bibitem[{{Amorisco} {et~al.}(2018){Amorisco}, {Monachesi}, {Agnello}, \&
  {White}}]{Amorisco2018}
{Amorisco}, N.~C., {Monachesi}, A., {Agnello}, A., \& {White}, S.~D.~M. 2018,
  \mnras, 475, 4235, \dodoi{10.1093/mnras/sty116}

\bibitem[{Bachl {et~al.}(2019)Bachl, Lindgren, Borchers, \& Illian}]{inlabru}
Bachl, F.~E., Lindgren, F., Borchers, D.~L., \& Illian, J.~B. 2019, Methods
  Ecol. Evol., 10, 760, \dodoi{10.1111/2041-210X.13168}

\bibitem[{Bertin \& Arnouts(1996)}]{Bertin1996}
Bertin, E., \& Arnouts, S. 1996, \aaps, 117, 393, \dodoi{10.1051/AAS:1996164}

\bibitem[{Binggeli {et~al.}(1984)Binggeli, Sandage, \& Tarenghi}]{Binggeli1984}
Binggeli, B., Sandage, A., \& Tarenghi, M. 1984, AJ, 89, 64,
  \dodoi{10.1086/113484}

\bibitem[{{Blakeslee}(1997)}]{Blakeslee1997}
{Blakeslee}, J.~P. 1997, \apjl, 481, L59, \dodoi{10.1086/310653}

\bibitem[{Bolin \& Lindgren(2015)}]{Bolin2015}
Bolin, D., \& Lindgren, F. 2015, J. R. Statist. Soc. B, 77, 85,
  \dodoi{10.1111/rssb.12055}

\bibitem[{Bolin \& Lindgren(2018)}]{Bolin2018}
---. 2018, J. Stat. Softw., 86, 1, \dodoi{10.18637/JSS.V086.I05}

\bibitem[{Brunzendorf \& Meusinger(1999)}]{Brunzendorf1999}
Brunzendorf, J., \& Meusinger, H. 1999, \aaps, 139, 141,
  \dodoi{10.1051/AAS:1999111}

\bibitem[{Burkert(2020)}]{Burkert2020}
Burkert, A. 2020, \apj, 904, 161, \dodoi{10.3847/1538-4357/abb242}

\bibitem[{{Burkert} \& {Forbes}(2020)}]{Burkert_Forbes2020}
{Burkert}, A., \& {Forbes}, D.~A. 2020, \aj, 159, 56,
  \dodoi{10.3847/1538-3881/ab5b0e}

\bibitem[{Coles \& Jones(1991)}]{Coles1991}
Coles, P., \& Jones, B. 1991, \mnras, 248, 1, \dodoi{10.1093/MNRAS/248.1.1}

\bibitem[{{Danieli} {et~al.}(2020){Danieli}, {van Dokkum}, {Abraham}, {Conroy},
  {Dolphin}, \& {Romanowsky}}]{Danieli+2020}
{Danieli}, S., {van Dokkum}, P., {Abraham}, R., {et~al.} 2020, \apjl, 895, L4,
  \dodoi{10.3847/2041-8213/ab8dc4}

\bibitem[{{Danieli} {et~al.}(2019){Danieli}, {van Dokkum}, {Conroy}, {Abraham},
  \& {Romanowsky}}]{Danieli2019}
{Danieli}, S., {van Dokkum}, P., {Conroy}, C., {Abraham}, R., \& {Romanowsky},
  A.~J. 2019, \apjl, 874, L12, \dodoi{10.3847/2041-8213/ab0e8c}

\bibitem[{{Danieli} {et~al.}(2022){Danieli}, {van Dokkum}, {Trujillo-Gomez},
  {Kruijssen}, {Romanowsky}, {Carlsten}, {Shen}, {Li}, {Abraham}, {Brodie},
  {Conroy}, {Gannon}, \& {Greco}}]{Danieli2022}
{Danieli}, S., {van Dokkum}, P., {Trujillo-Gomez}, S., {et~al.} 2022, \apjl,
  927, L28, \dodoi{10.3847/2041-8213/ac590a}

\bibitem[{Diggle(1985)}]{Diggle1985}
Diggle, P. 1985, J. R. Statist. Soc. C, 34, 138, \dodoi{10.2307/2347366}

\bibitem[{Diggle {et~al.}(2005)Diggle, Rowlingson, \& Su}]{Diggle2005}
Diggle, P., Rowlingson, B., \& Su, T.-l. 2005, Environmetrics, 16, 423,
  \dodoi{10.1002/ENV.712}

\bibitem[{Diggle {et~al.}(2013)Diggle, Moraga, Rowlingson, \&
  Taylor}]{Diggle2013}
Diggle, P.~J., Moraga, P., Rowlingson, B., \& Taylor, B.~M. 2013, Statist.
  Sci., 28, 542, \dodoi{10.1214/13-STS441}

\bibitem[{{Eadie} {et~al.}(2021){Eadie}, {Harris}, \&
  {Springford}}]{eadie+2021}
{Eadie}, G.~M., {Harris}, W.~E., \& {Springford}, A. 2021, arXiv e-prints,
  arXiv:2110.15376.
\newblock \doarXiv{2110.15376}

\bibitem[{Fawcett(2006)}]{Fawcett2006}
Fawcett, T. 2006, Pattern Recognit. Lett., 27, 861,
  \dodoi{10.1016/J.PATREC.2005.10.010}

\bibitem[{Forbes {et~al.}(2020)Forbes, Alabi, Romanowsky, Brodie, \&
  Arimoto}]{Forbes2020}
Forbes, D.~A., Alabi, A., Romanowsky, A.~J., Brodie, J.~P., \& Arimoto, N.
  2020, \mnras, 492, 487, \dodoi{10.1093/MNRAS/STAA180}

\bibitem[{{Forbes} {et~al.}(2020){Forbes}, {Dullo}, {Gannon}, {Couch},
  {Iodice}, {Spavone}, {Cantiello}, \& {Schipani}}]{Forbes+2020}
{Forbes}, D.~A., {Dullo}, B.~T., {Gannon}, J., {et~al.} 2020, \mnras, 494,
  5293, \dodoi{10.1093/mnras/staa1111}

\bibitem[{Forbes {et~al.}(2017)Forbes, Forbes, \& A.}]{Forbes2017}
Forbes, D.~A., Forbes, \& A., D. 2017, \mnras, 472, L104,
  \dodoi{10.1093/MNRASL/SLX148}

\bibitem[{{Forbes} {et~al.}(2019){Forbes}, {Gannon}, {Couch}, {Iodice},
  {Spavone}, {Cantiello}, {Napolitano}, \& {Schipani}}]{Forbes+2019}
{Forbes}, D.~A., {Gannon}, J., {Couch}, W.~J., {et~al.} 2019, \aap, 626, A66,
  \dodoi{10.1051/0004-6361/201935499}

\bibitem[{{Forbes} {et~al.}(2018){Forbes}, {Read}, {Gieles}, \&
  {Collins}}]{Forbes+2018}
{Forbes}, D.~A., {Read}, J.~I., {Gieles}, M., \& {Collins}, M. L.~M. 2018,
  \mnras, 481, 5592, \dodoi{10.1093/mnras/sty2584}

\bibitem[{Fuglstad {et~al.}(2018)Fuglstad, Simpson, Lindgren, \&
  Rue}]{Fuglstad2018}
Fuglstad, G.-A., Simpson, D., Lindgren, F., \& Rue, H. 2018, J. Am. Stat.
  Assoc., 114, 445, \dodoi{10.1080/01621459.2017.1415907}

\bibitem[{Gannon {et~al.}(2021)Gannon, Forbes, Romanowsky, Ferr{\'{e}}-Mateu,
  Couch, Brodie, Huang, Janssens, \& Okabe}]{Gannon2021}
Gannon, J.~S., Forbes, D.~A., Romanowsky, A.~J., {et~al.} 2021, \mnras, 000, 1,
  \dodoi{10.1093/mnras/stab3297}

\bibitem[{{Georgiev} {et~al.}(2010){Georgiev}, {Puzia}, {Goudfrooij}, \&
  {Hilker}}]{Georgiev10}
{Georgiev}, I.~Y., {Puzia}, T.~H., {Goudfrooij}, P., \& {Hilker}, M. 2010,
  \mnras, 406, 1967, \dodoi{10.1111/j.1365-2966.2010.16802.x}

\bibitem[{G{\'{o}}mez-Rubio \& Rue(2017)}]{Gomez-Rubio2017}
G{\'{o}}mez-Rubio, V., \& Rue, H. 2017, Stat. Comput., 28, 1033,
  \dodoi{10.1007/S11222-017-9778-Y}

\bibitem[{Greco {et~al.}(2018)Greco, Greene, Price-Whelan, Leauthaud, Huang,
  Goulding, Strauss, Komiyama, Lupton, Miyazaki, Takada, Tanaka, \&
  Usuda}]{Greco2018}
Greco, J.~P., Greene, J.~E., Price-Whelan, A.~M., {et~al.} 2018, \pasj, 70, 19,
  \dodoi{10.1093/PASJ/PSX051}

\bibitem[{Gudehus \& D.(1995)}]{Gudehus1995}
Gudehus, \& D. 1995, A\&A, 302, 21.
\newblock \url{https://ui.adsabs.harvard.edu/abs/1995A&A...302...21G/abstract}

\bibitem[{Harris(1991)}]{Harris1991}
Harris, W.~E. 1991, ARA\&A, 29, 543,
  \dodoi{10.1146/ANNUREV.AA.29.090191.002551}

\bibitem[{Harris {et~al.}(2017)Harris, Blakeslee, \& Harris}]{Harris+2017}
Harris, W.~E., Blakeslee, J.~P., \& Harris, G. L.~H. 2017, \apj, 836, 67,
  \dodoi{10.3847/1538-4357/836/1/67}

\bibitem[{{Harris} {et~al.}(2016){Harris}, {Blakeslee}, {Whitmore}, {Gnedin},
  {Geisler}, \& {Rothberg}}]{Harris+2016}
{Harris}, W.~E., {Blakeslee}, J.~P., {Whitmore}, B.~C., {et~al.} 2016, \apj,
  817, 58, \dodoi{10.3847/0004-637X/817/1/58}

\bibitem[{{Harris} {et~al.}(2013){Harris}, {Harris}, \& {Alessi}}]{Harris+2013}
{Harris}, W.~E., {Harris}, G. L.~H., \& {Alessi}, M. 2013, \apj, 772, 82,
  \dodoi{10.1088/0004-637X/772/2/82}

\bibitem[{Harris {et~al.}(2020)Harris, Brown, Durrell, Romanowsky, Blakeslee,
  Brodie, Janssens, Lisker, Okamoto, \& Wittmann}]{Harris2020}
Harris, W.~E., Brown, R.~A., Durrell, P.~R., {et~al.} 2020, \apj, 890, 105,
  \dodoi{10.3847/1538-4357/ab6992}

\bibitem[{Hu {et~al.}(2000)Hu, Barkana, \& Gruzinov}]{Hu2000}
Hu, W., Barkana, R., \& Gruzinov, A. 2000, Phys. Rev. Lett., 85, 1158,
  \dodoi{10.1103/PHYSREVLETT.85.1158}

\bibitem[{Hudson {et~al.}(1997)Hudson, Lucey, Smith, Steel, Hudson, Lucey,
  Smith, \& Steel}]{Hudson1997}
Hudson, M.~J., Lucey, J.~R., Smith, R.~J., {et~al.} 1997, MNRAS, 291, 488,
  \dodoi{10.1093/MNRAS/291.3.488}

\bibitem[{Hui {et~al.}(2017)Hui, Ostriker, Tremaine, \& Witten}]{Hui2017}
Hui, L., Ostriker, J.~P., Tremaine, S., \& Witten, E. 2017, Phys. Rev. D, 95,
  043541, \dodoi{10.1103/PHYSREVD.95.043541}

\bibitem[{Janssens {et~al.}(2019)Janssens, Abraham, Brodie, Forbes, \&
  Romanowsky}]{Janssens2019}
Janssens, S.~R., Abraham, R., Brodie, J., Forbes, D.~A., \& Romanowsky, A.~J.
  2019, \apj, 887, 92, \dodoi{10.3847/1538-4357/ab536c}

\bibitem[{{King}(1962)}]{King62}
{King}, I. 1962, \aj, 67, 471, \dodoi{10.1086/108756}

\bibitem[{{Larsen}(1999)}]{Larsen99}
{Larsen}, S.~S. 1999, \aaps, 139, 393, \dodoi{10.1051/aas:1999509}

\bibitem[{Li \& Barmby(2021)}]{Li2021}
Li, D., \& Barmby, P. 2021, \mnras, 501, 3472, \dodoi{10.1093/mnras/staa3908}

\bibitem[{Li {et~al.}(2012)Li, Brown, Gesink, \& Rue}]{Li2012}
Li, Y., Brown, P., Gesink, D.~C., \& Rue, H. 2012, Stat. Methods. Med. Res.,
  21, 479, \dodoi{10.1177/0962280212446326}

\bibitem[{Lim {et~al.}(2018)Lim, Peng, C{\^{o}}t{\'{e}}, Sales, den Brok,
  Blakeslee, \& Guhathakurta}]{Lim2018}
Lim, S., Peng, E.~W., C{\^{o}}t{\'{e}}, P., {et~al.} 2018, \apj, 862, 82,
  \dodoi{10.3847/1538-4357/AACB81}

\bibitem[{Lim {et~al.}(2020)Lim, C{\^{o}}t{\'{e}}, Peng, Ferrarese, Roediger,
  Durrell, Mihos, Wang, Gwyn, Cuillandre, Liu, S{\'{a}}nchez-Janssen, Toloba,
  Sales, Guhathakurta, Lan{\c{c}}on, \& Puzia}]{Lim2020}
Lim, S., C{\^{o}}t{\'{e}}, P., Peng, E.~W., {et~al.} 2020, \apj, 899, 69,
  \dodoi{10.3847/1538-4357/ABA433}

\bibitem[{Lindgren {et~al.}(2011)Lindgren, Rue, \&
  Lindstr{\"{o}}m}]{Lindgren2011}
Lindgren, F., Rue, H., \& Lindstr{\"{o}}m, J. 2011, J. R. Statist. Soc. B, 73,
  423, \dodoi{10.1111/j.1467-9868.2011.00777.x}

\bibitem[{Mart{\'{i}}nez-Delgado {et~al.}(2016)Mart{\'{i}}nez-Delgado,
  L{\"{a}}sker, Sharina, Toloba, Fliri, Beaton, Valls-Gabaud, Karachentsev,
  Chonis, Grebel, Forbes, Romanowsky, Gallego-Laborda, Teuwen,
  G{\'{o}}mez-Flechoso, Wang, Guhathakurta, Kaisin, \&
  Ho}]{Martinez-Delgado2016}
Mart{\'{i}}nez-Delgado, D., L{\"{a}}sker, R., Sharina, M., {et~al.} 2016, \aj,
  151, 96, \dodoi{10.3847/0004-6256/151/4/96}

\bibitem[{Mas {et~al.}(2013)Mas, Filho, Pontius, Guti{\'{e}}rrez, \&
  Rodrigues}]{Mas2013}
Mas, J.~F., Filho, B.~S., Pontius, R.~G., Guti{\'{e}}rrez, M.~F., \& Rodrigues,
  H. 2013, ISPRS Int. J. Geo-Inf., 2, 869, \dodoi{10.3390/IJGI2030869}

\bibitem[{Mei {et~al.}(2007)Mei, Blakeslee, Cote, Tonry, West, Ferrarese,
  Jordan, Peng, Anthony, \& Merritt}]{Mei_2007}
Mei, S., Blakeslee, J.~P., Cote, P., {et~al.} 2007, \apj, 655, 144,
  \dodoi{10.1086/509598}

\bibitem[{M{\o}ller(2003)}]{moller_2003}
M{\o}ller, J. 2003, Advances in Applied Probability, 35, 614–640,
  \dodoi{10.1239/aap/1059486821}

\bibitem[{M{\o}ller {et~al.}(1998)M{\o}ller, Syversveen, \&
  Waagepetersen}]{Moller1998}
M{\o}ller, J., Syversveen, A.~R., \& Waagepetersen, R.~P. 1998, \sjs, 25, 451,
  \dodoi{10.1111/1467-9469.00115}

\bibitem[{Neyman \& Scott(1958)}]{Neyman1958}
Neyman, J., \& Scott, E.~L. 1958, J. R. Statist. Soc. B, 20, 1,
  \dodoi{10.1111/j.2517-6161.1958.tb00272.x}

\bibitem[{Peng \& Lim(2016)}]{Peng2016}
Peng, E.~W., \& Lim, S. 2016, \apjl, 822, L31,
  \dodoi{10.3847/2041-8205/822/2/L31}

\bibitem[{{Peng} {et~al.}(2006){Peng}, {Jord{\'a}n}, {C{\^o}t{\'e}},
  {Blakeslee}, {Ferrarese}, {Mei}, {West}, {Merritt}, {Milosavljevi{\'c}}, \&
  {Tonry}}]{Peng+2006}
{Peng}, E.~W., {Jord{\'a}n}, A., {C{\^o}t{\'e}}, P., {et~al.} 2006, \apj, 639,
  95, \dodoi{10.1086/498210}

\bibitem[{{R Core Team}(2021)}]{Rcore}
{R Core Team}. 2021, R: A Language and Environment for Statistical Computing, R
  Foundation for Statistical Computing, Vienna, Austria.
\newblock \url{https://www.R-project.org/}

\bibitem[{Raynaud \& Nunan(2014)}]{Raynaud2014}
Raynaud, X., \& Nunan, N. 2014, PLOS ONE, 9, e87217,
  \dodoi{10.1371/JOURNAL.PONE.0087217}

\bibitem[{{Rom{\'a}n} {et~al.}(2019){Rom{\'a}n}, {Beasley}, {Ruiz-Lara}, \&
  {Valls-Gabaud}}]{Roman+2019}
{Rom{\'a}n}, J., {Beasley}, M.~A., {Ruiz-Lara}, T., \& {Valls-Gabaud}, D. 2019,
  \mnras, 486, 823, \dodoi{10.1093/mnras/stz835}

\bibitem[{Rue {et~al.}(2009)Rue, Martino, \& Chopin}]{Rue2009}
Rue, H., Martino, S., \& Chopin, N. 2009, J. R. Statist. Soc. B, 71, 319,
  \dodoi{10.1111/J.1467-9868.2008.00700.X}

\bibitem[{{Saifollahi} {et~al.}(2022){Saifollahi}, {Zaritsky}, {Trujillo},
  {Peletier}, {Knapen}, {Amorisco}, {Beasley}, \&
  {Donnerstein}}]{Saifollahi2022}
{Saifollahi}, T., {Zaritsky}, D., {Trujillo}, I., {et~al.} 2022, \mnras, 511,
  4633, \dodoi{10.1093/mnras/stac328}

\bibitem[{Samartsidis {et~al.}(2019)Samartsidis, Eickhoff, Eickhoff, Wager,
  Barrett, Atzil, Johnson, \& Nichols}]{Samartsidis2019}
Samartsidis, P., Eickhoff, C.~R., Eickhoff, S.~B., {et~al.} 2019, J. R.
  Statist. Soc. C, 68, 217, \dodoi{10.1111/RSSC.12295}

\bibitem[{Serra {et~al.}(2013)Serra, Saez, Mateu, Varga, Juan,
  D{\'{i}}az-{\'{A}}valos, \& Rue}]{Serra2013}
Serra, L., Saez, M., Mateu, J., {et~al.} 2013, Environ. Ecol. Stat., 21, 531,
  \dodoi{10.1007/S10651-013-0267-Y}

\bibitem[{S{\'{e}}rsic(1963)}]{Sersic1963}
S{\'{e}}rsic, J.~L. 1963, BAAA, 6, 41.
\newblock \url{https://ui.adsabs.harvard.edu/abs/1963BAAA....6...41S/abstract}

\bibitem[{Shen {et~al.}(2021)Shen, Danieli, van Dokkum, Abraham, Brodie,
  Conroy, Dolphin, Romanowsky, Kruijssen, \& Chowdhury}]{Shen2021}
Shen, Z., Danieli, S., van Dokkum, P., {et~al.} 2021, \apjl, 914, L12,
  \dodoi{10.3847/2041-8213/AC0335}

\bibitem[{Simpson {et~al.}(2015)Simpson, Illian, Lindgren, S{\o}rbye, \&
  Rue}]{Simpson2015}
Simpson, D., Illian, J.~B., Lindgren, F., S{\o}rbye, S.~H., \& Rue, H. 2015,
  Biometrika, 103, 49, \dodoi{10.1093/biomet/asv064}

\bibitem[{Simpson {et~al.}(2017)Simpson, Rue, Riebler, Martins, \&
  S{\o}rbye}]{Simpson2017}
Simpson, D., Rue, H., Riebler, A., Martins, T.~G., \& S{\o}rbye, S.~H. 2017,
  Statist. Sci., 32, 1, \dodoi{10.1214/16-STS576}

\bibitem[{{Spitler} \& {Forbes}(2009)}]{Spitler_Forbes2009}
{Spitler}, L.~R., \& {Forbes}, D.~A. 2009, \mnras, 392, L1,
  \dodoi{10.1111/j.1745-3933.2008.00567.x}

\bibitem[{Tempel {et~al.}(2016)Tempel, Stoica, Kipper, \& Saar}]{Tempel2016}
Tempel, E., Stoica, R.~S., Kipper, R., \& Saar, E. 2016, Astron. Comput, 16,
  17, \dodoi{10.1016/j.ascom.2016.03.004}

\bibitem[{{van Dokkum} {et~al.}(2016){van Dokkum}, {Abraham}, {Brodie},
  {Conroy}, {Danieli}, {Merritt}, {Mowla}, {Romanowsky}, \&
  {Zhang}}]{vanDokkum2016}
{van Dokkum}, P., {Abraham}, R., {Brodie}, J., {et~al.} 2016, \apjl, 828, L6,
  \dodoi{10.3847/2041-8205/828/1/L6}

\bibitem[{{van Dokkum} {et~al.}(2017){van Dokkum}, {Abraham}, {Romanowsky},
  {Brodie}, {Conroy}, {Danieli}, {Lokhorst}, {Merritt}, {Mowla}, \&
  {Zhang}}]{vanDokkum2017}
{van Dokkum}, P., {Abraham}, R., {Romanowsky}, A.~J., {et~al.} 2017, \apjl,
  844, L11, \dodoi{10.3847/2041-8213/aa7ca2}

\bibitem[{van Dokkum {et~al.}(2018{\natexlab{a}})van Dokkum, Danieli, Cohen,
  Merritt, Romanowsky, Abraham, Brodie, Conroy, Lokhorst, Mowla, O'Sullivan, \&
  Zhang}]{VanDokkum2018}
van Dokkum, P., Danieli, S., Cohen, Y., {et~al.} 2018{\natexlab{a}}, Nature,
  555, 629, \dodoi{10.1038/nature25767}

\bibitem[{van Dokkum {et~al.}(2018{\natexlab{b}})van Dokkum, Cohen, Danieli,
  Kruijssen, Romanowsky, Merritt, Abraham, Brodie, Conroy, Lokhorst, Mowla,
  O'Sullivan, \& Zhang}]{VanDokkum2018a}
van Dokkum, P., Cohen, Y., Danieli, S., {et~al.} 2018{\natexlab{b}}, \apj, 856,
  L30, \dodoi{10.3847/2041-8213/aab60b}

\bibitem[{van Dokkum {et~al.}(2019)van Dokkum, Wasserman, Danieli, Abraham,
  Brodie, Conroy, Forbes, Martin, Matuszewski, Romanowsky, \&
  Villaume}]{VanDokkum2019}
van Dokkum, P., Wasserman, A., Danieli, S., {et~al.} 2019, The Astrophysical
  Journal, 880, 91, \dodoi{10.3847/1538-4357/ab2914}

\bibitem[{{van Dokkum} {et~al.}(2022){van Dokkum}, {Shen}, {Keim},
  {Trujillo-Gomez}, {Danieli}, {Dutta Chowdhury}, {Abraham}, {Conroy},
  {Kruijssen}, {Nagai}, \& {Romanowsky}}]{vanDokkum2022}
{van Dokkum}, P., {Shen}, Z., {Keim}, M.~A., {et~al.} 2022, \nat, 605, 435,
  \dodoi{10.1038/s41586-022-04665-6}

\bibitem[{{van Dokkum} {et~al.}(2015){van Dokkum}, Abraham, Merritt, Zhang,
  Geha, \& Conroy}]{VanDokkum2015}
{van Dokkum}, P.~G., Abraham, R., Merritt, A., {et~al.} 2015, \apjl, 798, 45,
  \dodoi{10.1088/2041-8205/798/2/L45}

\bibitem[{Venables \& Ripley(2002)}]{spatial}
Venables, W.~N., \& Ripley, B.~D. 2002, Modern Applied Statistics with S, 4th
  edn. (New York: Springer).
\newblock \url{https://www.stats.ox.ac.uk/pub/MASS4/}

\bibitem[{Walker \& Pe{\~{n}}arrubia(2011)}]{Walker2011}
Walker, M.~G., \& Pe{\~{n}}arrubia, J. 2011, \apj, 742, 20,
  \dodoi{10.1088/0004-637X/742/1/20}

\bibitem[{{Wang} {et~al.}(2013){Wang}, {Peng}, {Blakeslee}, {C{\^o}t{\'e}},
  {Ferrarese}, {Jord{\'a}n}, {Mei}, \& {West}}]{Wang+2013}
{Wang}, Q., {Peng}, E.~W., {Blakeslee}, J.~P., {et~al.} 2013, \apj, 769, 145,
  \dodoi{10.1088/0004-637X/769/2/145}

\bibitem[{Wasserman {et~al.}(2019)Wasserman, van Dokkum, Romanowsky, Brodie,
  Danieli, Forbes, Abraham, Martin, Matuszewski, Villaume, Tamanas, \&
  Profumo}]{Wasserman2019}
Wasserman, A., van Dokkum, P., Romanowsky, A.~J., {et~al.} 2019, \apj, 885,
  155, \dodoi{10.3847/1538-4357/ab3eb9}

\bibitem[{Wickham {et~al.}(2019)Wickham, Averick, Bryan, Chang, McGowan,
  François, Grolemund, Hayes, Henry, Hester, Kuhn, Pedersen, Miller, Bache,
  Müller, Ooms, Robinson, Seidel, Spinu, Takahashi, Vaughan, Wilke, Woo, \&
  Yutani}]{tidyverse}
Wickham, H., Averick, M., Bryan, J., {et~al.} 2019, JOSS, 4, 1686,
  \dodoi{10.21105/joss.01686}

\bibitem[{Wittmann {et~al.}(2019)Wittmann, Kotulla, Lisker, Grebel, Conselice,
  Janz, \& Penny}]{Wittmann2019}
Wittmann, C., Kotulla, R., Lisker, T., {et~al.} 2019, \apjs, 245, 10,
  \dodoi{10.3847/1538-4365/AB4998}

\bibitem[{Wittmann {et~al.}(2017)Wittmann, Lisker, Tilahun, Grebel, Conselice,
  Penny, Janz, Gallagher, Kotulla, \& McCormac}]{Wittmann2017}
Wittmann, C., Lisker, T., Tilahun, L.~A., {et~al.} 2017, \mnras, 470, 1512,
  \dodoi{10.1093/mnras/stx1229}

\bibitem[{Yagi {et~al.}(2016)Yagi, Koda, Komiyama, \& Yamanoi}]{Yagi2016}
Yagi, M., Koda, J., Komiyama, Y., \& Yamanoi, H. 2016, \apjs, 225, 11,
  \dodoi{10.3847/0067-0049/225/1/11}

\end{thebibliography}
\bibliographystyle{aasjournal}

\appendix

\section{Poisson Point Process}\label{sec:ppp}

Consider any Borel-measurable set $B \subset S \subset \mathbb{R}^2$, let $\mathbf{X}$ denote a point process defined on $S$ and $N(B) \in \mathbb{N}$ denote the random variable counting how many points of $\mathbf{X}$ there are in $B$. Then $\mathbf{X}$ is a Poisson point process (PPP) with intensity function $\lambda(s)$, denoted by $\mathbf{X} \sim \text{PPP}(\lambda(s))$, if it satisfies the following conditions:
\begin{enumerate}
    \item $N(B) \sim \text{Poisson}(\mu(B))$ where $\mu(B) = \int_B\lambda(s)ds$. That is the number of points within $B$ is a Poisson random variable with mean $\mu(B)$.
    \item For any Borel-measurable sets $A, B \subset S \subset \mathbb{R}^2$ such that $A\cap B = \emptyset$, $N(A)$ and $N(B)$ are independent. 
\end{enumerate}
The above definition implies some important properties about PPP. Firstly, if we assume that the intensity function $\lambda(s)$ is constant, say $\lambda(s) \equiv \lambda > 0$, then condition 1 from above implies $\mathbb{E}(N(B)) = \mu(B) = \lambda |B|$ where $|\cdot|$ is the area of $B$. It is then easy to see that $\lambda$ here is the familiar mean surface number density in astronomy. 

Secondly, assume again that $\lambda(s) \equiv \lambda$, if we want to simulate a PPP in $S$ with intensity $\lambda$, we only need to generate a number from a Poisson random variable with parameter $\lambda|S|$, say $n$, then generate $n$ points uniformly across $S$. On another note, for the simulation of the GC system of UDG carried out in Section \ref{sec:simulation}, given the intensity function $\lambda(s)$ (a S\'{e}rsic profile) and the number of GCs ($N_{\rm GC}$) we want to generate, the locations of GCs can be generated by randomly sampling $N_{\rm GC}$ number of points from the probability density function $\pi(s) = \lambda(s)/\int_S \lambda(s)ds$.

The last and the most important property of a PPP is that the occurrence of a point has absolutely no effect on the occurrence of any other points. It is due to this last property that a PPP with constant intensity function is used as a reference point process to represent the notion of \textit{complete spatial randomness}.

\section{Computation of the Quantiles of Posterior Marginal Distribution of \texorpdfstring{$\mathcal{U}(s)$}{Lg}}\label{sec:quantile}
The marginal distribution of a random field $\mathcal{U}(s)$ is effectively its one-point distribution function in astronomy. Hence, if $\mathcal{U}(s)$ is a zero mean Gaussian process with marginal variance $\sigma^2$, its marginal distribution is then $\mathcal{N}(0, \sigma^2)$. Thus, \cite{Diggle2005} first suggested to compute the quantiles of the posterior marginal distribution of $\mathcal{U}(s)$ by assuming $\mathcal{U}(s)|\mathbf{x}$ approximately follows $\mathcal{N}(0,\sigma^2)$ where $\sigma$ is a point estimate from the posterior distribution of $\sigma$. We then choose our quantiles from $\mathcal{N}(0,\sigma^2)$. However, this procedure is only suitable for visualization purposes, and is not suited for accurate numerical analysis since it is a frequentist approach imposed on a fully Bayesian framework. Moreover, even though the prior distribution of $\mathcal{U}(s)$ is a Gaussian process, the posterior distribution $\mathcal{U}(s)|\mathbf{x}$ is not, and it is in fact analytically intractable. This means that, theoretically, the true posterior marginal distribution $\mathcal{U}(s)|\mathbf{x}$ is unknown. Hence, the accuracy of the approximation of $\mathcal{U}(s)|\mathbf{x}$ by a normal distribution is not guaranteed. However, a workaround is by the following identity:
\begin{equation}\label{posterior_U}
    \pi(\mathcal{U}(s)|\mathbf{x}) = \int \pi(\mathcal{U}(s)|\sigma,\mathbf{x})\pi(\sigma|\mathbf{x})d\sigma.
\end{equation}
The workaround method is using the fact that $\mathcal{U}(s)|\sigma,\mathbf{x}$ can be very well-approximated by a Gaussian process through INLA. This means that the marginal distribution for $\mathcal{U}(s)|\sigma,\mathbf{x}$ is now well-approximated by $\mathcal{N}(0, \sigma^2)$. Therefore, equation \ref{posterior_U} means that the posterior marginal of $\mathcal{U}(s)|\mathbf{x}$ is well-approximated by a mixture distribution of $\mathcal{N}(0, \sigma^2)$ where $\sigma$ is generated from the posterior distribution of $\sigma$. 

Based on the previous argument, a principled framework we adopt is the following: we draw a sample $(\sigma_1, \dots, \sigma_n)$ from the posterior distribution $\pi(\sigma|\mathbf{x})$. For each of the individual $\sigma_i, i \in \{1, \dots, n\}$, we generate a random value from $\mathcal{N}(0, \sigma_i^2)$. From this sample of a mixture of normal distributions, we compute the empirical $p$-th quantile, $Q(p)$. This is then the actual $p$-th quantile of the posterior marginal distribution $\mathcal{U}(s)|\mathbf{x}$.

\clearpage
\section{Detection Results for Pointings with Detected UDGs}\label{sec:detect_figs}

\subsection{Pointings with Contaminating Bright Galaxies}

\begin{figure*}[ht]
    \gridline{\fig{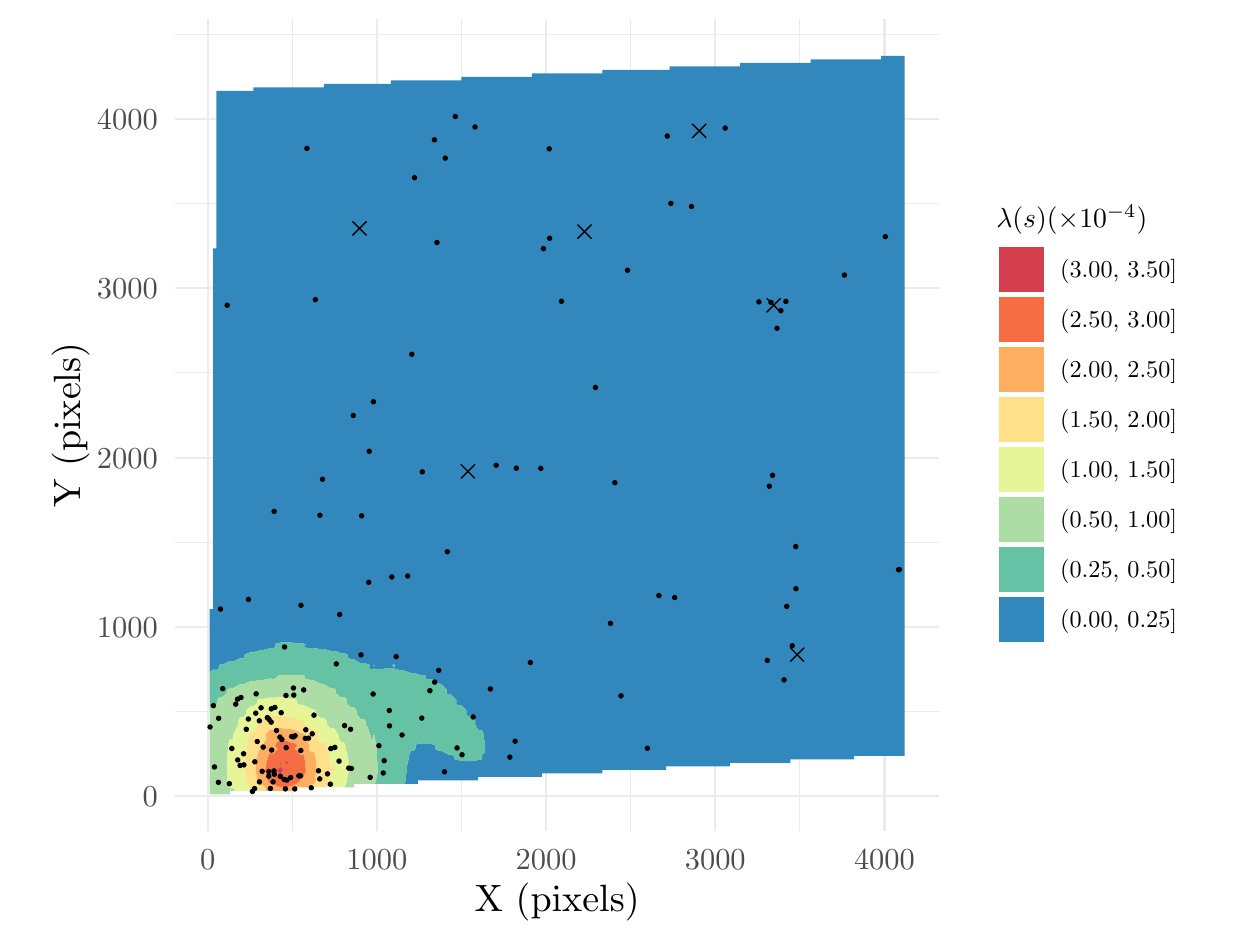}{0.5\textwidth}{(a)}}
    \gridline{\fig{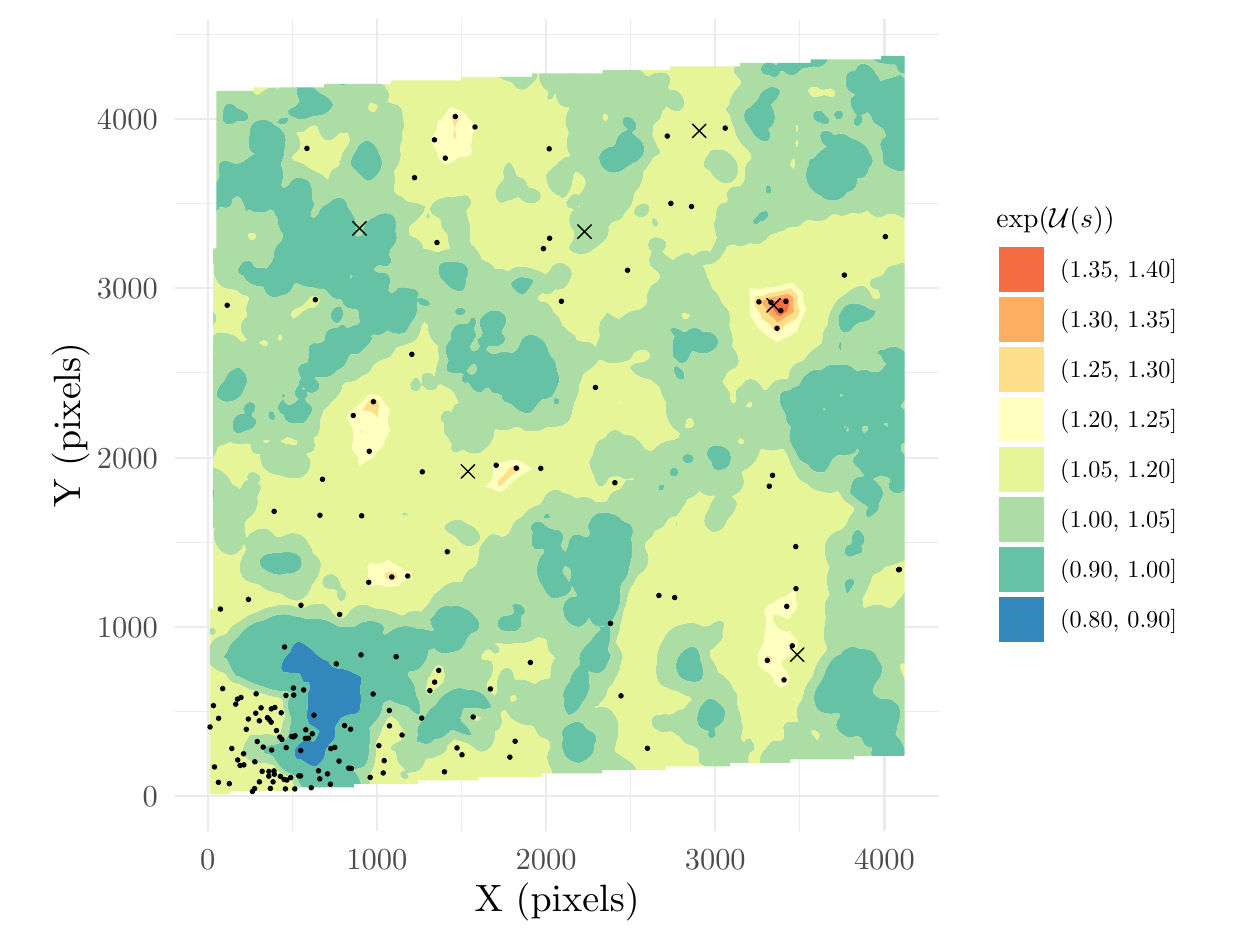}{0.5\textwidth}{(b)}
    \fig{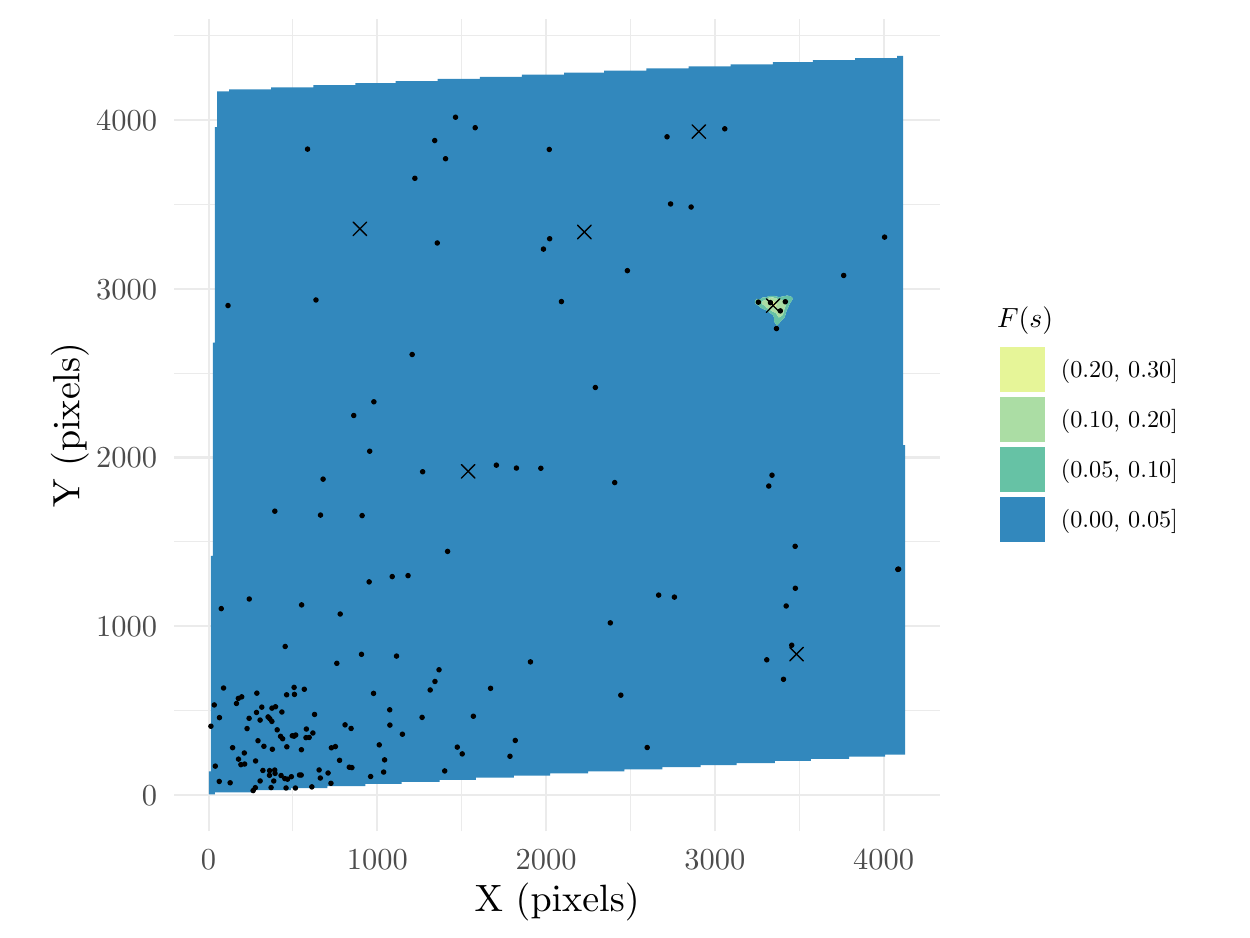}{0.5\textwidth}{(c)}}
    \caption{Posterior results for V8-WFC3 pointing. (a) Posterior mean intensity $\lambda(s)$; (b) Posterior mean spatial random effect $\exp(\mathcal{U}(s))$; (c) Excursion function $F_C(s)$ with $C = Q(0.95)$. Black points are locations of GC candidates. Black crosses are the locations of previously known UDG candidates.}
    \label{fig:v8wfc3_results}
\end{figure*}

\begin{figure*}[ht]
    \gridline{\fig{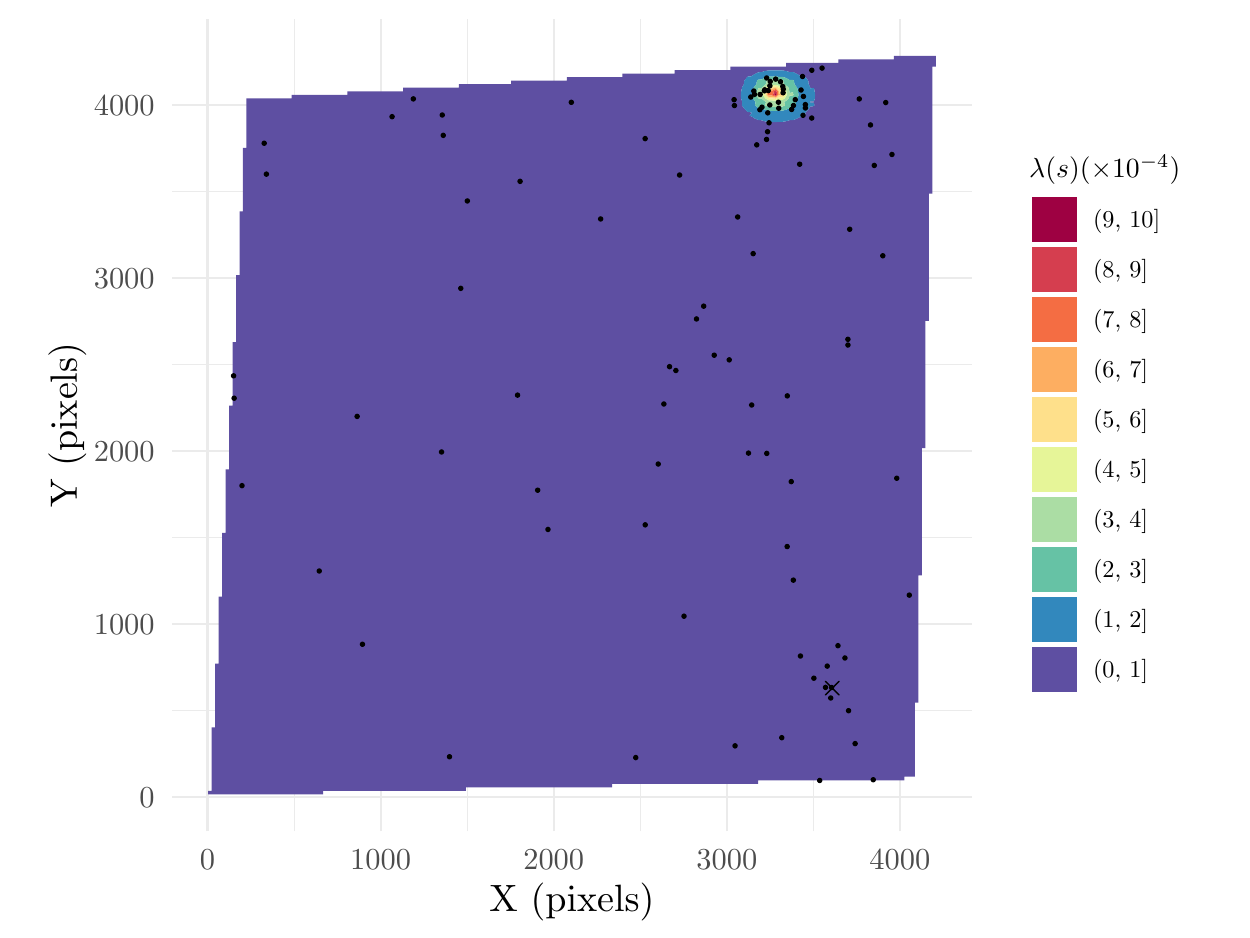}{0.5\textwidth}{(a)}}
    \gridline{\fig{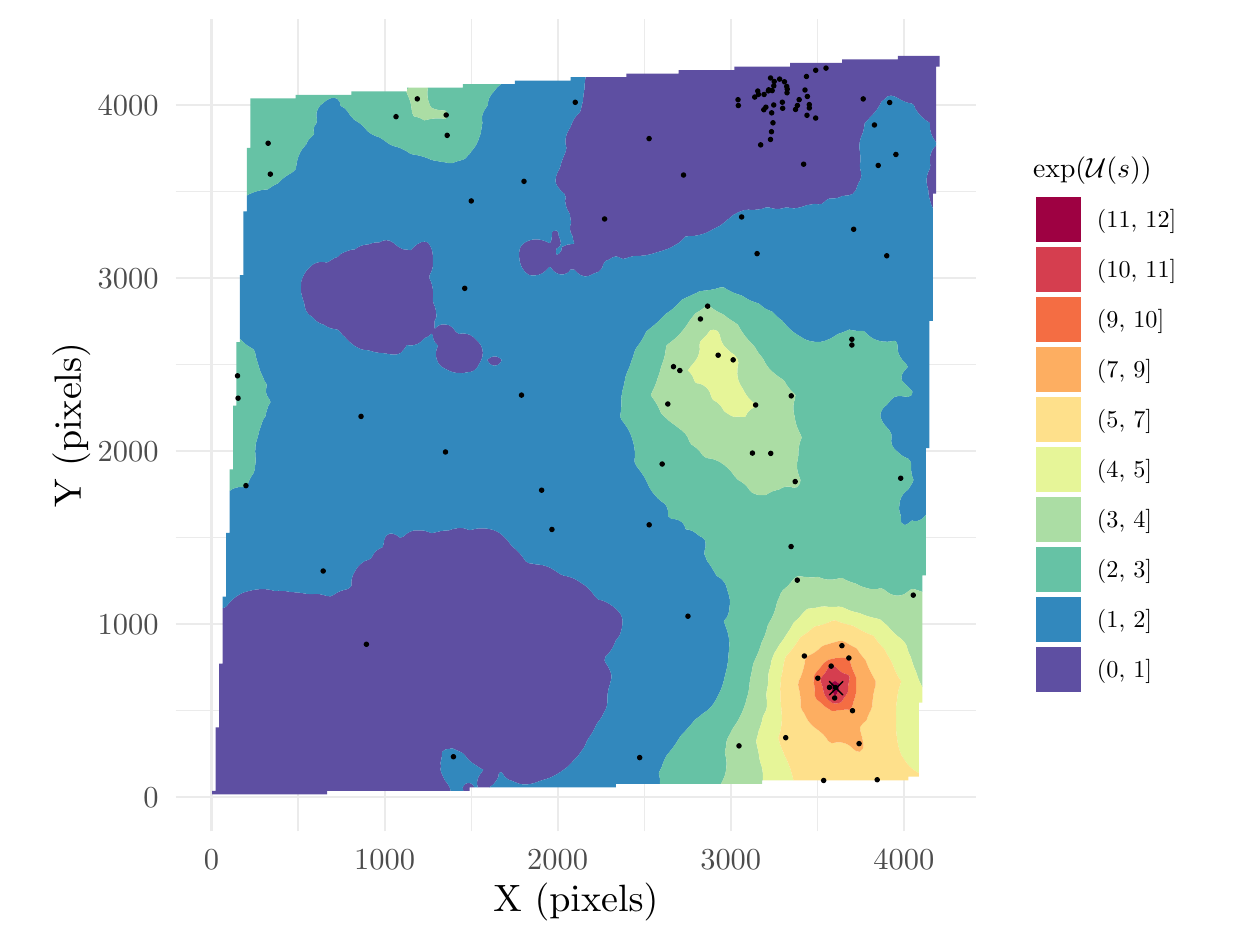}{0.5\textwidth}{(b)}
    \fig{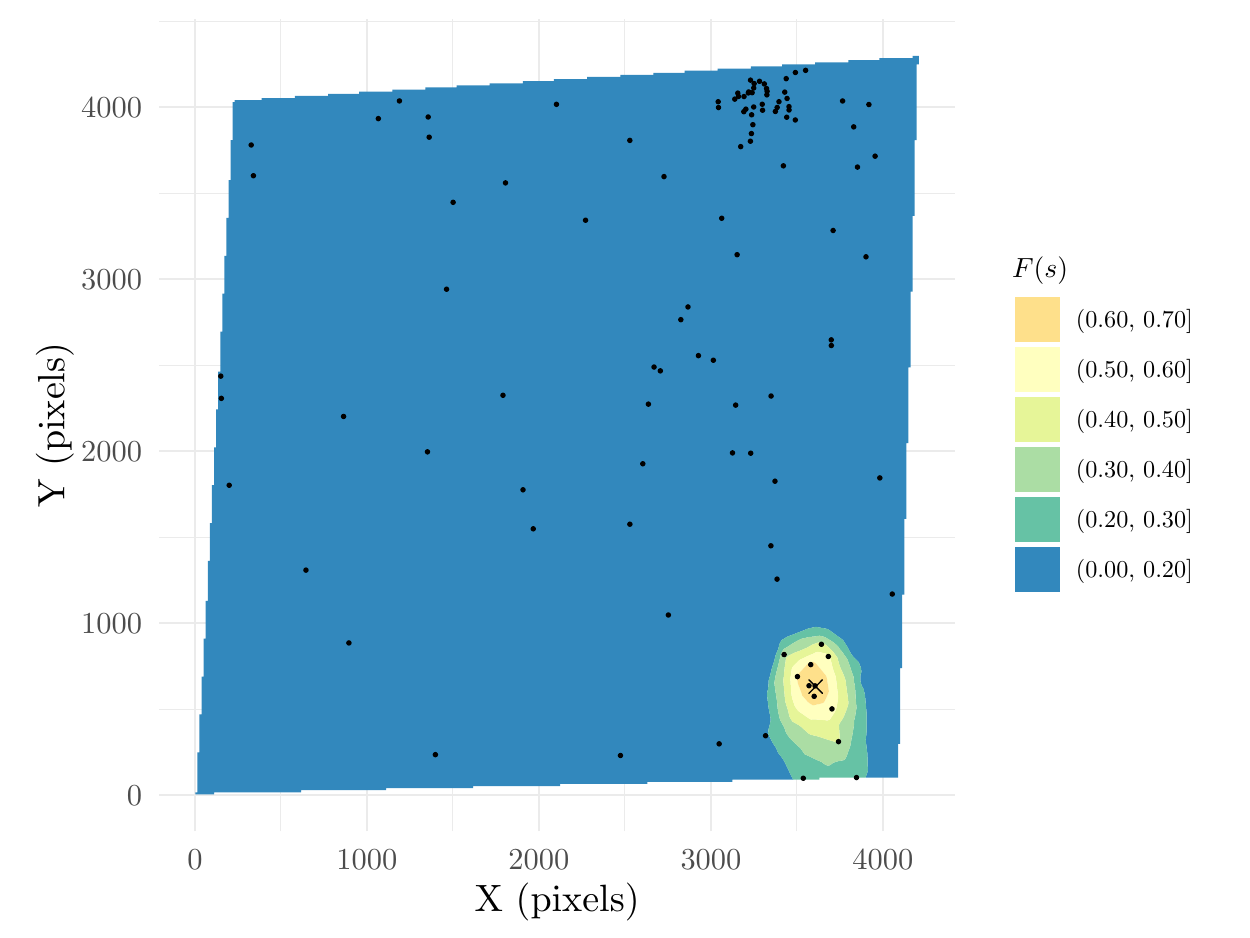}{0.5\textwidth}{(c)}}
    \caption{Posterior results for V10-ACS pointing. (a) Posterior mean intensity $\lambda(s)$; (b) Posterior mean spatial random effect $\exp(\mathcal{U}(s))$; (c) Excursion function $F_C(s)$ with $C = Q(0.95)$. Black points are locations of GC candidates. Black cross is the location of previously known UDG candidate.}
    \label{fig:v10acs_results}
\end{figure*}

\begin{figure*}[ht]
    \gridline{\fig{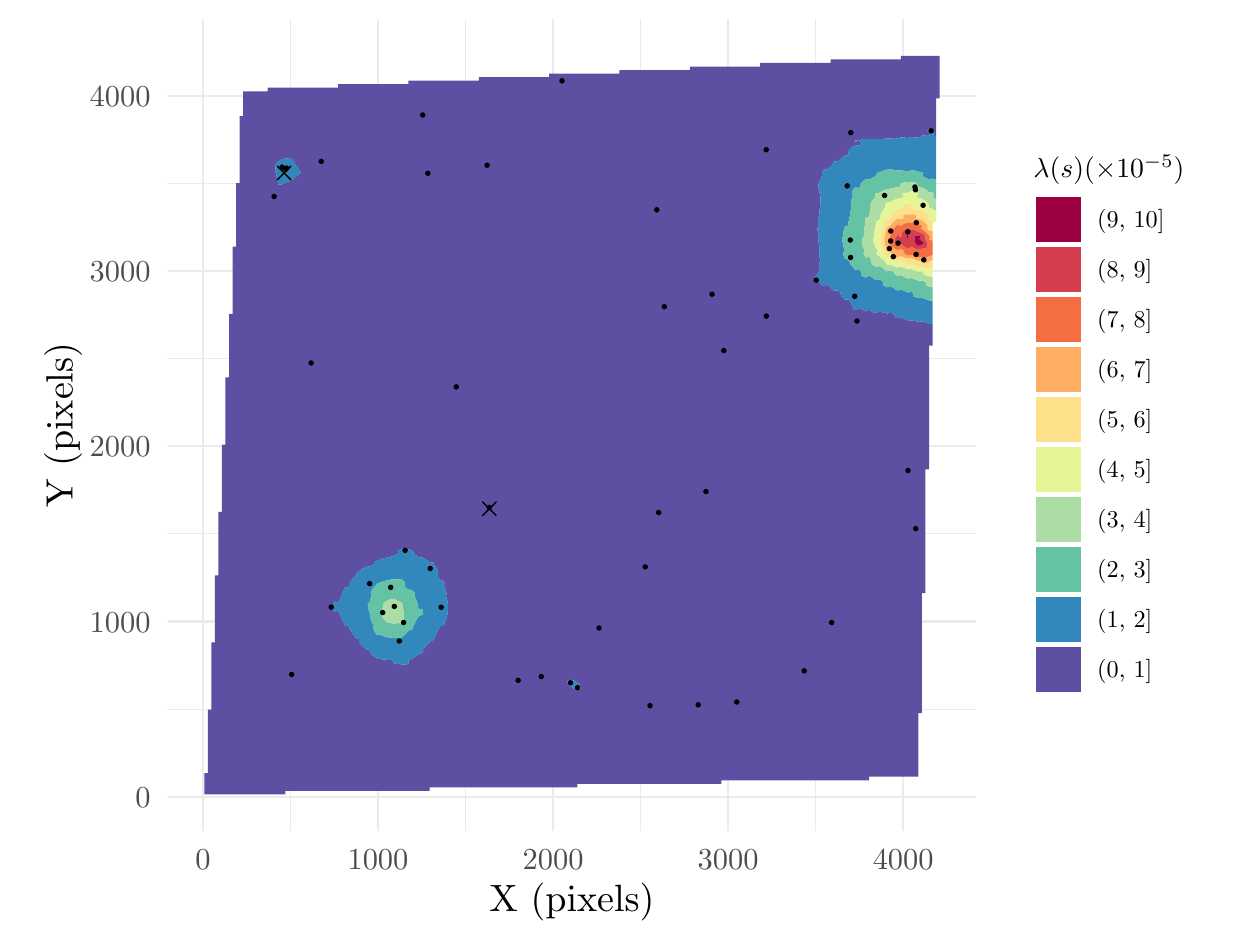}{0.5\textwidth}{(a)}}
    \gridline{\fig{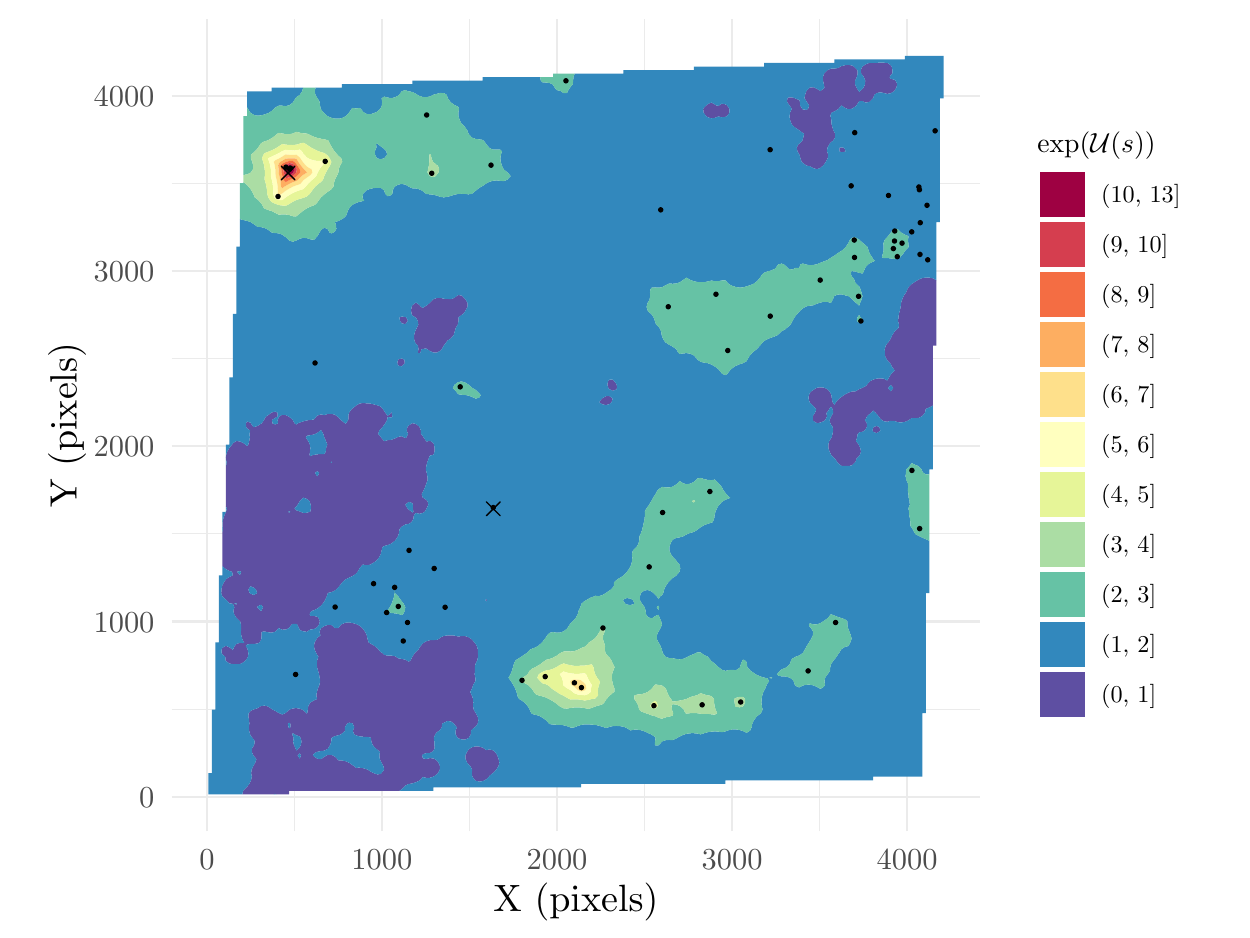}{0.5\textwidth}{(b)}
    \fig{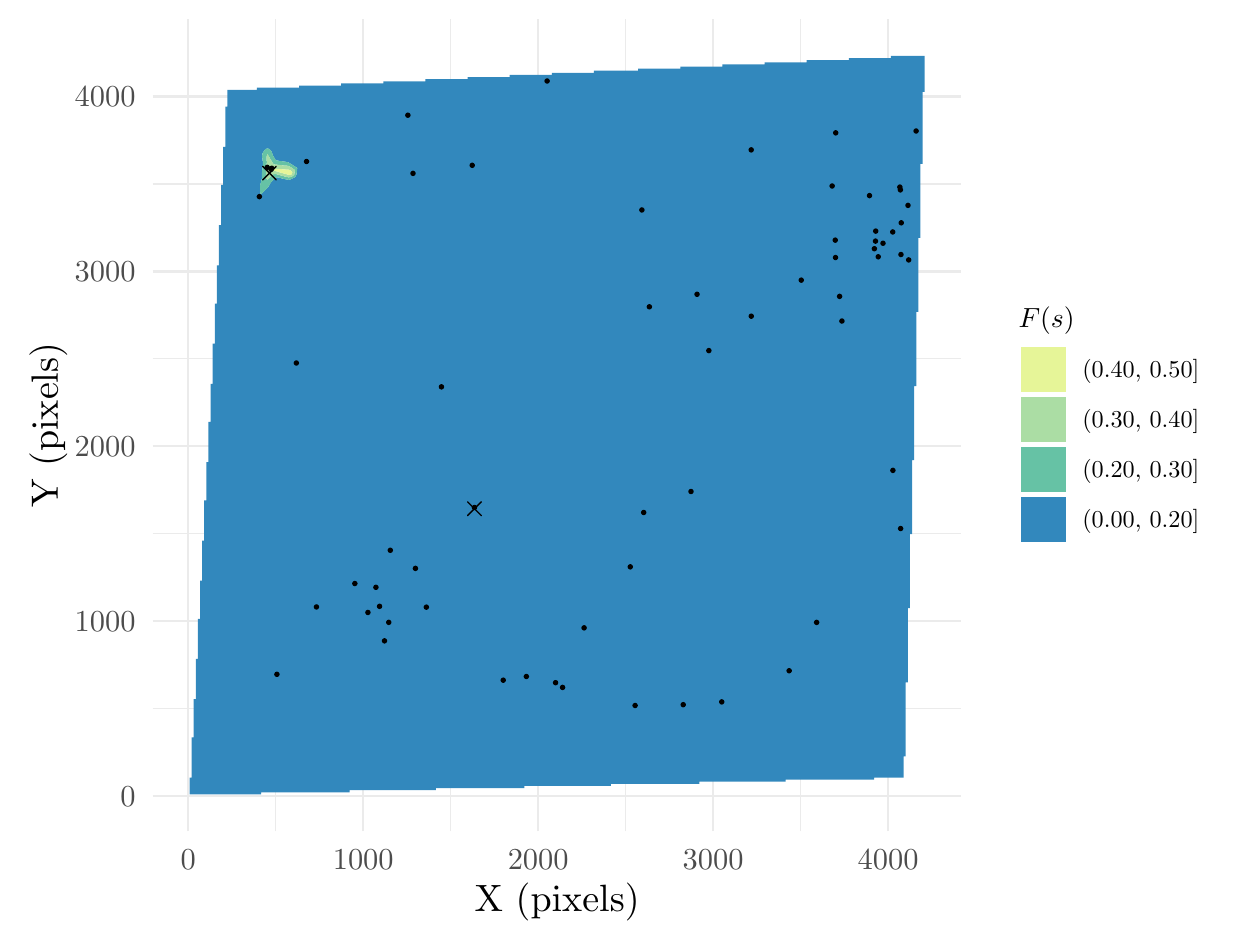}{0.5\textwidth}{(c)}}
    \caption{Posterior results for V15-ACS pointing. (a) Posterior mean intensity $\lambda(s)$; (b) Posterior mean spatial random effect $\exp(\mathcal{U}(s))$; (c) Excursion function $F_C(s)$ with $C = Q(0.95)$. Black points are locations of GC candidates. Black crosses are the locations of previously known UDG candidates.}
    \label{fig:v15acs_results}
\end{figure*}

\clearpage
\subsection{Pointings without Contaminating Bright Galaxies}

\begin{figure*}[ht]
    \gridline{\fig{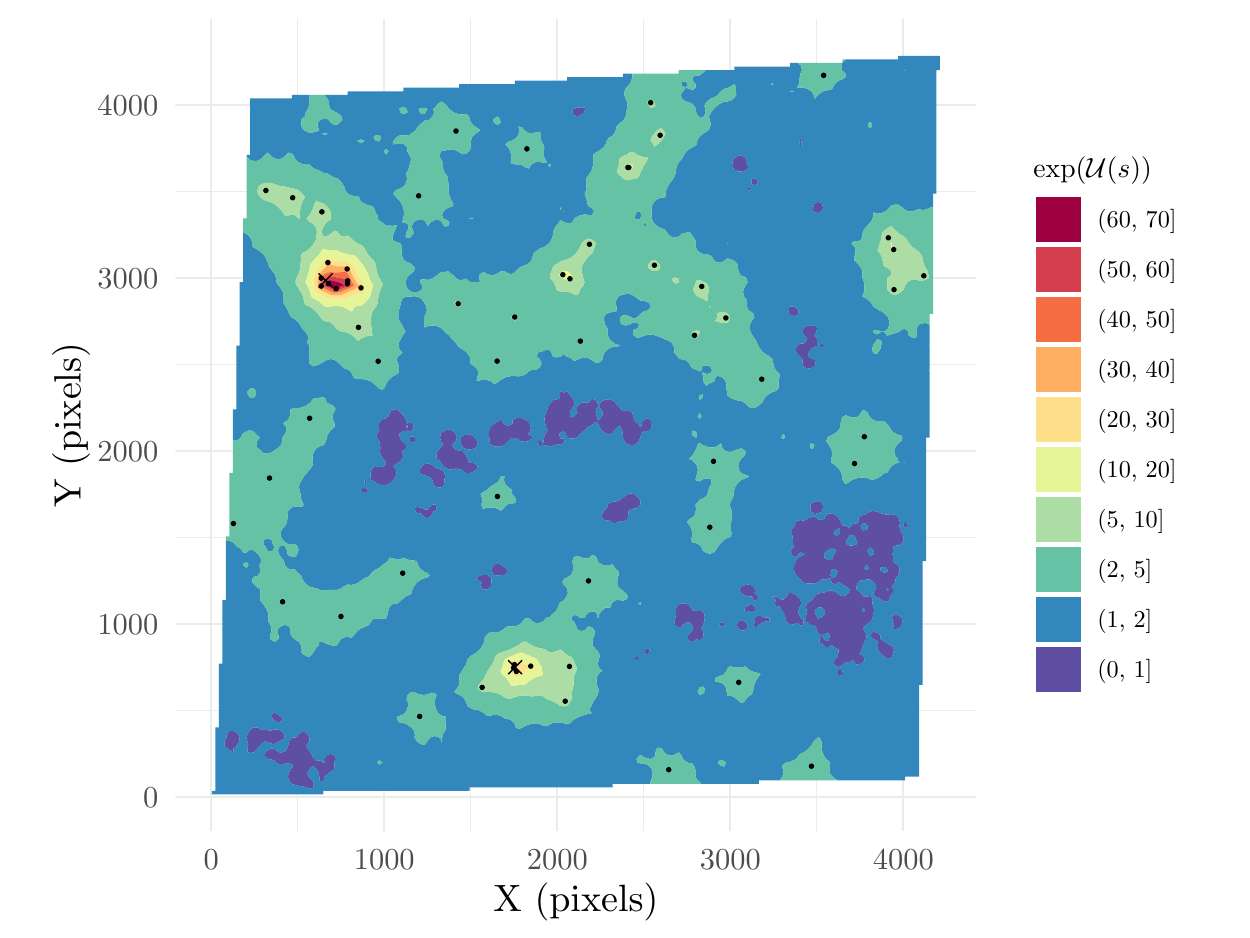}{0.5\textwidth}{(a)}
    \fig{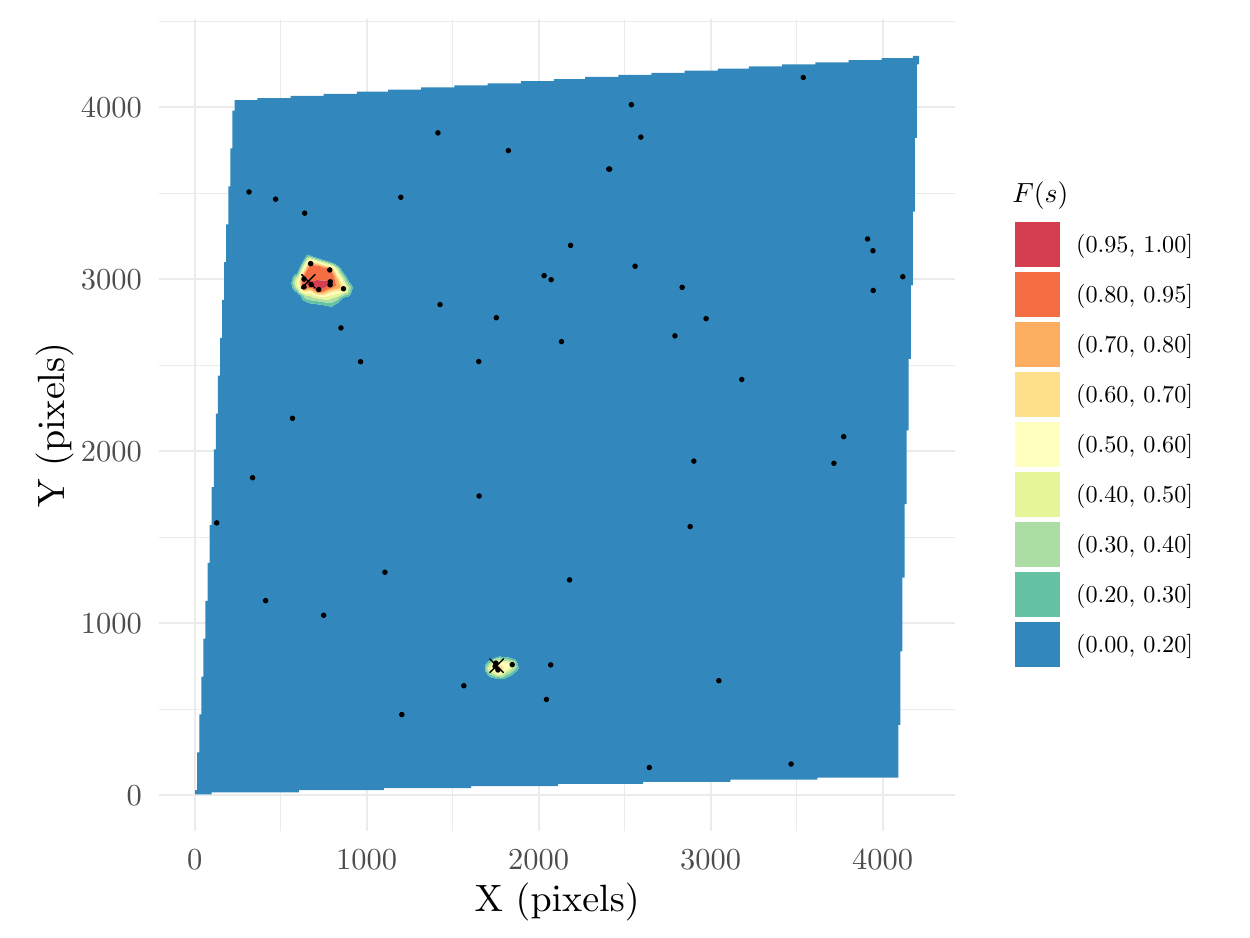}{0.5\textwidth}{(b)}}
    \caption{Posterior results for V7-ACS pointing. (a) Posterior mean spatial random effect $\exp(\mathcal{U}(s))$; (b) Excursion function $F_C(s)$ with $C = Q(0.95)$. Black points are locations of GC candidates. Black cross is the location of previously known UDG candidate.}
    \label{fig:v7acs_results}
\end{figure*}

\begin{figure*}[ht]
    \gridline{\fig{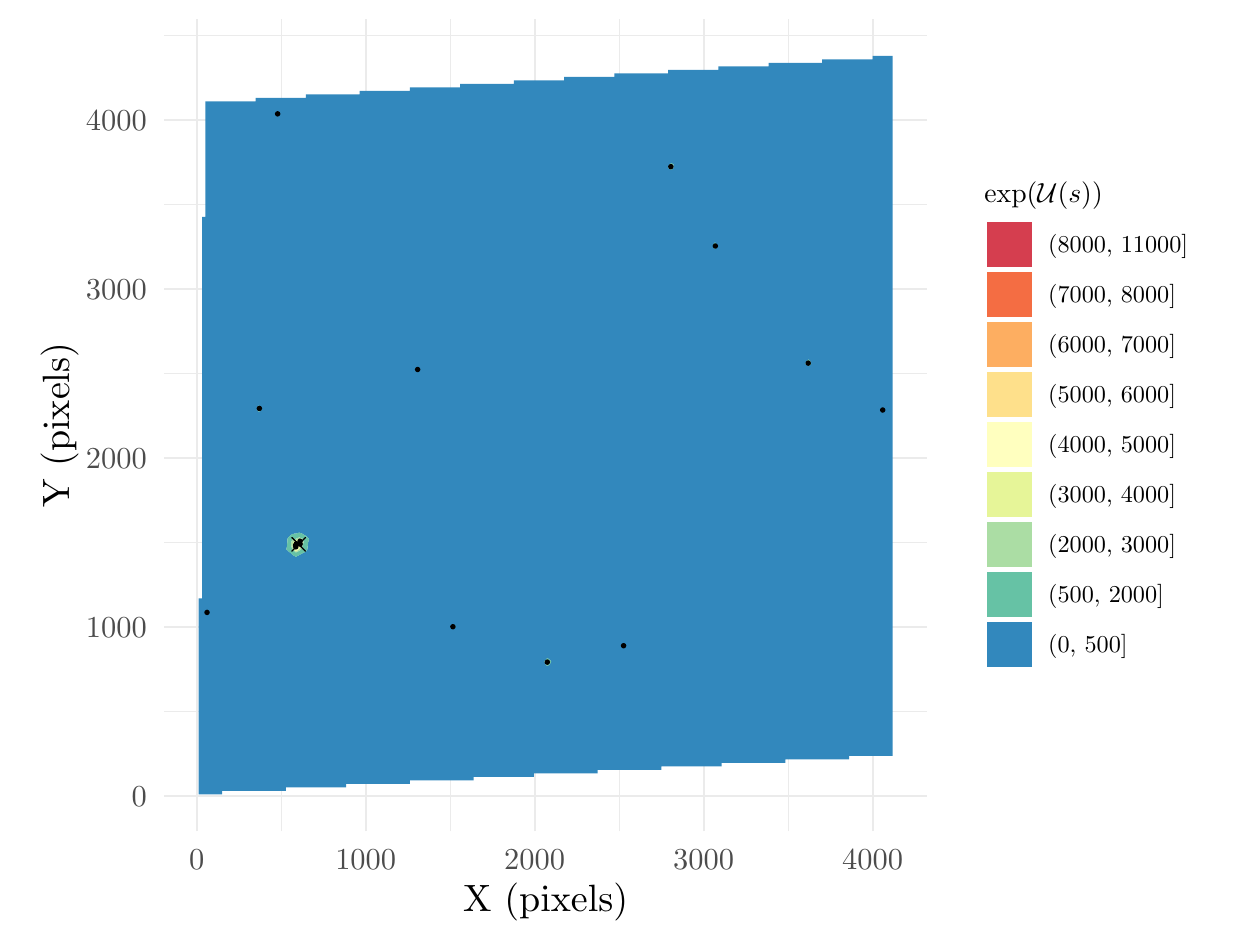}{0.5\textwidth}{(a)}
    \fig{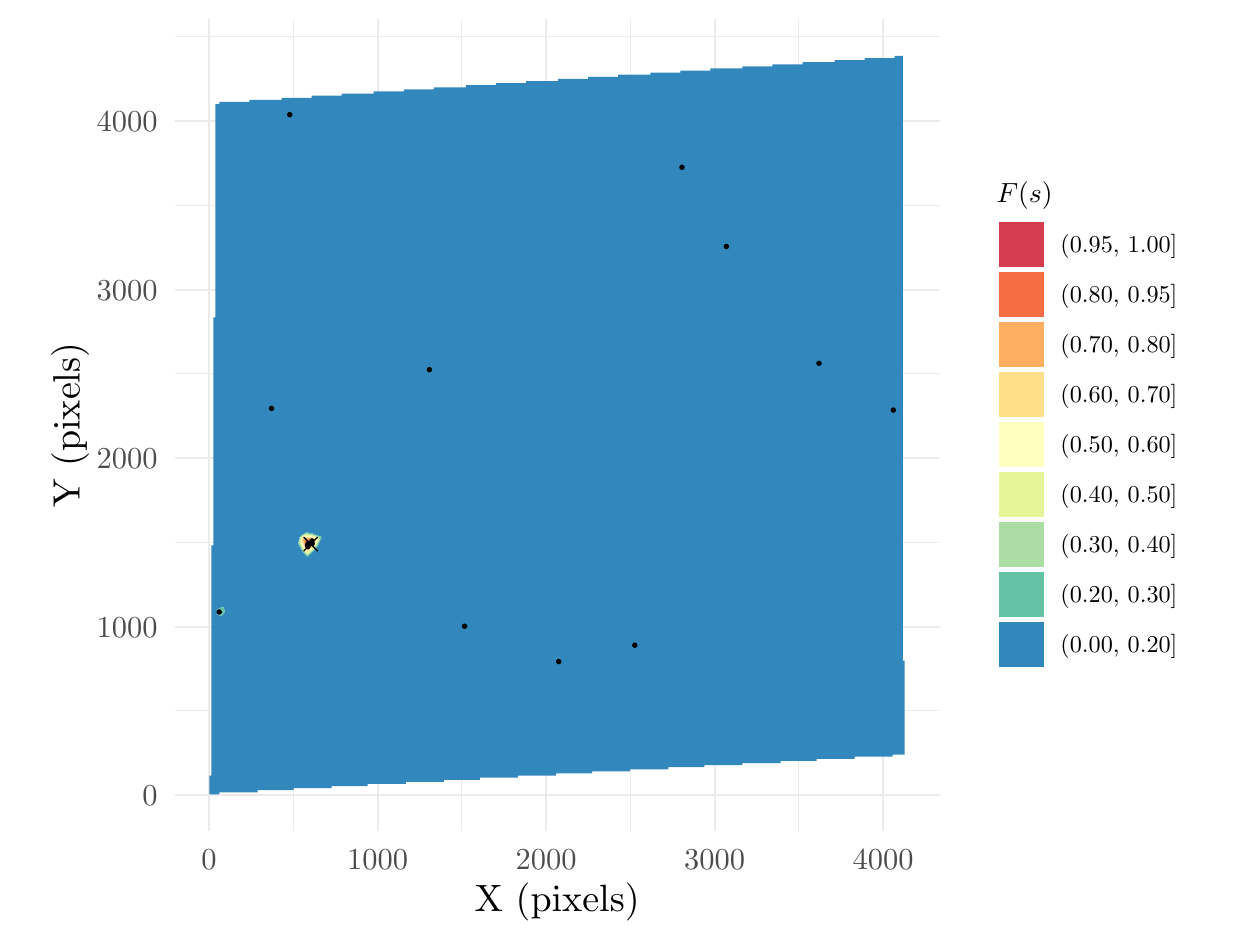}{0.5\textwidth}{(b)}}
    \caption{Posterior results for V7-WFC3 pointing. (a) Posterior mean spatial random effect $\exp(\mathcal{U}(s))$; (b) Excursion function $F_C(s)$ with $C = Q(0.95)$. Black points are locations of GC candidates. Black cross is the location of previously known UDG candidate.}
    \label{fig:v7wfc3_results}
\end{figure*}

\begin{figure*}[ht]
    \gridline{\fig{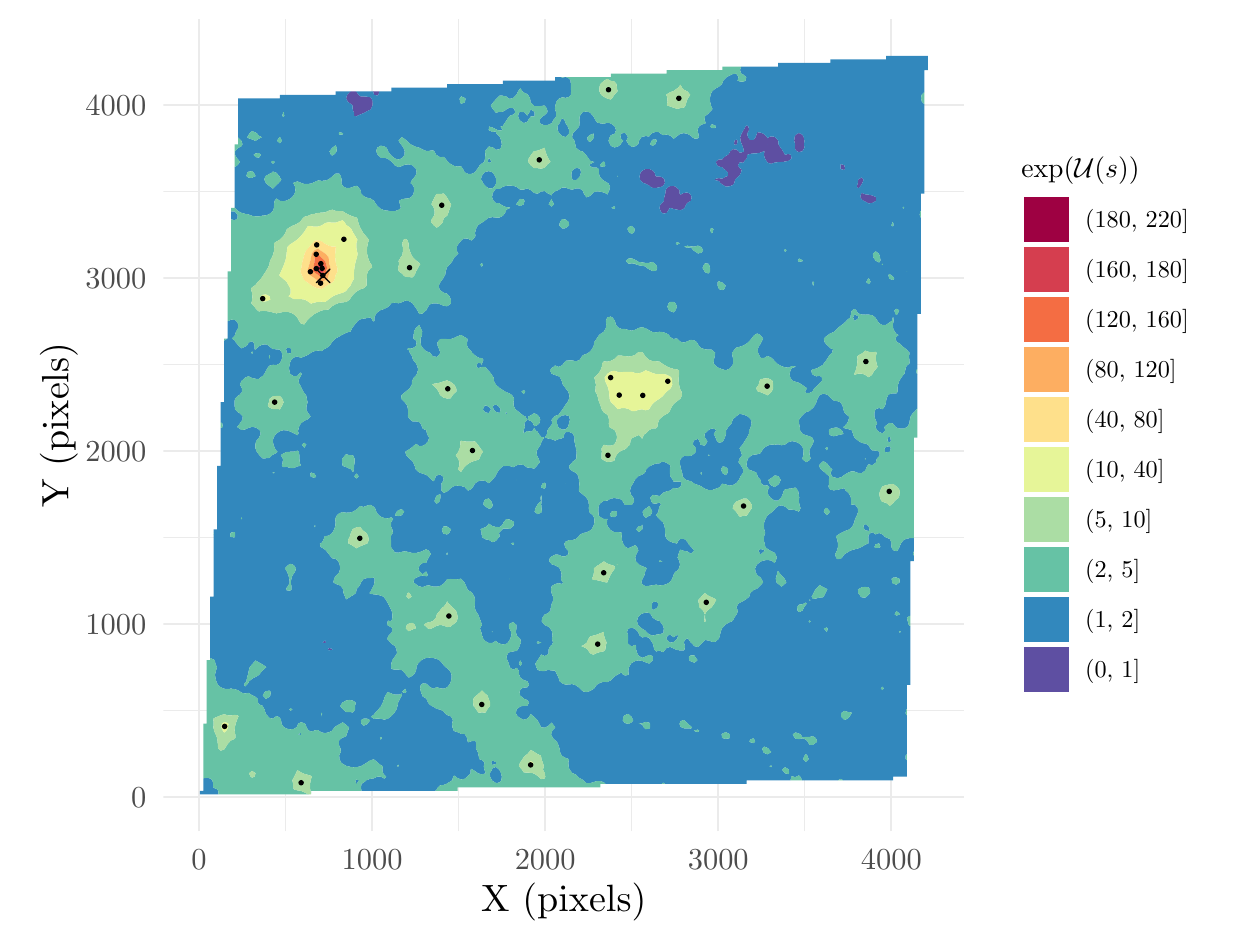}{0.5\textwidth}{(a)}
    \fig{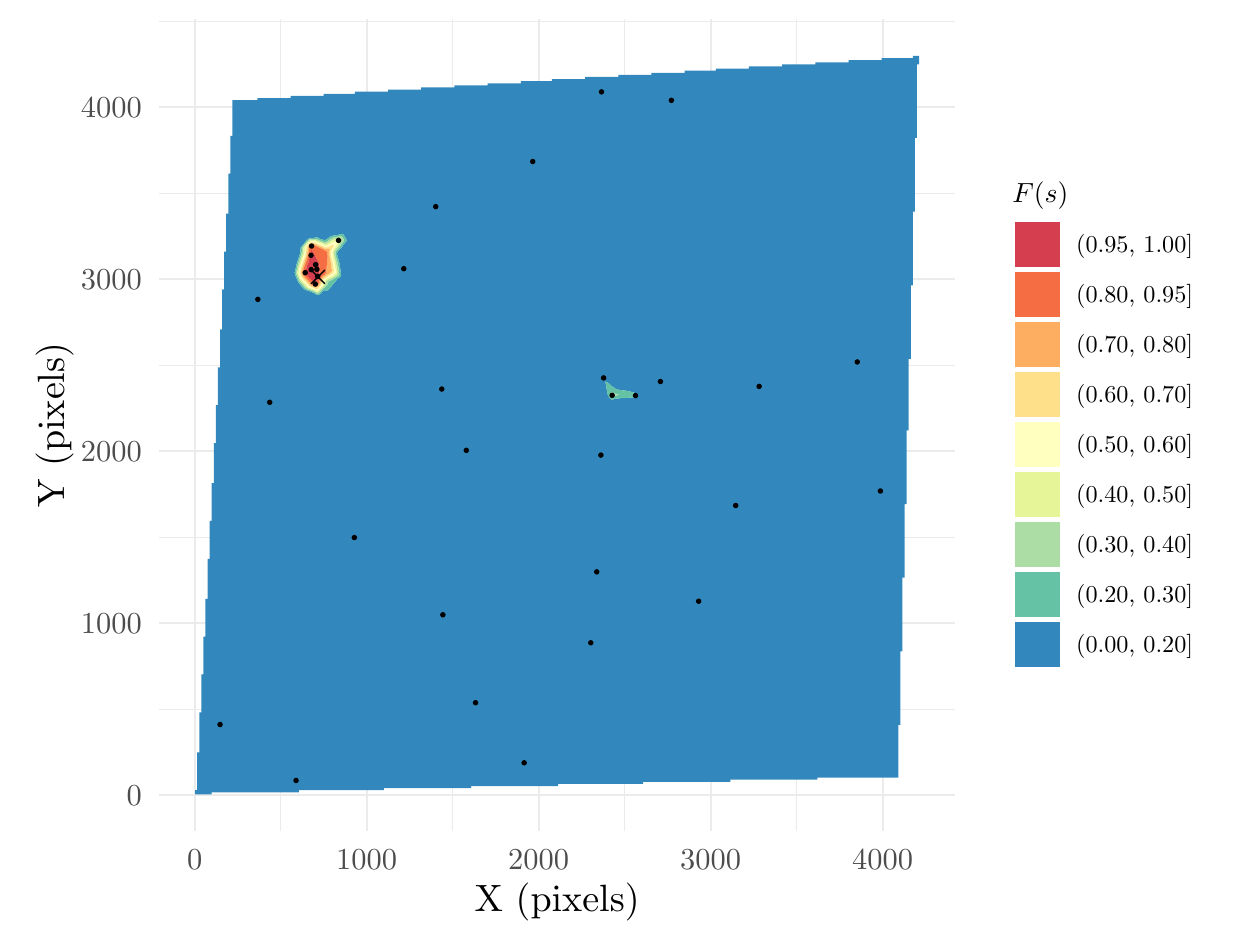}{0.5\textwidth}{(b)}}
    \caption{Posterior results for V9-ACS pointing. (a) Posterior mean spatial random effect $\exp(\mathcal{U}(s))$; (b) Excursion function $F_C(s)$ with $C = Q(0.95)$. Black points are locations of GC candidates. Black cross is the location of previously known UDG candidate.}
    \label{fig:v9acs_results}
\end{figure*}

\begin{figure*}[ht]
    \gridline{\fig{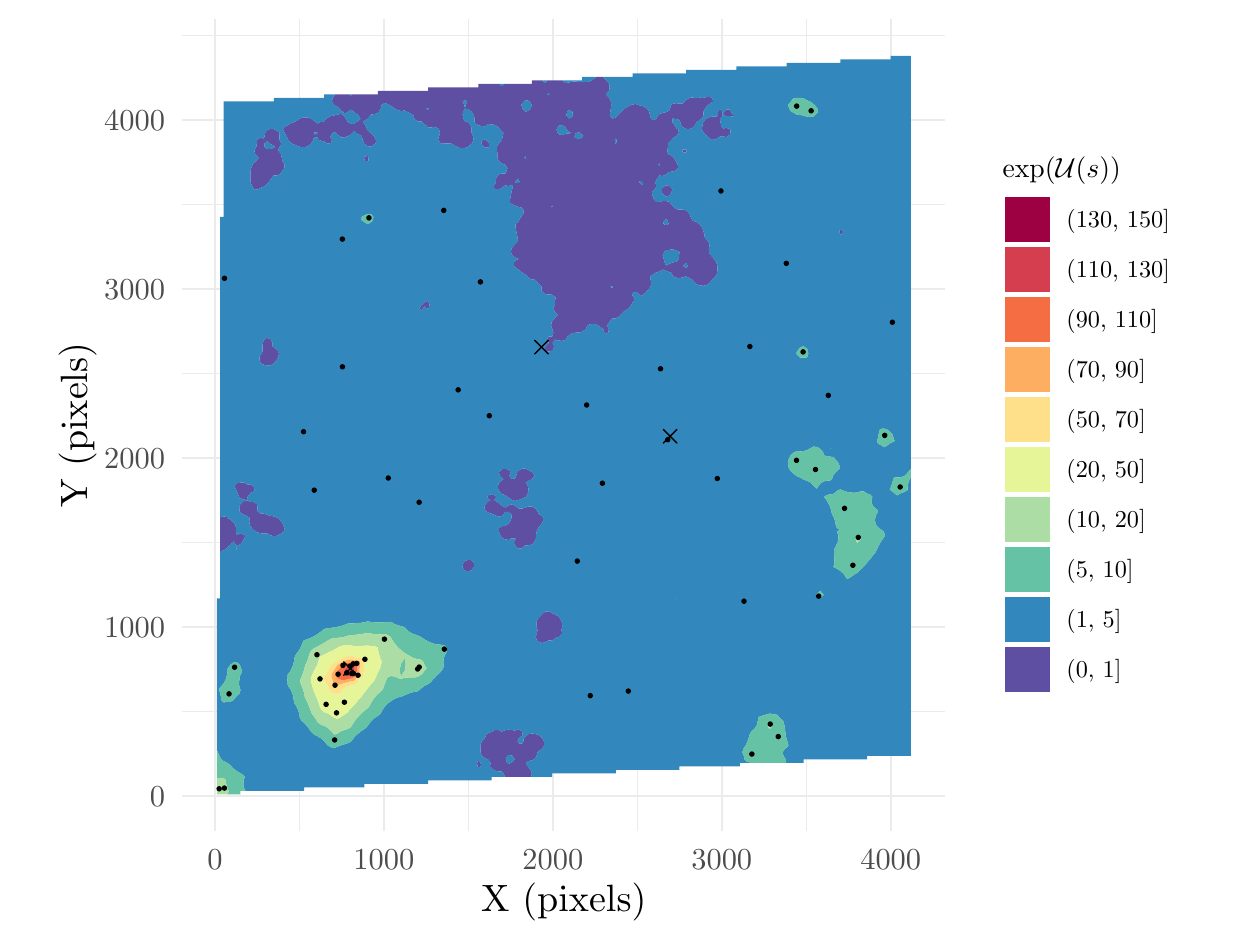}{0.5\textwidth}{(a)}
    \fig{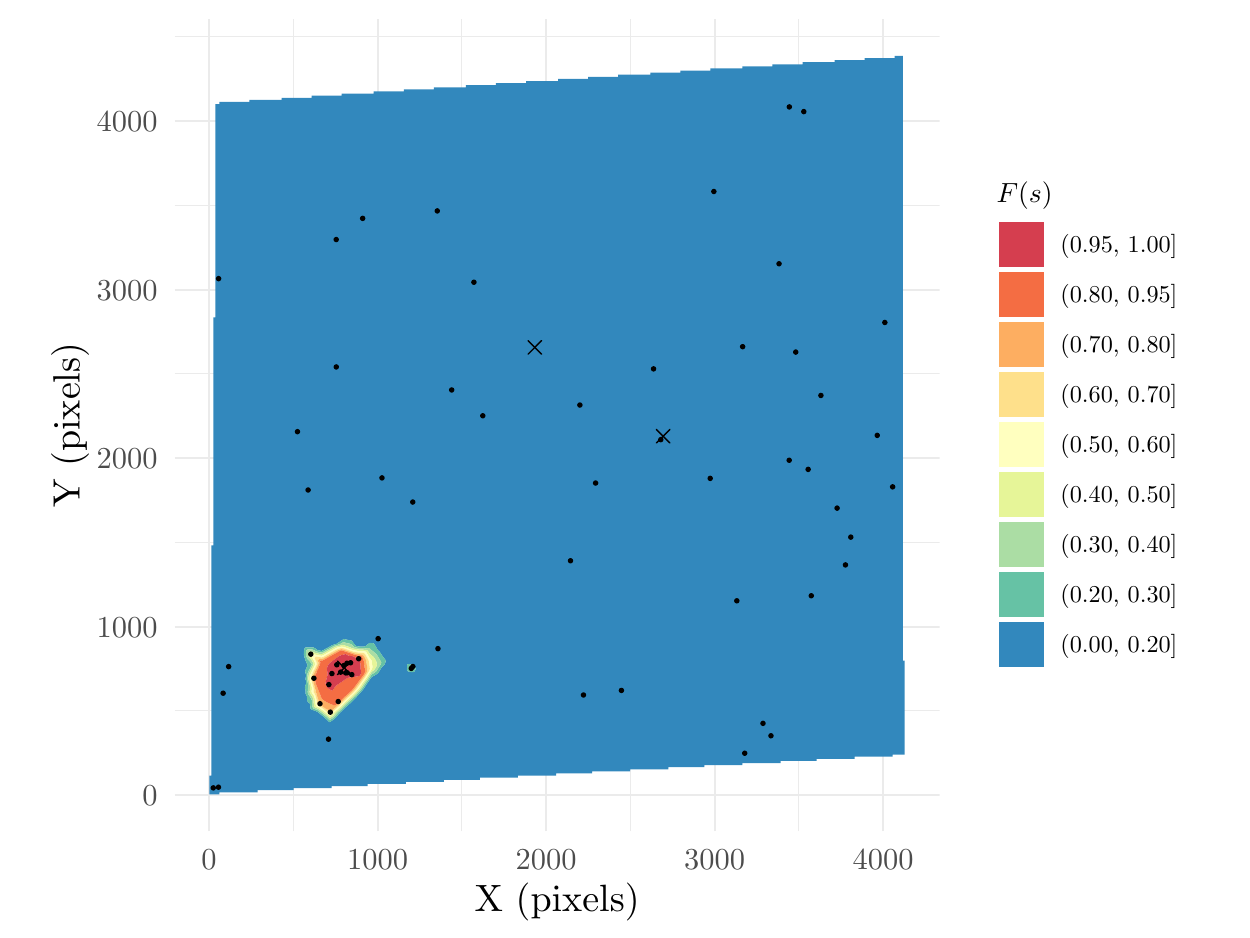}{0.5\textwidth}{(b)}}
    \caption{Posterior results for V10-WFC3 pointing. (a) Posterior mean spatial random effect $\exp(\mathcal{U}(s))$; (b) Excursion function $F_C(s)$ with $C = Q(0.95)$. Black points are locations of GC candidates. Black crosses are the locations of previously known UDG candidates.}
    \label{fig:v10wfc3_results}
\end{figure*}

\newpage
\section{Prior Distribution Parameter for \texorpdfstring{$R_k$}{Lg}}\label{sec:Rk}

\begin{deluxetable*}{lcclccc}[ht]

\tablecaption{Galaxy ID, coordinates, visit pointings, half-light radius $R_e$, and mean of prior distribution for $R_k$. Galaxies listed here are all contaminating galaxies in pointings with detected UDGs. \label{tab:Rk_prior}}
\tablenum{4}

\tablehead{\colhead{ID} & \colhead{R.A.} & \colhead{Dec.} & \colhead{Pointing} & \colhead{$R_e$} & \colhead{Mean $R_k (\times R_e)$}\\ 
\colhead{} & \colhead{(J2000)} & \colhead{(J2000)} & \colhead{} & \colhead{(kpc)} & \colhead{(kpc)}} 
\startdata
IC~312 & 03 18 08.41 & +41 45 15.75 & V8-ACS & 2.0 & 10.0 ($5\times$) \\
SDSS J031813.09+414809.0 & 03 18 13.20 & +41 48 09.17 & V8-ACS & 1.3 & 3.9 ($3\times$) \\
LEDA 12319 \tablenotemark{a} & 03 18 38.91 & +41 40 08.19 & V8-WFC3 & 2.2 & 11.0 ($5\times$) \\
LEDA 12315 & 03 18 35.39 & 	+41 40 01.03 & V8-WFC3 & 1.0 & 3.0 ($3\times$) \\
$\left[\rm BM99\right]$-317 \tablenotemark{b} & 03 20 19.11 & +41 42 42.90 & V10-ACS & 3.4 & 10.2 ($3\times$) \\
PGC 012437 & 03 19 52.97 & +41 18 08.52 & V11-ACS & 2.6 & 13.0 ($5\times$) \\
UGC 02673 & 03 20 01.64 & +41 15 04.15 & V11-ACS & 3.3 & 16.5 ($5\times$) \\
PCC 3164 & 03 18 52.25 & +41 16 02.94 & V15-ACS & 3.5 & 10.5  ($3\times$) \\
LEDA 12356 & 03 19 04.30 & +41 13 59.00 & V15-ACS & 2.0 & 10.0 ($5\times$)
\enddata
\tablenotetext{a}{This galaxy is a dE but a simple inspection by eye shows its GC system is much more extended than its light profile. Hence, a five-times ratio is adopted.}
\tablenotetext{b}{[BM99]-NNN catalog was obtained by \cite{Brunzendorf1999}.}
\end{deluxetable*}


\end{CJK*}
\end{document}